%%%%%%%%%%%%%%%%%%%%%%%%%%%%%%%%%%%%%%%%%%%%%
%%%%%%%%%%%%%%%%%%  tex macros for preprints, cm version %%%%%%%%%%%%%%
%         (P. Ginsparg <ginsparg@lanl.gov>, last updated 7/94)
%                if confused, type `b' in response to query 
%           hypertex extensions (still provisional), 7/26/94
%
%---------------------------------------------------------------------%
%\input hyperbasics %comment out this line to restore non-hyper functionality
%
%% site dependent options:
%% \unredoffs and \redoffs define horizontal and vertical offsets
%% respectively for unreduced and reduced modes. \speclscape defines
%% the \special{} call that sets printer to landscape (sideways) mode.
%% from standard set below, leave uncommented as appropriate or redefine
%
%%% next 400dpi
\def\unredoffs{} \def\redoffs{\voffset=-.31truein\hoffset=-.48truein}
\def\speclscape{}
%\def\speclscape{\special{papersize=11in,8.5in}}
%
%%% apple lw
%\def\unredoffs{} \def\redoffs{\voffset=-.31truein\hoffset=-.59truein}
%\def\speclscape{\special{ps: landscape}}
%
%%% qms lasergrafix:
%\def\unredoffs{} \def\redoffs{\voffset=-.4truein\hoffset=.125truein}
%\def\speclscape{\special{qms: landscape}}
%
%%% saclay A4 paper:
%\def\unredoffs{\hoffset-.14truein\voffset-.2truein}
%\def\redoffs{\voffset=-.45truein\hoffset=-.21truein}
%\def\speclscape{\special{landscape}}
%
%---------------------------------------------------------------------%
%
\newbox\leftpage \newdimen\fullhsize \newdimen\hstitle \newdimen\hsbody
\tolerance=1000\hfuzz=2pt
\catcode`\@=11 % This allows us to modify PLAIN macros.
\ifx\hyperdef\UNd@FiNeD\def\hyperdef#1#2#3#4{#4}\def\hyperref#1#2#3#4{#4}\fi
\def\bigans{b }
\def\answ{b }
%\message{ big or little (b/l)? }\read-1 to\answ
%
\ifx\answ\bigans\message{(This will come out unreduced.}
\magnification=1200\unredoffs\baselineskip=16pt plus 2pt minus 1pt
\hsbody=\hsize \hstitle=\hsize %take default values for unreduced format
\else\message{(This will be reduced.} \let\l@r=L
\magnification=1000\baselineskip=16pt plus 2pt minus 1pt \vsize=7truein
\redoffs \hstitle=8truein\hsbody=4.75truein\fullhsize=10truein\hsize=\hsbody
\output={\ifnum\pageno=0 %%% This is the HUTP version
  \shipout\vbox{\speclscape{\hsize\fullhsize\makeheadline}
    \hbox to \fullhsize{\hfill\pagebody\hfill}}\advancepageno
  \else
  \almostshipout{\leftline{\vbox{\pagebody\makefootline}}}\advancepageno
  \fi}
\def\almostshipout#1{\if L\l@r \count1=1 \message{[\the\count0.\the\count1]}
      \global\setbox\leftpage=#1 \global\let\l@r=R
 \else \count1=2
  \shipout\vbox{\speclscape{\hsize\fullhsize\makeheadline}
      \hbox to\fullhsize{\box\leftpage\hfil#1}}  \global\let\l@r=L\fi}
\fi
%---------------------------------------------------------------------
%
\newcount\yearltd\yearltd=\year\advance\yearltd by -2000

\def\Title#1#2{\nopagenumbers\abstractfont\hsize=\hstitle\rightline{#1}%
\vskip 1in\centerline{\titlefont #2}\abstractfont\vskip .5in\pageno=0}
\def\Date#1{\vfill\leftline{#1}\tenpoint\supereject\global\hsize=\hsbody%
\footline={\hss\tenrm\hyperdef\hypernoname{page}\folio\folio\hss}}%
% (restores pagenumbers)
%
%       use following instead of \Date on the preliminary draft,
%       puts date/time on each page in big mode, writes labels in margins

\def\draftmode{\message{ DRAFTMODE }\def\draftdate{{\rm preliminary draft:
\number\month/\number\day/{0}\number\yearltd\ \ \hourmin}}%
\headline={\hfil\draftdate}\writelabels\baselineskip=20pt plus 2pt minus 2pt
 {\count255=\time\divide\count255 by 60 \xdef\hourmin{\number\count255}
  \multiply\count255 by-60\advance\count255 by\time
  \xdef\hourmin{\hourmin:\ifnum\count255<10 0\fi\the\count255}}}
%       use \nolabels to get rid of eqn, ref, and fig labels in draft mode
\def\nolabels{\def\wrlabeL##1{}\def\eqlabeL##1{}\def\reflabeL##1{}}
\def\writelabels{\def\wrlabeL##1{\leavevmode\vadjust{\rlap{\smash%
{\line{{\escapechar=` \hfill\rlap{\sevenrm\hskip.03in\string##1}}}}}}}%
\def\eqlabeL##1{{\escapechar-1\rlap{\sevenrm\hskip.05in\string##1}}}%
\def\reflabeL##1{\noexpand\llap{\noexpand\sevenrm\string\string\string##1}}}
\nolabels
%
% tagged sec numbers
\global\newcount\secno \global\secno=0
\global\newcount\meqno \global\meqno=1
\def\s@csym{}
\def\newsec#1{\global\advance\secno by1%
{\toks0{#1}\message{(\the\secno. \the\toks0)}}%
%\ifx\answ\bigans \vfill\eject \else \bigbreak\bigskip \fi  %if desired
\global\subsecno=0\eqnres@t\let\s@csym\secsym\xdef\secn@m{\the\secno}\noindent
{\bf\hyperdef\hypernoname{section}{\the\secno}{\the\secno.} #1}%
\writetoca{{\string\hyperref{}{section}{\the\secno}{\it\the\secno.}} {{\it #1} }}%
\par\nobreak\medskip\nobreak}
\def\eqnres@t{\xdef\secsym{\the\secno.}\global\meqno=1\bigbreak\bigskip}
\def\sequentialequations{\def\eqnres@t{\bigbreak}}\xdef\secsym{}
\global\newcount\subsecno \global\subsecno=0
\def\subsec#1{\global\advance\subsecno by1%
{\toks0{#1}\message{(\s@csym\the\subsecno. \the\toks0)}}%
\ifnum\lastpenalty>9000\else\bigbreak\fi       \global\subsubsecno=0
\noindent{\it\hyperdef\hypernoname{subsection}{\secn@m.\the\subsecno}%
{\secn@m.\the\subsecno.} #1}\writetoca{\string\quad
{\string\hyperref{}{subsection}{\secn@m.\the\subsecno}{\secn@m.\the\subsecno.}}
{#1}}\par\nobreak\medskip\nobreak}
\def\appendix#1#2{\global\meqno=1\global\subsecno=0\xdef\secsym{\hbox{#1.}}%
\bigbreak\bigskip\noindent{\bf Appendix \hyperdef\hypernoname{appendix}{#1}%
{#1.} #2}{\toks0{(#1. #2)}\message{\the\toks0}}%
\xdef\s@csym{#1.}\xdef\secn@m{#1}%
\writetoca{\string\hyperref{}{appendix}{#1}{{\it Appendix} {\it #1.}} {\it #2}}%
\par\nobreak\medskip\nobreak}
%
%       \eqn\label{a+b=c}	gives displayed equation, numbered
%				consecutively within sections.
%     \eqnn and \eqna define labels in advance (of eqalign?)
%
\def\checkm@de#1#2{\ifmmode{\def\f@rst##1{##1}\hyperdef\hypernoname{equation}%
{#1}{#2}}\else\hyperref{}{equation}{#1}{#2}\fi}
\def\eqnn#1{\DefWarn#1\xdef #1{(\noexpand\relax\noexpand\checkm@de%
{\s@csym\the\meqno}{\secsym\the\meqno})}%
\wrlabeL#1\writedef{#1\leftbracket#1}\global\advance\meqno by1}
\def\f@rst#1{\c@t#1a\em@ark}\def\c@t#1#2\em@ark{#1}
\def\eqna#1{\DefWarn#1\wrlabeL{#1$\{\}$}%
\xdef #1##1{(\noexpand\relax\noexpand\checkm@de%
{\s@csym\the\meqno\noexpand\f@rst{##1}}{\hbox{$\secsym\the\meqno##1$}})}
\writedef{#1\numbersign1\leftbracket#1{\numbersign1}}\global\advance\meqno by1}
\def\eqn#1#2{\DefWarn#1%
\xdef #1{(\noexpand\hyperref{}{equation}{\s@csym\the\meqno}%
{\secsym\the\meqno})}$$#2\eqno(\hyperdef\hypernoname{equation}%
{\s@csym\the\meqno}{\secsym\the\meqno})\eqlabeL#1$$%
\writedef{#1\leftbracket#1}\global\advance\meqno by1}
\def\xeqn{\expandafter\xe@n}\def\xe@n(#1){#1}
\def\xeqna#1{\expandafter\xe@n#1}
\def\eqns#1{(\e@ns #1{\hbox{}})}
\def\e@ns#1{\ifx\UNd@FiNeD#1\message{eqnlabel \string#1 is undefined.}%
\xdef#1{(?.?)}\fi{\let\hyperref=\relax\xdef\next{#1}}%
\ifx\next\em@rk\def\next{}\else%
\ifx\next#1\xeqn#1\else\def\n@xt{#1}\ifx\n@xt\next#1\else\xeqna#1\fi
\fi\let\next=\e@ns\fi\next}

\def\DefWarn#1{\ifx\UNd@FiNeD#1\else
\immediate\write16{*** WARNING: the label \string#1 is already defined ***}\fi}
%
%			 footnotes
\newskip\footskip\footskip14pt plus 1pt minus 1pt %sets footnote baselineskip
\def\footnotefont{\ninepoint}\def\f@t#1{\footnotefont #1\@foot}
\def\f@@t{\baselineskip\footskip\bgroup\footnotefont\aftergroup\@foot\let\next}
\setbox\strutbox=\hbox{\vrule height9.5pt depth4.5pt width0pt}
\global\newcount\ftno \global\ftno=0
\def\foot{\global\advance\ftno by1\def\foot@rg{\hyperref{}{footnote}%
{\the\ftno}{\the\ftno}\xdef\foot@rg{\noexpand\hyperdef\noexpand\hypernoname%
{footnote}{\the\ftno}{\the\ftno}}}\footnote{$^{\foot@rg}$}}
%
%say \footend to put footnotes at end
%will cause problems if \ref used inside \foot, instead use \nref before
\newwrite\ftfile
\def\footend{\def\foot{\global\advance\ftno by1\chardef\wfile=\ftfile
%%$^{\the\ftno}$\ifnum\ftno=1\immediate\openout\ftfile=\jobname.fts\fi%
\hyperref{}{footnote}{\the\ftno}{$^{\the\ftno}$}%
\ifnum\ftno=1\immediate\openout\ftfile=\jobname.fts\fi%
\immediate\write\ftfile{\noexpand\smallskip%
%%\noexpand\item{f\the\ftno:\ }\pctsign}\findarg}%
\noexpand\item{\noexpand\hyperdef\noexpand\hypernoname{footnote}
{\the\ftno}{f\the\ftno}:\ }\pctsign}\findarg}%
\def\footatend{\vfill\eject\immediate\closeout\ftfile{\parindent=20pt
\centerline{\bf Footnotes}\nobreak\bigskip\input \jobname.fts }}}
\def\footatend{}
%
%     \ref\label{text}
% generates a number, assigns it to \label, generates an entry.
% To list the refs on a separate page,  \listrefs
%
\global\newcount\refno \global\refno=1
\newwrite\rfile
\def\ref{[\hyperref{}{reference}{\the\refno}{\the\refno}]\nref}
\def\nref#1{\DefWarn#1%
\xdef#1{[\noexpand\hyperref{}{reference}{\the\refno}{\the\refno}]}%
\writedef{#1\leftbracket#1}%
\ifnum\refno=1\immediate\openout\rfile=\jobname.refs\fi
\chardef\wfile=\rfile\immediate\write\rfile{\noexpand\item{[\noexpand\hyperdef%
\noexpand\hypernoname{reference}{\the\refno}{\the\refno}]\ }%
\reflabeL{#1\hskip.31in}\pctsign}\global\advance\refno by1\findarg}
%	horrible hack to sidestep tex \write limitation
\def\findarg#1#{\begingroup\obeylines\newlinechar=`\^^M\pass@rg}
{\obeylines\gdef\pass@rg#1{\writ@line\relax #1^^M\hbox{}^^M}%
\gdef\writ@line#1^^M{\expandafter\toks0\expandafter{\striprel@x #1}%
\edef\next{\the\toks0}\ifx\next\em@rk\let\next=\endgroup\else\ifx\next\empty%
\else\immediate\write\wfile{\the\toks0}\fi\let\next=\writ@line\fi\next\relax}}
\def\striprel@x#1{} \def\em@rk{\hbox{}}
\def\lref{\begingroup\obeylines\lr@f}
\def\lr@f#1#2{\DefWarn#1\gdef#1{\let#1=\UNd@FiNeD\ref#1{#2}}\endgroup\unskip}

\def\addref#1{\immediate\write\rfile{\noexpand\item{}#1}} %now unnecessary
\def\listrefs{\footatend\vfill\supereject\immediate\closeout\rfile\writestoppt
\baselineskip=\footskip\centerline{{\bf References}}\bigskip{\parindent=20pt%
\frenchspacing\escapechar=` \input \jobname.refs\vfill\eject}\nonfrenchspacing}
\def\startrefs#1{\immediate\openout\rfile=\jobname.refs\refno=#1}
\def\xref{\expandafter\xr@f}\def\xr@f[#1]{#1}
\def\refs#1{\count255=1[\r@fs #1{\hbox{}}]}
\def\r@fs#1{\ifx\UNd@FiNeD#1\message{reflabel \string#1 is undefined.}%
\nref#1{need to supply reference \string#1.}\fi%
\vphantom{\hphantom{#1}}{\let\hyperref=\relax\xdef\next{#1}}%
\ifx\next\em@rk\def\next{}%
\else\ifx\next#1\ifodd\count255\relax\xref#1\count255=0\fi%
\else#1\count255=1\fi\let\next=\r@fs\fi\next}
%

%
% this is ugly, but moore insists
\newwrite\ffile\global\newcount\figno \global\figno=1
\def\fig{fig.~\hyperref{}{figure}{\the\figno}{\the\figno}\nfig}
\def\nfig#1{\DefWarn#1%
\xdef#1{fig.~\noexpand\hyperref{}{figure}{\the\figno}{\the\figno}}%
\writedef{#1\leftbracket fig.\noexpand~\xfig#1}%
\ifnum\figno=1\immediate\openout\ffile=\jobname.figs\fi\chardef\wfile=\ffile%
{\let\hyperref=\relax
\immediate\write\ffile{\noexpand\medskip\noexpand\item{Fig.\ %
\noexpand\hyperdef\noexpand\hypernoname{figure}{\the\figno}{\the\figno}. }
\reflabeL{#1\hskip.55in}\pctsign}}\global\advance\figno by1\findarg}
\def\listfigs{\vfill\eject\immediate\closeout\ffile{\parindent40pt
\baselineskip14pt\centerline{{\bf Figure Captions}}\nobreak\medskip
\escapechar=` \input \jobname.figs\vfill\eject}}
\def\xfig{\expandafter\xf@g}\def\xf@g fig.\penalty\@M\ {}
\def\figs#1{figs.~\f@gs #1{\hbox{}}}
\def\f@gs#1{{\let\hyperref=\relax\xdef\next{#1}}\ifx\next\em@rk\def\next{}\else
\ifx\next#1\xfig #1\else#1\fi\let\next=\f@gs\fi\next}
\def\figin{\epsfcheck\figin}\def\figins{\epsfcheck\figins}
\def\epsfcheck{\ifx\epsfbox\UNd@FiNeD
\message{(NO epsf.tex, FIGURES WILL BE IGNORED)}
\gdef\figin##1{\vskip2in}\gdef\figins##1{\hskip.5in}% blank space instead
\else\message{(FIGURES WILL BE INCLUDED)}%
\gdef\figin##1{##1}\gdef\figins##1{##1}\fi}
\def\DefWarn#1{}
\def\figinsert{\goodbreak\midinsert}
\def\ifig#1#2#3{\DefWarn#1\xdef#1{Fig.~\noexpand\hyperref{}{figure}%
{\the\figno}{\the\figno}}\writedef{#1\leftbracket fig.\noexpand~\xfig#1}%
\figinsert\figin{\centerline{#3}}\medskip\centerline{\vbox{\baselineskip12pt
\advance\hsize by -1truein\noindent\wrlabeL{#1=#1}\footnotefont%
{\bf Fig.~\hyperdef\hypernoname{figure}{\the\figno}{\the\figno}:} #2}}
\bigskip\endinsert\global\advance\figno by1}
\newwrite\lfile
{\escapechar-1\xdef\pctsign{\string\%}\xdef\leftbracket{\string\{}
\xdef\rightbracket{\string\}}\xdef\numbersign{\string\#}}
\def\writedefs{\immediate\openout\lfile=\jobname.defs \def\writedef##1{%
{\let\hyperref=\relax\let\hyperdef=\relax\let\hypernoname=\relax
 \immediate\write\lfile{\string\def\string##1\rightbracket}}}}%
\def\writestop{\def\writestoppt{\immediate\write\lfile{\string\pageno
 \the\pageno\string\startrefs\leftbracket\the\refno\rightbracket
 \string\def\string\secsym\leftbracket\secsym\rightbracket
 \string\secno\the\secno\string\meqno\the\meqno}\immediate\closeout\lfile}}
\def\writestoppt{}\def\writedef#1{}
\def\seclab#1{\DefWarn#1%
\xdef #1{\noexpand\hyperref{}{section}{\the\secno}{\the\secno}}%
\writedef{#1\leftbracket#1}\wrlabeL{#1=#1}}
\def\subseclab#1{\DefWarn#1%
\xdef #1{\noexpand\hyperref{}{subsection}{\secn@m.\the\subsecno}%
{\secn@m.\the\subsecno}}\writedef{#1\leftbracket#1}\wrlabeL{#1=#1}}
\def\applab#1{\DefWarn#1%
\xdef #1{\noexpand\hyperref{}{appendix}{\secn@m}{\secn@m}}%
\writedef{#1\leftbracket#1}\wrlabeL{#1=#1}}
\newwrite\tfile \def\writetoca#1{}
\def\leaderfill{\leaders\hbox to 1em{\hss.\hss}\hfill}
%	use this to write file with table of contents
\def\writetoc{\immediate\openout\tfile=\jobname.toc
   \def\writetoca##1{{\edef\next{\write\tfile{\noindent ##1
   \string\leaderfill {\string\hyperref{}{page}{\noexpand\number\pageno}%
                       {\noexpand\number\pageno}} \par}}\next}}}
%       and this lists table of contents on second pass
\newread\ch@ckfile
\def\listtoc{\immediate\closeout\tfile\immediate\openin\ch@ckfile=\jobname.toc
\ifeof\ch@ckfile\message{no file \jobname.toc, no table of contents this pass}%
\else\closein\ch@ckfile\centerline{\bf Contents}\nobreak\medskip%
{\baselineskip=18pt  % \footnotefont
\parskip=2pt\catcode`\@=12\input\jobname.toc
\catcode`\@=12\bigbreak\bigskip}\fi}
\catcode`\@=12 % at signs are no longer letters
%
%	Unpleasantness in calling in abstract and title fonts
\edef\tfontsize{\ifx\answ\bigans scaled\magstep3\else scaled\magstep4\fi}
\font\titlerm=cmr10 \tfontsize \font\titlerms=cmr7 \tfontsize
\font\titlermss=cmr5 \tfontsize \font\titlei=cmmi10 \tfontsize
\font\titleis=cmmi7 \tfontsize \font\titleiss=cmmi5 \tfontsize
\font\titlesy=cmsy10 \tfontsize \font\titlesys=cmsy7 \tfontsize
\font\titlesyss=cmsy5 \tfontsize \font\titleit=cmti10 \tfontsize
\skewchar\titlei='177 \skewchar\titleis='177 \skewchar\titleiss='177
\skewchar\titlesy='60 \skewchar\titlesys='60 \skewchar\titlesyss='60
\def\titlefont{\def\rm{\fam0\titlerm}% switch to title font
\textfont0=\titlerm \scriptfont0=\titlerms \scriptscriptfont0=\titlermss
\textfont1=\titlei \scriptfont1=\titleis \scriptscriptfont1=\titleiss
\textfont2=\titlesy \scriptfont2=\titlesys \scriptscriptfont2=\titlesyss
\textfont\itfam=\titleit \def\it{\fam\itfam\titleit}\rm}
 \ifx\answ\bigans\else scaled\magstep1\fi
\ifx\answ\bigans\def\abstractfont{\tenpoint}\else
\font\absit=cmti10 scaled \magstep1
\font\abssl=cmsl10 scaled \magstep1
\font\absrm=cmr10 scaled\magstep1 \font\absrms=cmr7 scaled\magstep1
\font\absrmss=cmr5 scaled\magstep1 \font\absi=cmmi10 scaled\magstep1
\font\absis=cmmi7 scaled\magstep1 \font\absiss=cmmi5 scaled\magstep1
\font\abssy=cmsy10 scaled\magstep1 \font\abssys=cmsy7 scaled\magstep1
\font\abssyss=cmsy5 scaled\magstep1 \font\absbf=cmbx10 scaled\magstep1
\skewchar\absi='177 \skewchar\absis='177 \skewchar\absiss='177
\skewchar\abssy='60 \skewchar\abssys='60 \skewchar\abssyss='60
\def\abstractfont{\def\rm{\fam0\absrm}% switch to abstract font
\textfont0=\absrm \scriptfont0=\absrms \scriptscriptfont0=\absrmss
\textfont1=\absi \scriptfont1=\absis \scriptscriptfont1=\absiss
\textfont2=\abssy \scriptfont2=\abssys \scriptscriptfont2=\abssyss
\textfont\itfam=\absit \def\it{\fam\itfam\absit}\def\footnotefont{\tenpoint}%
\textfont\slfam=\abssl \def\sl{\fam\slfam\abssl}%
\textfont\bffam=\absbf \def\bf{\fam\bffam\absbf}\rm}\fi
\def\tenpoint{\def\rm{\fam0\tenrm}% switch back to 10-point type
\textfont0=\tenrm \scriptfont0=\sevenrm \scriptscriptfont0=\fiverm
\textfont1=\teni  \scriptfont1=\seveni  \scriptscriptfont1=\fivei
\textfont2=\tensy \scriptfont2=\sevensy \scriptscriptfont2=\fivesy
\textfont\itfam=\tenit \def\it{\fam\itfam\tenit}\def\footnotefont{\ninepoint}%
\textfont\bffam=\tenbf \def\bf{\fam\bffam\tenbf}\def\sl{\fam\slfam\tensl}\rm}
\font\ninerm=cmr9 \font\sixrm=cmr6 \font\ninei=cmmi9 \font\sixi=cmmi6
\font\ninesy=cmsy9 \font\sixsy=cmsy6 \font\ninebf=cmbx9
\font\nineit=cmti9 \font\ninesl=cmsl9 \skewchar\ninei='177
\skewchar\sixi='177 \skewchar\ninesy='60 \skewchar\sixsy='60
\def\ninepoint{\def\rm{\fam0\ninerm}% switch to footnote font
\textfont0=\ninerm \scriptfont0=\sixrm \scriptscriptfont0=\fiverm
\textfont1=\ninei \scriptfont1=\sixi \scriptscriptfont1=\fivei
\textfont2=\ninesy \scriptfont2=\sixsy \scriptscriptfont2=\fivesy
\textfont\itfam=\ninei \def\it{\fam\itfam\nineit}\def\sl{\fam\slfam\ninesl}%
\textfont\bffam=\ninebf \def\bf{\fam\bffam\ninebf}\rm}
%
%---------------------------------------------------------------------
%
\def\noblackbox{\overfullrule=0pt}
\hyphenation{anom-aly anom-alies coun-ter-term coun-ter-terms}
\def\inv{^{\raise.15ex\hbox{${\scriptscriptstyle -}$}\kern-.05em 1}}

\def\Dsl{\,\raise.15ex\hbox{/}\mkern-13.5mu D} %this one can be subscripted
\def\dsl{\raise.15ex\hbox{/}\kern-.57em\partial}

\def\tr{{\rm tr}} \def\Tr{{\rm Tr}}
 %pound sterling
\def\lspace{\ifx\answ\bigans{}\else\qquad\fi}
\def\lbspace{\ifx\answ\bigans{}\else\hskip-.2in\fi} % $$\lbspace...$$
\def\boxeqn#1{\vcenter{\vbox{\hrule\hbox{\vrule\kern3pt\vbox{\kern3pt
	\hbox{${\displaystyle #1}$}\kern3pt}\kern3pt\vrule}\hrule}}}
\def\mbox#1#2{\vcenter{\hrule \hbox{\vrule height#2in
		\kern#1in \vrule} \hrule}}  %e.g. \mbox{.1}{.1}
%	matters of taste
%\def\tilde{\widetilde} \def\bar{\overline} \def\hat{\widehat}
%
% some sample definitions
  %     curly letters

\def\vev#1{\langle #1 \rangle}

\def\darr#1{\raise1.5ex\hbox{$\leftrightarrow$}\mkern-16.5mu #1}
 %pound sterling

 %puts a small half in a displayed eqn
\def\roughly#1{\raise.3ex\hbox{$#1$\kern-.75em\lower1ex\hbox{$\sim$}}}

%%%%%%%%%%%%%%%%%%%%%%%%%%%%%%%%%%%%%%%%%%%%%%%%%%%%%%%%%%%%%%%%%%%%%
%%%%%%%%%%%%%%%   Subsubsection  %%%%%%%%%%%%%%%%%%%%%%%%%%%%%%%%%%%%
%%%%%%%%%%%%%%%%%%%%%%%%%%%%%%%%%%%%%%%%%%%%%%%%%%%%%%%%%%%%%%%%%%%%%
\global\newcount\subsubsecno \global\subsubsecno=0
\def\subsubsec#1{\global\advance\subsubsecno by1%
{\toks0{#1}\message{(\the\secno\the\subsecno\the\subsubsecno. \the\toks0)}}%
\ifnum\lastpenalty>9000\else\bigbreak\fi
\noindent{\it\hyperdef\hypernoname{subsubsection}{\the\secno.\the\subsecno\the\subsubsecno}%
{\the\secno.\the\subsecno.\the\subsubsecno.} #1}
%%% Add Subsubsections to Index
%% \writetoca{\string\quad{\string\hyperref{}{subsubsection}{\the\secno\the\subsecno\the
%%\subsubsecno}{\baselineskip=9pt\it\the\secno.\the\subsecno.\the\subsubsecno.}}
%% {\baselineskip=9pt\it\ #1}}
\par\nobreak\medskip\nobreak}
%%%%%%%%%%%%%%%%%%%%%%%%%%%%%%%%%%%%%%%%%%%%%%%%%%%%%%%%%%%%%%%%%%%%%
%%%%%%%%%%%%%%%%%%%%%%%%%%%%%%%%%%%%%%%%%%%%%%%%%%%%%%%%%%%%%%%%%%%
%%%%%% BOX
%%%%%%%%%%%%%%%%%%%%%%%%%%%%%%%%%%%%%%%%%%
\def\boxit#1{\vbox{\hrule\hbox{\vrule\kern8pt
\vbox{\hbox{\kern8pt}\hbox{\vbox{#1}}\hbox{\kern8pt}}
\kern8pt\vrule}\hrule}}
\def\mathboxit#1{\vbox{\hrule\hbox{\vrule\kern8pt\vbox{\kern8pt
\hbox{$\displaystyle #1$}\kern8pt}\kern8pt\vrule}\hrule}}
%%%%%%%%%%%%%%%%%%%%%%%%%%%%%%%%%%%%%%%%%%%%%%%%%%%%%%%%%%%%%%%%%%%
%%%%%%%%%%%%%%%%%%%%%%%%%%%%%%%%%%%%%%%%%%%%%%%%%%%%%%%%%%%%%%%%
%%%%%   Dirac-Slash
%%%%%%%%%%%%%%%%%%%%%%%%%%%%%%%%%%%%%%%%%%%%%%%%%%%%%%%%%%%%%%%%
\def\slashchar#1{\setbox0=\hbox{$#1$}           % set a box for #1
   \dimen0=\wd0                                 % and get its size
   \setbox1=\hbox{/} \dimen1=\wd1               % get size of /
   \ifdim\dimen0>\dimen1                        % #1 is bigger
      \rlap{\hbox to \dimen0{\hfil/\hfil}}      % so center / in box
      #1                                        % and print #1
   \else                                        % / is bigger
      \rlap{\hbox to \dimen1{\hfil$#1$\hfil}}   % so center #1
      /                                         % and print /
   \fi}
%%%%%%%%%%%%%%%%%%%%%%%%%%%%%%%%%%%%%%%%%%%%%%%%%%%%%%%%%%%%%%%%%
%%%%%%%%%%%%%%%%%%%%%%%%%%%%%%%%%%%%%%%%%%%%%%%%%%%%%%%%%%%
%  To produce a box for a Dalembertian (adapted from p. 320 of TeXbook):
\def\sqr#1#2{{\vcenter{\vbox{\hrule height.#2pt
         \hbox{\vrule width.#2pt height#1pt \kern#1pt
            \vrule width.#2pt}
         \hrule height.#2pt}}}}

%%%%%%%%%%%%%%%%%%%%%%%%%%%%%%%%%%%%%%%%%%%%%%%%%%%%%%%%%%%

% \draftmode
\noblackbox %%%%%%%%%%%%%%%%% Lineskip
%%%%%%%%%%%%%%%%%%%%%%%%%%%%%%%%%%%%%%
\ifx\answ\bigans
\magnification=1200\baselineskip=14pt plus 2pt minus 1pt
\else\baselineskip=16pt % plus 2pt minus 1pt % 32 lines in l-format
\fi

%%%%%%%%%%%% Local definitons %%%%%%%%%%%%%%%%%%%%%%%%%%%%%%%%%%%%

\def\gs{g_{\rm string}}

\def\ap{\alpha'}
\def\Ms{M_{\rm string}}

\def\gev{{\rm GeV}}

\def\cf{{cf.\ }}
\def\ie{{i.e.\ }}
\def\eg{{e.g.\ }}
\def\eqq{{Eq.\ }}
\def\eqqs{{Eqs.\ }}
\def\th{\theta}

\def\eps{\epsilon}
\def\al{\alpha}

\def\si{\sigma}

\def\Si{{\Sigma}}
\def\ov{{\overline}}

\def\comment#1{{}}

\def\FF#1#2{{_#1F_#2}}

%New definitions
\def\ov {\overline}
\def\1{{\bf 1}}
\def\3{{\bf 3}}
\def\2{{\bf 2}}
\def\4{{\bf 4}}
\def\5{{\bf 5}}
\def\6{{\bf 6}}
\def\8{{\bf 8}}

%%%%%%%%%%%%%%%%%%%%%%%%%%%%%%%%%%%%%%%%%%%%%%%%%%%%%%%%%%%%%%%%
%%%%%   Dirac-Slash
%%%%%%%%%%%%%%%%%%%%%%%%%%%%%%%%%%%%%%%%%%%%%%%%%%%%%%%%%%%%%%%%
\def\slashchar#1{\setbox0=\hbox{$#1$}           % set a box for #1
 \dimen0=\wd0                                 % and get its size
 \setbox1=\hbox{/} \dimen1=\wd1               % get size of /
 \ifdim\dimen0>\dimen1                        % #1 is bigger
    \rlap{\hbox to \dimen0{\hfil/\hfil}}      % so center / in box
    #1                                        % and print #1
 \else                                        % / is bigger
    \rlap{\hbox to \dimen1{\hfil$#1$\hfil}}   % so center #1
    /                                         % and print /
 \fi}
%%%%%%%%%%%%%%%%%%%%%%%%%%%%%%%%%%%%%%%%%%%%%%%%%%%%%%%%%%%%%%%%%
%%%%%%%%%%%%%%%%%%%%%%%%%%%%%%%%%%%%%%%%%%%%%%%%%%%%%%%%%%%%%%%%%%%
%%%%%% BOX
%%%%%%%%%%%%%%%%%%%%%%%%%%%%%%%%%%%%%%%%%%
\def\boxit#1{\vbox{\hrule\hbox{\vrule\kern8pt
\vbox{\hbox{\kern8pt}\hbox{\vbox{#1}}\hbox{\kern8pt}}
\kern8pt\vrule}\hrule}}
\def\mathboxit#1{\vbox{\hrule\hbox{\vrule\kern8pt\vbox{\kern8pt
\hbox{$\displaystyle #1$}\kern8pt}\kern8pt\vrule}\hrule}}
%%%%%%%%%%%%%%%%%%%%%%%%%%%%%%%%%%%%%%%%%%%%%%%%%%%%%%%%%%%%%%%%%%%
%%%%% Referencing  %%%%%%%%%%%%%%%%%%%%%%%%%%%%%%%%%%%%%%%%%%%%%%
%%%%%%%%%%%%%%%%%%%%%%%%%%%%%%%%%%%%%%%%%%%%%%%%%%%%%%%%%%%%%%%%%
\newif\ifnref
\def\rrr#1#2{\relax\ifnref\nref#1{#2}\else\ref#1{#2}\fi}
\def\ldf#1#2{\begingroup\obeylines
\gdef#1{\rrr{#1}{#2}}\endgroup\unskip}
\def\nrf#1{\nreftrue{#1}\nreffalse}

\def\multrefiv#1#2#3#4{\nrf{#1#2#3#4}\refs{#1{--}#4}}

\def\doubref#1#2{\refs{{#1},{#2} }}

\def\multrefix#1#2#3#4#5#6#7#8#9{\nrf{#1#2#3#4#5#6#7#8#9}\refs{#1{--}#9}}
\nreffalse

\def\lref{\ldf}
%%%%%%%%%%%%%%%%%%%%%%%%%%%%%%%%%%%%%%%%%%%%%%%%%%%%%%%%%%%%%%%%%%
%%%%%%%%%%%%%%%%%   Stuff for Figures  %%%%%%%%%%%%%%%%%%%%%%%%%%%
%%%%%%%%%%%%%%%%%%%%%%%%%%%%%%%%%%%%%%%%%%%%%%%%%%%%%%%%%%%%%%%%%%

\input epsf
\def\figin{\epsfcheck\figin}\def\figins{\epsfcheck\figins}
\def\epsfcheck{\ifx\epsfbox\UnDeFiNeD
\message{(NO epsf.tex, FIGURES WILL BE IGNORED)}
\gdef\figin##1{\vskip2in}\gdef\figins##1{\hskip.5in}% blank space instead
\else\message{(FIGURES WILL BE INCLUDED)}%
\gdef\figin##1{##1}\gdef\figins##1{##1}\fi}
\def\DefWarn#1{}
\def\figinsert{\goodbreak\midinsert}
\def\ifig#1#2#3{\DefWarn#1\xdef#1{Figure~\the\figno}
\writedef{#1\leftbracket fig.\noexpand~\the\figno}%
\figinsert\figin{\centerline{#3}}\medskip\centerline{\vbox{\baselineskip12pt
\advance\hsize by -1truein\noindent\footnotefont\centerline{{\bf
Fig.~\the\figno}\ #2}}}
\bigskip\endinsert\global\advance\figno by1}
%%%%%%%%  Second line in figure caption
\def\iifig#1#2#3#4{\DefWarn#1\xdef#1{Figure~\the\figno}
\writedef{#1\leftbracket fig.\noexpand~\the\figno}%
\figinsert\figin{\centerline{#4}}\medskip\centerline{\vbox{\baselineskip12pt
\advance\hsize by -1truein\noindent\footnotefont\centerline{{\bf
Fig.~\the\figno}\ #2}}}\smallskip\centerline{\vbox{\baselineskip12pt
\advance\hsize by -1truein\noindent\footnotefont\centerline{\ \ \ #3}}}
\bigskip\endinsert\global\advance\figno by1}

%%%%%%%%%%%%%%%%%%%%%%%%%%%%%%%%%%%%%%%%%%%%%%%%%%%%%%%%%%%%%%%%%%%%%
%%%%%%%%%%%%%%%   Standard alltime definitions   %%%%%%%%%%%%%%%%%%%%
%%%%%%%%%%%%%%%%%%%%%%%%%%%%%%%%%%%%%%%%%%%%%%%%%%%%%%%%%%%%%%%%%%%%%

\def\tilde{\widetilde}

\def\h {{1\over 2}}

\def\o {\over}
\def\fc#1#2{{#1 \o #2}}

\def\IZ{ {\bf Z}}\def\IQ{{\bf Q}}

\def\IR{ {\bf R}}
\def\hat{\widehat}
      % For Eisenstein E2
  % For Polylogarithm

\def\br{\hfill\break}
\def\tr {{\rm tr}}

\def\lf {\left}
\def\ri {\right}
\def\ra {\rightarrow}
\def\lra {\longrightarrow}
\def\re {{\rm Re}}
\def\im {{\rm Im}}
\def\p {\partial}

 \def\Oc {{\cal O}}
\def\Lc {{\cal L}} 
\def\Mc {{\cal M}}

\def\Ic {{\cal I}} 
\def\Kc {{\cal K}} 
  \def\Kc{{\cal K}}
%%%%%%%%%%%%%%%%%%%%%%%%%%%%%%%%%%%%%%%%%%%%%%%%%%%%%%%%%%%%%%%%%%%
%%%%%%%%%%%%%%%%%%%%%%%%%%%%%%%%%%%%%%%%%%%%%%%%%%%%%%%%%%%%%%%%%%%

\lref\AntoniadisQM{
  I.~Antoniadis, E.~Kiritsis and T.N.~Tomaras,
  ``A D-brane alternative to unification,''
  Phys.\ Lett.\  B {\bf 486}, 186 (2000)
  [arXiv:hep-ph/0004214];\br
  %%CITATION = PHLTA,B486,186;%%
I.~Antoniadis, E.~Kiritsis, J.~Rizos and T.N.~Tomaras,
  ``D-branes and the standard model,''
  Nucl.\ Phys.\  B {\bf 660}, 81 (2003)
  [arXiv:hep-th/0210263].
  %%CITATION = NUPHA,B660,81;%%
}

\lref\KiritsisAJ{
  E.~Kiritsis and P.~Anastasopoulos,
  ``The anomalous magnetic moment of the muon in the D-brane realization of
  the standard model,''
  JHEP {\bf 0205}, 054 (2002)
  [arXiv:hep-ph/0201295];\br
  %%CITATION = JHEPA,0205,054;%%
D.M.~Ghilencea, L.E.~Ibanez, N.~Irges and F.~Quevedo,
``TeV-scale Z' bosons from D-branes,''
JHEP {\bf 0208}, 016 (2002)
[arXiv:hep-ph/0205083].
%%CITATION = JHEPA,0208,016;%%
}

\lref\CorianoJS{C.~Coriano, N.~Irges and E.~Kiritsis,
  ``On the effective theory of low scale orientifold string vacua,''
  Nucl.\ Phys.\  B {\bf 746}, 77 (2006)
  [arXiv:hep-ph/0510332];\br
  %%CITATION = NUPHA,B746,77;%%
S.A.~Abel, M.D.~Goodsell, J.~Jaeckel, V.V.~Khoze and A.~Ringwald,
``Kinetic Mixing of the Photon with Hidden U(1)s in String Phenomenology,''
arXiv:0803.1449 [hep-ph].
%%CITATION = ARXIV:0803.1449;%%
}

\lref\ChialvaGT{
  D.~Chialva, R.~Iengo and J.G.~Russo,
  ``Cross sections for production of closed superstrings at high energy
  colliders in brane world models,''
  Phys.\ Rev.\  D {\bf 71}, 106009 (2005)
  [arXiv:hep-ph/0503125].
  %%CITATION = PHRVA,D71,106009;%%
}

\lref\barger{V.D. Barger and R.J.N. Phillips,
{\it Collider Physics}, Westview Press (1996).}

\lref\JSCH{M.B.~Green and J.H.~Schwarz,
``Supersymmetrical Dual String Theory. 2. Vertices And Trees,''
Nucl.\ Phys.\ B {\bf 198}, 252 (1982);\br
%%CITATION = NUPHA,B198,252;%%}
J.H.~Schwarz,
``Superstring Theory,''
Phys.\ Rept.\  {\bf 89}, 223 (1982).
%%CITATION = PRPLC,89,223;%%
}

\lref\report{R.~Blumenhagen, B.~K\"ors, D.~L\"ust and S.~Stieberger,
``Four-dimensional String Compactifications with D-Branes, Orientifolds   and Fluxes,''
Phys.\ Rept.\  {\bf 445}, 1 (2007)
[arXiv:hep-th/0610327].
%%CITATION = PRPLC,445,1;%%
}

\lref\AnchordoquiHI{
  L.A.~Anchordoqui, H.~Goldberg and T.R.~Taylor,
``Decay widths of lowest massive Regge excitations of open strings,''
  arXiv:0806.3420 [hep-ph].}

\lref\AbelFK{
  S.A.~Abel, M.~Masip and J.~Santiago,
  ``Flavour changing neutral currents in intersecting brane models,''
  JHEP {\bf 0304}, 057 (2003)
  [arXiv:hep-ph/0303087];\br
  %%CITATION = JHEPA,0304,057;%%
S.A.~Abel, O.~Lebedev and J.~Santiago,
``Flavour in intersecting brane models and bounds on the string scale,''
  Nucl.\ Phys.\  B {\bf 696}, 141 (2004)
  [arXiv:hep-ph/0312157].
  %%CITATION = NUPHA,B696,141;%%
}

\lref\Dan{
D.~Oprisa and S.~Stieberger,
``Six gluon open superstring disk amplitude, multiple hypergeometric  series
and Euler-Zagier sums,''
arXiv:hep-th/0509042.
%%CITATION = HEP-TH/0509042;%%
}

\lref\AnchordoquiAC{
L.A.~Anchordoqui, H.~Goldberg, S.~Nawata and T.R.~Taylor,
``Direct photons as probes of low mass strings at the LHC,''
arXiv:0804.2013 [hep-ph].
%%CITATION = ARXIV:0804.2013;%%
}

\lref\AnastasopoulosDA{
P.~Anastasopoulos, T.P.T.~Dijkstra, E.~Kiritsis and A.N.~Schellekens,
``Orientifolds, hypercharge embeddings and the standard model,''
Nucl.\ Phys.\  B {\bf 759}, 83 (2006)
[arXiv:hep-th/0605226].
%%CITATION = NUPHA,B759,83;%%
}

\lref\AnchordoquiDA{
L.A.~Anchordoqui, H.~Goldberg, S.~Nawata and T.R.~Taylor,
``Jet signals for low mass strings at the LHC,''
Phys.\ Rev.\ Lett.\  {\bf 100}, 171603 (2008)
[arXiv:0712.0386 [hep-ph]].
%%CITATION = PRLTA,100,171603;%%
}

\lref\CremadesDH{
D. Cremades, L.E.~Ibanez and F.~Marchesano,
``Standard model at intersecting D5-branes: Lowering the string scale,''
Nucl.\ Phys.\  B {\bf 643}, 93 (2002)
[arXiv:hep-th/0205074];\br
%%CITATION = NUPHA,B643,93;%%
C.~Kokorelis,
``Exact standard model structures from intersecting D5-branes,''
  Nucl.\ Phys.\  B {\bf 677}, 115 (2004)
  [arXiv:hep-th/0207234].
  %%CITATION = NUPHA,B677,115;%%
}

\lref\BergWT{
M.~Berg, M.~Haack and E.~Pajer,
``Jumping Through Loops: On Soft Terms from Large Volume Compactifications,''
JHEP {\bf 0709}, 031 (2007)
[arXiv:0704.0737 [hep-th]].
%%CITATION = JHEPA,0709,031;%%
}

\lref\AAB{
  E.~Accomando, I.~Antoniadis and K.~Benakli,
``Looking for TeV-scale strings and extra-dimensions,''
  Nucl.\ Phys.\  B {\bf 579}, 3 (2000)
  [arXiv:hep-ph/9912287].
  %%CITATION = NUPHA,B579,3;%%
}

\lref\Nima{I.~Antoniadis, N.~Arkani-Hamed, S.~Dimopoulos and G.R.~Dvali,
``New dimensions at a millimeter to a Fermi and superstrings at a TeV,''
Phys.\ Lett.\  B {\bf 436}, 257 (1998)
[arXiv:hep-ph/9804398].
%%CITATION = PHLTA,B436,257;%%
}

\lref\BlumenhagenTN{
R.~Blumenhagen, M.~Cvetic, F.~Marchesano and G.~Shiu,
``Chiral D-brane models with frozen open string moduli,''
JHEP {\bf 0503}, 050 (2005)
[arXiv:hep-th/0502095].
%%CITATION = JHEPA,0503,050;%%
}

\lref\Dudas{E.~Dudas and J.~Mourad,
``String theory predictions for future accelerators,''
  Nucl.\ Phys.\  B {\bf 575}, 3 (2000)
  [arXiv:hep-th/9911019].
  %%CITATION = NUPHA,B575,3;%%
}

\lref\AntoniadisEW{
I.~Antoniadis,
``A Possible new dimension at a few TeV,''
Phys.\ Lett.\  B {\bf 246}, 377 (1990).
%%CITATION = PHLTA,B246,377;%%
}

\lref\KW{I.R.~Klebanov and E.~Witten,
``Proton decay in intersecting D-brane models,''
Nucl.\ Phys.\  B {\bf 664}, 3 (2003)
[arXiv:hep-th/0304079].
%%CITATION = NUPHA,B664,3;%%
}

\lref\LustFI{
D.~L\"ust, S.~Reffert and S.~Stieberger,
``Flux-induced soft supersymmetry breaking in chiral type IIb  orientifolds
with D3/D7-branes,''
Nucl.\ Phys.\  B {\bf 706}, 3 (2005)
[arXiv:hep-th/0406092].
%%CITATION = NUPHA,B706,3;%%
}

\lref\BlumenhagenSM{
R.~Blumenhagen, S.~Moster and E.~Plauschinn,
``Moduli Stabilisation versus Chirality for MSSM like Type IIB
Orientifolds,''
JHEP {\bf 0801}, 058 (2008)
[arXiv:0711.3389 [hep-th]].
%%CITATION = JHEPA,0801,058;%%
}

\lref\BalasubramanianZX{
V.~Balasubramanian, P.~Berglund, J.P.~Conlon and F.~Quevedo,
``Systematics of moduli stabilisation in Calabi-Yau flux
compactifications,''
JHEP {\bf 0503}, 007 (2005)
[arXiv:hep-th/0502058].
%%CITATION = JHEPA,0503,007;%%
}

\lref\ConlonXV{
J.P.~Conlon, C.H.~Kom, K.~Suruliz, B.C.~Allanach and F.~Quevedo,
``Sparticle Spectra and LHC Signatures for Large Volume String
Compactifications,''
JHEP {\bf 0708}, 061 (2007)
[arXiv:0704.3403 [hep-ph]].
%%CITATION = JHEPA,0708,061;%%
}

\lref\ConlonKI{
J.P.~Conlon, F.~Quevedo and K.~Suruliz,
``Large-volume flux compactifications: Moduli spectrum and D3/D7 soft
supersymmetry breaking,''
JHEP {\bf 0508}, 007 (2005)
[arXiv:hep-th/0505076].
%%CITATION = JHEPA,0508,007;%%
}

\lref\BeasleyKW{
  C.~Beasley, J.J.~Heckman and C.~Vafa,
  ``GUTs and Exceptional Branes in F-theory~-~I,''
  arXiv:0802.3391 [hep-th];
  %%CITATION = ARXIV:0802.3391;%%
  ``GUTs and Exceptional Branes in F-theory~-~II: Experimental Predictions,''
  arXiv:0806.0102 [hep-th].
  %%CITATION = ARXIV:0806.0102;%%
}

\lref\BerensteinNK{
  D.~Berenstein, V.~Jejjala and R.G.~Leigh,
  ``The standard model on a D-brane,''
  Phys.\ Rev.\ Lett.\  {\bf 88}, 071602 (2002)
  [arXiv:hep-ph/0105042].
  %%CITATION = PRLTA,88,071602;%%
}

\lref\banksi{T.~Banks, L.J.~Dixon, D.~Friedan and E.J.~Martinec,
``Phenomenology and Conformal Field Theory Or Can String Theory Predict the
Weak Mixing Angle?,''
Nucl.\ Phys.\  B {\bf 299}, 613 (1988)
%%CITATION = NUPHA,B299,613;%%
}
\lref\banksii{T.~Banks and L.J.~Dixon,
``Constraints on String Vacua with Space-Time Supersymmetry,''
Nucl.\ Phys.\  B {\bf 307}, 93 (1988).
%%CITATION = NUPHA,B307,93;%%
}

\lref\JOE{
J. Polchinski, "String Theory'', Sections 6 \& 12, Cambridge University Press 1998.}

\lref\flt{
S.~Ferrara, D.~L\"ust and S.~Theisen,
``World Sheet Versus Spectrum Symmetries In Heterotic And Type II Superstrings,''
Nucl.\ Phys.\  B {\bf 325}, 501 (1989).
%%CITATION = NUPHA,B325,501;%%
}

\lref\LMRS{D.~L\"ust, P.~Mayr, R.~Richter and S.~Stieberger,
``Scattering of gauge, matter, and moduli fields from intersecting  branes,''
Nucl.\ Phys.\  B {\bf 696}, 205 (2004)
[arXiv:hep-th/0404134].
%%CITATION = NUPHA,B696,205;%%
}

\lref\groupf{
T.~van Ritbergen, A.N.~Schellekens and J.A.M.~Vermaseren,
``Group theory factors for Feynman diagrams,''
Int.\ J.\ Mod.\ Phys.\  A {\bf 14}, 41 (1999)
[arXiv:hep-ph/9802376].}

\lref\Richter{M.~Cvetic and R.~Richter,
``Proton decay via dimension-six operators in intersecting D6-brane models,''
Nucl.\ Phys.\  B {\bf 762}, 112 (2007)
[arXiv:hep-th/0606001].
%%CITATION = NUPHA,B762,112;%%
}

\lref\Chemtob{M.~Chemtob,
``Nucleon decay in gauge unified models with intersecting D6-branes,''
Phys.\ Rev.\  D {\bf 76}, 025002 (2007)
[arXiv:hep-ph/0702065].
%%CITATION = PHRVA,D76,025002;%%
}

\lref\Cvetic{M.~Cvetic and I.~Papadimitriou,
``Conformal field theory couplings for intersecting D-branes on orientifolds,''
Phys.\ Rev.\  D {\bf 68}, 046001 (2003)
[Erratum-ibid.\  D {\bf 70}, 029903 (2004)]
[arXiv:hep-th/0303083].
%%CITATION = PHRVA,D68,046001;%%
}

\lref\phencvetic{M.~Cvetic, P.~Langacker and G.~Shiu,
``Phenomenology of a three-family standard-like string model,''
Phys.\ Rev.\ D {\bf 66}, 066004 (2002)
[arXiv:hep-ph/0205252].}

\lref\tye{G.~Shiu and S.H.~Tye,
``TeV scale superstring and extra dimensions,''
Phys.\ Rev.\ D {\bf 58}, 106007 (1998)
[arXiv:hep-th/9805157].
%%CITATION = HEP-TH 9805157;%%
}

\lref\CIM{D.~Cremades, L.E.~Ibanez and F.~Marchesano,
``SUSY quivers, intersecting branes and the modest hierarchy problem,''
JHEP {\bf 0207}, 009 (2002)
[arXiv:hep-th/0201205].
%%CITATION = HEP-TH 0201205;%%
}

\lref\STi{S.~Stieberger and T.R.~Taylor,
``Amplitude for N-gluon superstring scattering,''
Phys.\ Rev.\ Lett.\  {\bf 97}, 211601 (2006)
[arXiv:hep-th/0607184].
%%CITATION = PRLTA,97,211601;%%
}

\lref\STii{S.~Stieberger and T.R.~Taylor,
``Multi-gluon scattering in open superstring theory,''
Phys.\ Rev.\  D {\bf 74}, 126007 (2006)
[arXiv:hep-th/0609175].
%%CITATION = PHRVA,D74,126007;%%
}

\lref\STiii{S.~Stieberger and T.R.~Taylor,
``Supersymmetry Relations and MHV Amplitudes in Superstring Theory,''
Nucl.\ Phys.\  B {\bf 793}, 83 (2008)
[arXiv:0708.0574 [hep-th]].
%%CITATION = NUPHA,B793,83;%%
}

\lref\STiv{S.~Stieberger and T.R.~Taylor,
``Complete Six-Gluon Disk Amplitude in Superstring Theory,''
Nucl.\ Phys.\  B {\bf 801}, 128 (2008)
[arXiv:0711.4354 [hep-th]].
%%CITATION = ARXIV:0711.4354;%%
}

\lref\MeadeSZ{
P.~Meade and L.~Randall,
``Black Holes and Quantum Gravity at the LHC,''
  JHEP {\bf 0805}, 003 (2008)
  [arXiv:0708.3017 [hep-ph]].
  %%CITATION = JHEPA,0805,003;%%
}

\lref\HorowitzNW{
G.T.~Horowitz and J.~Polchinski,
``A correspondence principle for black holes and strings,''
Phys.\ Rev.\  D {\bf 55} 6189 (1997)
[arXiv:hep-th/9612146].
%%CITATION = PHRVA,D55,6189;%%
}
\lref\DimoL{
S.~Dimopoulos and G.L.~Landsberg,
``Black holes at the LHC,''
Phys.\ Rev.\ Lett.\  {\bf 87}, 161602 (2001)
[arXiv:hep-ph/0106295].
%%CITATION = PRLTA,87,161602;%%
}
\lref\GTh{
S.B.~Giddings and S.D.~Thomas,
``High energy colliders as black hole factories: The end of short  distance physics,''
Phys.\ Rev.\  D {\bf 65}, 056010 (2002)
[arXiv:hep-ph/0106219].
%%CITATION = PHRVA,D65,056010;%%
}

\lref\Benakli{I.~Antoniadis, K.~Benakli and A.~Laugier,
``Contact interactions in D-brane models,''
  JHEP {\bf 0105}, 044 (2001)
  [arXiv:hep-th/0011281].
  %%CITATION = JHEPA,0105,044;%%
}

\lref\Narain{E.~Gava, K.S.~Narain and M.H.~Sarmadi,
``On the bound states of p- and (p+2)-branes,''
  Nucl.\ Phys.\  B {\bf 504}, 214 (1997)
  [arXiv:hep-th/9704006].
  %%CITATION = NUPHA,B504,214;%%
}

\lref\CullenEF{
  S.~Cullen, M.~Perelstein and M.E.~Peskin,
  ``TeV strings and collider probes of large extra dimensions,''
  Phys.\ Rev.\  D {\bf 62}, 055012 (2000)
  [arXiv:hep-ph/0001166].
  %%CITATION = PHRVA,D62,055012;%%
}

\lref\MalyshevZZ{
  D.~Malyshev and H.~Verlinde,
  ``D-branes at Singularities and String Phenomenology,''
  Nucl.\ Phys.\ Proc.\ Suppl.\  {\bf 171}, 139 (2007)
  [arXiv:0711.2451 [hep-th]].
  %%CITATION = NUPHZ,171,139;%%
}

\lref\VerlindeJR{
  H.~Verlinde and M.~Wijnholt,
  ``Building the standard model on a D3-brane,''
  JHEP {\bf 0701}, 106 (2007)
  [arXiv:hep-th/0508089].
  %%CITATION = JHEPA,0701,106;%%
}

\lref\EJSS{
S.~Stieberger, D.~Jungnickel, J.~Lauer and M.~Spalinski,
"Yukawa couplings for bosonic ${\bf Z}_N$ orbifolds:
Their moduli and twisted sector dependence,''
Mod.\ Phys.\ Lett.\ A {\bf 7}, 3059 (1992)
[arXiv:hep-th/9204037];\br
%%CITATION = HEP-TH 9204037;%%
J.~Erler, D.~Jungnickel, M.~Spalinski and S.~Stieberger,
"Higher twisted sector couplings of ${\bf Z}_N$ orbifolds,''
Nucl.\ Phys.\ B {\bf 397}, 379 (1993)
[arXiv:hep-th/9207049].
%%CITATION = HEP-TH 9207049;%%
}

\lref\BKM{T.T.~Burwick, R.K.~Kaiser and H.F.~M\"uller,
"General Yukawa Couplings Of Strings On ${\bf Z}_N$ Orbifolds,''
Nucl.\ Phys.\ B {\bf 355}, 689 (1991).
%%CITATION = NUPHA,B355,689;%%
}

\lref\Abelii{
S.A.~Abel and A.W.~Owen,
``N-point amplitudes in intersecting brane models,''
  Nucl.\ Phys.\  B {\bf 682}, 183 (2004)
  [arXiv:hep-th/0310257].
  %%CITATION = NUPHA,B682,183;%%
}

\lref\Abeli{
S.A.~Abel and A.W.~Owen,
``Interactions in intersecting brane models,''
  Nucl.\ Phys.\  B {\bf 663}, 197 (2003)
  [arXiv:hep-th/0303124].
  %%CITATION = NUPHA,B663,197;%%
}

\lref\Berenstein{
  D.~Berenstein,
``Possible exotic stringy signatures at the LHC,''
  arXiv:0803.2545 [hep-th].
  %%CITATION = ARXIV:0803.2545;%%
}

\lref\FBM{
D.~Cremades, L.E.~Ibanez and F.~Marchesano,
``Yukawa couplings in intersecting D-brane models,''
JHEP {\bf 0307}, 038 (2003)
[arXiv:hep-th/0302105];
%%CITATION = JHEPA,0307,038;%%
``Computing Yukawa couplings from magnetized extra dimensions,''
JHEP {\bf 0405}, 079 (2004)
[arXiv:hep-th/0404229].
%%CITATION = JHEPA,0405,079;%%
}

%\ArkaniHamedRS
\lref\ArkaniHamedRS{
N.~Arkani-Hamed, S.~Dimopoulos and G.R.~Dvali,
``The hierarchy problem and new dimensions at a millimeter,''
Phys.\ Lett.\  B {\bf 429}, 263 (1998)
[arXiv:hep-ph/9803315].
%%CITATION = PHLTA,B429,263;%%
}

\lref\Ignatios{I.~Antoniadis,
``The physics of extra dimensions,''
Lect.\ Notes Phys.\  {\bf 720}, 293 (2007)
[arXiv:hep-ph/0512182].
%%CITATION = LNPHA,720,293;%%
}

\lref\Jaxo{D.~Binosi and L.~Theussl,
``JaxoDraw: A graphical user interface for drawing Feynman diagrams,''
Comput.\ Phys.\ Commun.\  {\bf 161}, 76 (2004)
[arXiv:hep-ph/0309015].
%%CITATION = HEP-PH 0309015;%%
}

\lref\aglnst{L. Anchordoqui, H. Goldberg, D. L\"ust, S. Nawata, S. Stieberger,
T.R. Taylor, "Dijet signals for low mass strings at the LHC," MPP--2008--86, 
LMU--ASC 42/08, to appear.}

%%%%%%%%%%%%%%%%%%%%%%%%%%%%%%%%%%%%%%%%%%%%%%%%%%%%%%%%%%%%%%%%%%%%%%%%%%%%%%%
\Title{\vbox{\rightline{MPP--2008--05}\rightline{LMU--ASC 38/08}}}
{\vbox{\centerline{The LHC String Hunter's Companion}}}
%\smallskip
\centerline{Dieter L\"ust$^{a,b}$,\ \ Stephan Stieberger$^{a}$,\ \
Tomasz R. Taylor$^{a,b,c}$}
\bigskip
\centerline{\it $^a$ Max--Planck--Institut f\"ur Physik,
Werner--Heisenberg--Institut,}
\centerline{\it 80805 M\"unchen, Germany}
\vskip8pt
\centerline{\it $^b$ Arnold--Sommerfeld--Center for Theoretical Physics,}
\centerline{\it Ludwig--Maximilians--Universit\"at M\"unchen, 80333 M\"unchen, Germany}
\vskip8pt
\centerline{\it $^c$ Department of Physics, Northeastern University,
Boston, MA 02115, USA}
%\medskip
\bigskip
\centerline{\bf Abstract}
\vskip8pt
\noindent
The mass scale of fundamental strings can be as low as few
TeV$\!/$c$^2$ provided that spacetime extends into large extra dimensions.
We discuss the phenomenological aspects of weakly coupled low mass string theory
related to experimental searches for physics beyond the Standard Model at the
Large Hadron Collider (LHC). We consider the extensions of the Standard Model
based on open strings ending on D-branes, with
gauge bosons due to strings attached to stacks of D-branes and chiral matter
due to strings stretching between intersecting D-branes.
We focus on the model-independent, universal features of low mass string theory.
We compute, collect and tabulate the full-fledged string amplitudes describing
all $2\to 2$ parton scattering subprocesses
at the leading order of string perturbation theory.
We cast our results in a form suitable for the implementation of stringy
partonic cross sections in the LHC data analysis. The amplitudes involving
four gluons as well as those with two gluons plus two quarks  do not depend on
the compactification details and are completely model-independent.
They exhibit resonant behavior at the parton center of mass energies equal to
the masses of Regge resonances.  The existence of these resonances is the
primary signal of string physics and should be easy to detect. On the other
hand, the four-fermion processes like quark-antiquark scattering include also
the exchanges of heavy Kaluza-Klein and winding states,
whose details depend
on the form of  internal geometry. They could be used as ``precision tests''
in order to distinguish between various compactification scenarios.

\Date{}
\noindent

\goodbreak

\listtoc
\writetoc
\break
%%%%%%%%%%%%%%%%%%%%%%%%%%%%%%%%%%%%%%%%%%%%%%%%%%%%%%%%%%%%%%%%%%%%%%%%%%%%%%%
\newsec{Introduction}

The Standard Model (SM) of particle physics is a well established quantum field theory
that describes the spectrum and the interactions of elementary particles to high accuracy
and
in excellent  agreement with almost all experiments. Only astrophysical observations
provide indirect experimental evidence for new physics beyond the SM in the form
of not yet directly observed dark matter particles. However at the conceptual level,
there exist several unsolved problems, which strongly hint at new physics beyond the SM.
Probably the  most mysterious puzzle is the hierarchy problem, namely the question
why the Planck mass $M_{\rm Planck}\simeq 10^{19}\ {\rm GeV}$ is huge compared
to the electroweak scale $M_{EW}$:
\eqn\hierarchy{
M_{\rm Planck}\gg M_{EW}\, ?}
In fact, there are some good reasons to believe that the resolution of the
hierarchy problem lies in new
physics around the TeV mass scale. The LHC collider at CERN is designed to
discover new physics
precisely in this energy range, hopefully giving important clues about the
nature of dark matter and perhaps at the same time about the solution of the
hierarchy problem. In fact, there are at least three, not necessarily mutually
exclusive scenarios, offered as solutions of the hierarchy problem:
\vskip0.3cm

\noindent \hskip1cm $\bullet$\  Low energy supersymmetry at around 1 TeV.

\vskip0.2cm

\noindent \hskip1cm $\bullet$\
New strong dynamics at around 1 TeV (technicolor, little Higgs models, etc).

\vskip0.2cm

\noindent \hskip1cm $\bullet$\ Low energy scale
for (quantum) gravity and large extra dimensions at few TeVs.

\vskip0.3cm\noindent
In the latter scenario, the observed weakness of gravity at energies below few
TeVs is due to the existence
of large extra dimensions \doubref\ArkaniHamedRS\Nima.
In string theory, extra dimensions appear naturally, therefore it is an
obvious question to ask whether they can be large enough to accommodate such 
new physics at few TeVs.
This is possible only if the intrinsic
scale of string excitations, called the string mass $M_{\rm string}$ is
also of order few TeVs.
In this case a whole tower of infinite  string excitations
 will open up at the string mass threshold, and new particles follow the well known Regge
trajectories of vibrating strings,
\eqn\regge{
j=j_0+\alpha' M^2\ ,}
with the spin $j$ and $\alpha'$ the Regge slope parameter that determines the fundamental
string  mass scale $M_{\rm string}^2=\alpha'^{-1}$.
In this work, we discuss the phenomenological aspects of low mass string
theory related to experimental searches for physics beyond the SM at  the
LHC. We focus on its model-independent, universal features  that can be
observed and tested at the LHC.

Let us list what kind of string signatures from a low string scale and from
large
extra dimensions can be
possibly expected  at the LHC:

\vskip0.3cm

\noindent $\bullet$ The discovery of new exotic particles around $M_{\rm
string}$. For
example, many string models predict the existence of new, massive $Z'$ gauge
bosons
from additional $U(1)$ gauge symmetries
(see e.g. \KiritsisAJ).
They can also have an interesting effect, since they mix with the standard
photon by
their kinetic energies (see e.g. \CorianoJS).

\vskip0.2cm

\noindent $\bullet$ The discovery of quantum gravity effects in the form of
mini black
holes (see e.g. \refs{\HorowitzNW,\MeadeSZ}).

\vskip0.2cm

\noindent $\bullet$ The discovery of string Regge excitations with masses of
order $M_{\rm string}$.
These stringy states will lead to new contributions to SM scattering
processes, which can be measurable at LHC in case of low string scale.
Furthermore, there are the Kaluza--Klein (KK) 
and winding excitations along the small internal
dimensions, i.e.
KK and winding excitations of the
SM fields. Their masses depend
on the internal volumes\foot{In fact, there may be another kind of
KK excitations, namely  along the large extra dimensions, \ie KK
excitations in the gravitational (bulk) sector
of the theory. Their masses can be as low as $10^{-3}$ eV. However, KK modes
from the bulk couple only at one--loop (annulus) to SM fields resulting in
a suppression by a factor of $\gs\sim g^2$ compared to tree--level 
processes on the brane.
Hence contributions from KK modes of the bulk are less relevant than those
from string Regge excitations, \cf also the discussion in \CullenEF.},
and they should be also near the string scale $M_{\rm string}$.

\vskip0.3cm\noindent
It is precisely this last item, which we want to discuss in this work.
So let us be more specific with what we mean
by new contributions to SM processes  from a low string scale
$M_{\rm string}$. Namely these are the
$\alpha'$-contributions to ordinary SM processes like the scattering of quarks
and gluons into SM fields. As already mentioned, at tree--level the
$\alpha'$-corrections to SM processes are due to the exchanges
of massive string excitations encompassing all Regge recurrences.
In addition, there are contributions of KK states and winding modes
with their spectrum depending on the form of extra-dimensional geometry.
However a class of amplitudes, e.g. the $N$--gluon amplitudes of QCD, receive  universal
$\alpha'$-corrections only, which are insensitive to the details of
specific compactifications and to the extent of supersymmetry preserved in four 
dimensions \multrefiv\Dan\STi\STii\STiv.
Similarly, also other SM processes that involve quarks, leptons and other gauge
bosons receive $\alpha'$--corrections, leading to characteristic deviations from the SM
predictions and hence can be
tested at the LHC.
An important step in this direction has already been undertaken
in \refs{\AAB,\CullenEF} where the effects of string resonances have 
been pointed out as an important signal of string physics.
Other string four--point amplitudes that involve four SM fermions, relevant to Yukawa
couplings and possibly leading
to proton decay and FCNC in specific models,
have been computed and analyzed in
\multrefix\KW\Cvetic\AbelFK\Abeli\Abelii\LMRS\FBM\Richter\Chemtob.
More recently, in \AnchordoquiDA\ and \AnchordoquiAC\
the string effects in the process $gg\rightarrow~g\gamma$ have been considered.
In the present work we systematically investigate all possible string tree--level
$\alpha'$-corrections to SM
processes that involve quarks, leptons and
SM gauge bosons, as they arise in intersecting D-brane compactifications of
orientifold
models\foot{In contrast to Ref. \CullenEF\ we are describing the SM fermions,
which are located at D-brane intersections,  by boundary changing operators in
the open string CFT. This leads to  different $\alpha'$ corrections in the fermion
scattering amplitudes.}.
We work in a model--independent way and essentially only need the local information about
how the SM is realized on type IIA/IIB intersecting D-branes. Some of the
processes
exist already at tree level
in the SM, and hence the tree level SM background must be subtracted from the
string
corrections. Other
processes like
$gg\rightarrow g\gamma$ or $gg\rightarrow \gamma\gamma$
do not at all exist in the SM at tree--level and can be viewed as the
``smoking guns'' for
D-brane string compactifications with a low string scale and large extra
dimensions.
However let us remind
that it requires much   fortune
to see these string effects at the LHC along the lines discussed in this work.
In particular we need a low string scale, large extra dimensions and also weak
string
coupling in order for our calculations to be reliable and
testable at the LHC.

The present work is organized as follows. In the next Section we discuss some
general aspects and the basic
setting of string compactifications with a low string scale and large extra
dimensions.
Then, in Section three,
we recall how the SM can be constructed from type IIA/IIB D-branes on
orientifolds.
We do not discuss fully consistent global orientifold models,
which lead to the SM, but we rather focus on the local, intersecting
D-brane configurations that realize the SM by open strings.
Eventually the local D-brane systems
have to be included into a compact manifold in order to obtain a fully
consistent
orientifold compactifications.
However, as we shall argue, the local D-brane systems are sufficient to
compute
all relevant tree level
scattering amplitudes among the SM open string excitations. In Section four we
discuss how large extra dimensions can be realized in Calabi--Yau (CY) orientifolds
and how the local, SM
D-brane system has to be embedded into a large volume CY space. To some extent
we follow the recent constructions for large volume compactifications by
\BalasubramanianZX\ using the
``Swiss cheese" CY spaces, however with the
difference that in \BalasubramanianZX\ the string scale is around an
intermediate
scale of $10^{11}$ GeV
and supersymmetry is broken at the TeV scale, whereas in our case $M_{\rm string}$ is
at few TeVs and
low energy supersymmetry is not needed. In Section five we present a complete
calculation of all possible four-point string scattering amplitudes of gauge
and SM matter fields. We analyze string corrections to scattering processes that
involve quarks and gluons,
since they are the most relevant processes for the LHC.
The computations 
of the scattering of four gauge bosons and of the scattering of two gauge
bosons and two matter
fermions is performed in a model independent and
universal way. Our results hold for all compactifications, even for those that break 
supersymmetry.
The poles of the respective amplitudes are due to
the exchanges of massless gauge bosons and universal string Regge excitations only.
On the other
hand, the amplitudes that involve four matter fields  depend on the details of the 
D-brane geometry, and how the D-branes are embedded into the compact CY space. Here also
modes of the internal
geometry can be exchanged during the four fermion scattering processes. Finally, 
in Section six, we compute the squared moduli of all
amplitudes, sum over polarizations and colors of final particles and average
over polarization and colors of incident particles, as needed for the
unpolarized parton
cross sections. The results are presented in  Tables.

In another publication \aglnst\ written in collaboration with Luis
Anchordoqui, Haim Goldberg and Satoshi Nawata, we use our results to analyze the
dijet signals for low mass strings at the LHC.

\newsec{Physics of large extra dimensions and low string scale}

Large extra dimensions are a very appealing solution to the hierarchy problem \Nima.
The gravitational and gauge interactions are unified at the electroweak scale
and the observed weakness of gravity at lower energies is due to the existence
of large extra dimensions. Gravitons may scatter into the extra space and
by this the gravitational coupling constant is decreased to its observed value.
Extra dimensions arise naturally in string theory. Hence, one obvious question is how
to embed the above scenario into string theory and how to compute
cross sections.

\subsec{Planck mass and gauge couplings in D--brane compactifications}

Here we discuss the gravitational and gauge couplings
in orientifold compactifications.
In the following we consider the type II superstring compactified on
a six--dimensional compactification manifold.
In addition, we consider a D$p$--brane wrapped on a $p-3$--cycle with the
remaining four dimensions extended into the uncompactified space--time.
We have $d_\parallel=p-3$ internal directions
parallel to the D$p$--brane world volume
and $d_\perp=9-p$ internal directions transverse to the D$p$--brane world volume.
Let us denote the radii (in the string frame) of the parallel directions by
$R^\parallel_i\ ,i=1,\ldots,d_\parallel$ and the
radii of the transverse directions by $R^\perp_j,\ j=1,\ldots,d_\perp$.
The generic setup is displayed in Figure 1.
\ifig\anton{D--brane setup with $d_\parallel$ parallel and $d_\perp$
transverse internal directions (from Ref. \Ignatios).}
{\epsfxsize=0.65\hsize\epsfbox{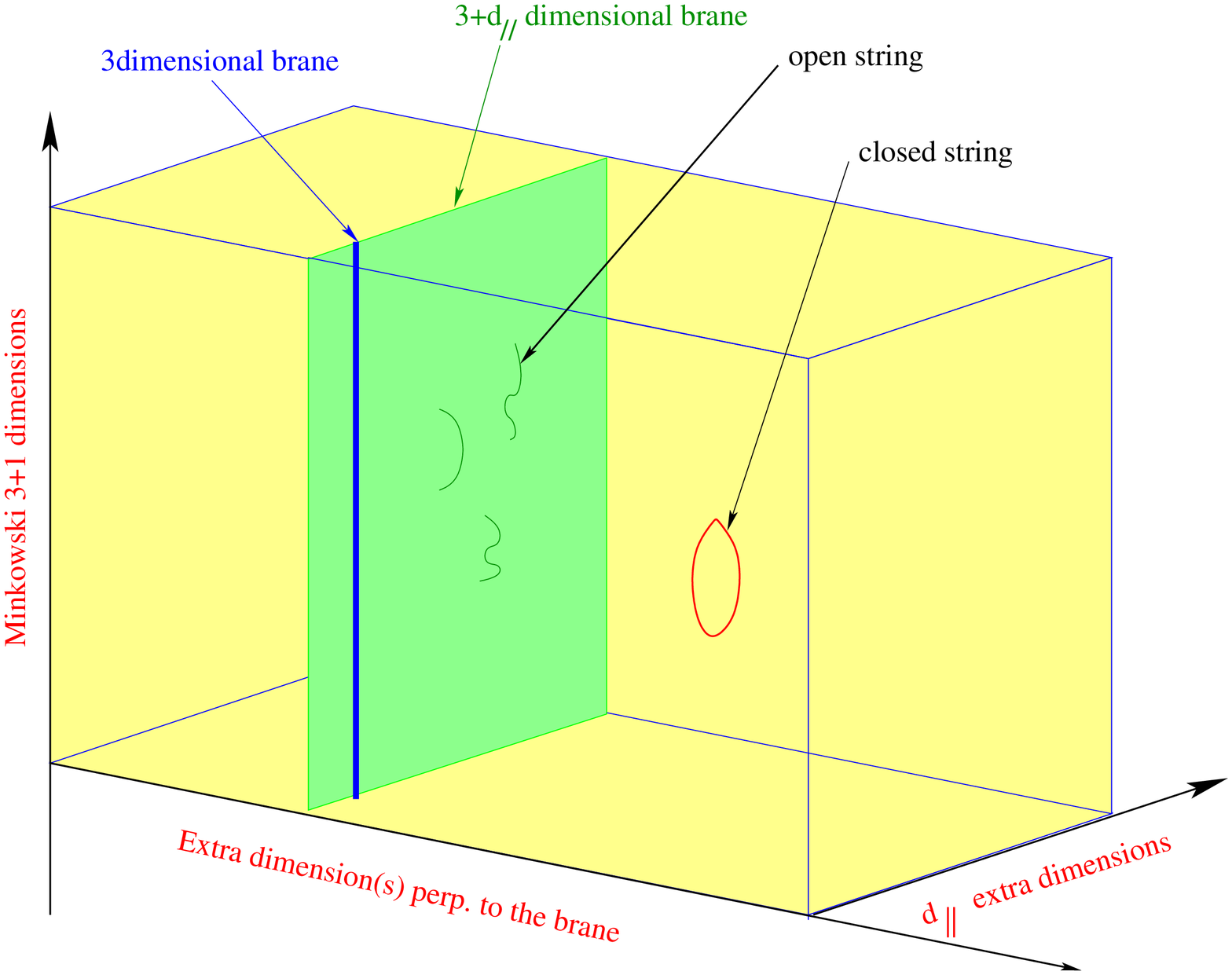}}

\noindent

While the gauge interactions are localized
on the D--brane world volume, the gravitational interactions are also spread into the
transverse space. This gives qualitatively different quantities for their
couplings.
In $D=4$ we obtain for the Planck mass
($\alpha'=M_{\rm string}^{-2}$)~\LMRS
\eqn\Planckmass{
M_{\rm Planck}^2=8\ e^{-2\phi_{10}}\ M_{\rm string}^8\ \fc{V_6}{(2\pi)^{6}}\ ,}
where the internal six-dimensional (string frame) volume $V_6$ is expressed
in terms of the parallel and transversal
radii as
%\foot{{\bf This confuses the experimentalist and should go away:}
%Using dimensionless radii $r_i=R_i/\sqrt{\alpha'}$ and a
%redefined string scale
%$M_s=e^{-\phi_{10}}M_{\rm string}$ this relation reads $M_{\rm
%Planck}=8M_s\sqrt V'$, where $V'=\prod r_i$.}
\eqn\vol{
V_6=(2\pi)^6\ \prod_{i=1}^{d_\parallel} R^\parallel_i\ \prod_{j=1}^{d_\perp} R^\perp_j\ .}
The dilaton field $\phi_{10}$ is related to the $D=10$
type II string coupling constant through $g_{\rm string}=e^{\phi_{10}}$.
The gravitational coupling constant follows from \Planckmass\ through the
relation $G_N=M_{\rm Planck}^{-2}$.
On the other hand, in type II superstring theory
the gauge theory on the D--brane world--volume has the gauge coupling:
\eqn\gaugeAB{
g_{Dp}^{-2}=(2\pi)^{-1}\ \ap^{\fc{3-p}{2}}\ e^{-\phi_{10}}\
\prod_{i=1}^{d_\parallel}R_i^\parallel\ .}
In \gaugeAB\ each factor $i$ accounts for an $1$--cycle wrapped along the $i$--th
coordinate segment.
While the size of the gauge couplings is determined by the size of the parallel
dimensions, the strength of gravity is influenced by all directions.

\subsec{Large extra dimensions and low string scale}

{}From \Planckmass\ and the gauge coupling \gaugeAB\ we
may deduce a relation between the Planck mass $M_{\rm Planck}$,
the string mass $M_{\rm string}$ and the sizes $R_j$ of the compactified
internal directions. For type II we obtain\foot{The discussion takes over to
type
I superstring theory.
The type I theory may be obtained from type IIB by an orientifold projection.
The world--volume gauge theory on the D--brane sitting on the orientifold plane
becomes then $SO(2N)$ or $USp(N)$.
In that case, all gauge couplings, derived in the following for $U(N)$ gauge groups,
have to be multiplied by a factor of $2$, \ie $g^2_{Dp,SO(2N)}=2g_{Dp}^2$ \JOE.}:
\eqn\relation{
g_{Dp}^2\ M_{\rm Planck}=2^{5/2}\pi\ M_{\rm string}^{7-p} \
\lf(\prod\limits_{j=1}^{d_\perp} R_j^\perp\ri)^\h\
\lf(\prod\limits_{i=1}^{d_\parallel} R_i^\parallel\ri)^{-1/2}\ .}
Hence, by enlarging some of the transverse compactification
radii $R_j^\perp$ the string scale has to become lower in order to achieve the 
correct Planck mass ($p<7$).
This is to be contrasted with a theory of closed (heterotic) strings only. In that case
the relation between the Planck mass and the string scale does not depend on
the volume. It is given by the relation
$M_{\rm string}=g_{\rm string}\ M_{\rm Planck}$, which requires
a high string scale $M_{\rm string} \sim 10^{17} \gev$ for the correct Planck mass.

A priori, there are no compelling reasons why the string mass scale 
$M_{\rm string}$ should be much lower than the
Planck mass. In the large volume compactifications
of \refs{\BalasubramanianZX,\ConlonKI,\ConlonXV} it was shown that that one
can indeed stabilize moduli in such a way that the string scale $M_{\rm string}$
is at intermediate energies of about $10^{11-12}~{\rm GeV}$. Then
the internal CY volume $V_6$ is of order $V_6M_{\rm string}^6={\cal O}(10^{16})$.
The motivation for
this scenario is to obtain a supersymmetry breaking scale around $1~{\rm TeV}$, since
one derives the following relation for the gravitino mass:
\eqn\gravitino{
m_{3/2}\sim {\Ms^2\over M_{\rm Planck}}\ .}
However, giving up the requirement of supersymmetry at the TeV scale, one is free to
consider
CY manifolds with much larger volume.
In fact, if it happens for $M_{\rm string}$ to be within the range of LHC energies,
not too far beyond
1 TeV, string theory can be tested.
In this case the CY volume is as large as $V_6M_{\rm string}^6
={\cal O}(10^{32})$. Of course one has
to find
scalar potentials with minima that lead to such big internal volumes.

Some spectacular signatures are expected near the string mass threshold.
They are related  to the production of virtual or real string Regge
excitations of mass of order $\Ms$
and to the effects of strongly coupled gravity like the production and decays
of microscopic black holes
\refs{\DimoL,\GTh}. The reason why gravity is expected to become 
strong at energies comparable
to the string mass is the inevitable presence of Kaluza Klein excitations of 
gravitons and other particles propagating in the bulk of large extra
dimensions, with the (model-dependent)
masses expected in the range from $10^{-3}$ eV order to 1 MeV order. 
Although ordinary matter
particles couple to these excitations very weakly, with the strength determined by the
Newton's constant, the combined effect of a large number of virtual 
Kaluza Klein gravitons
is to increase the strength of gravitational forces at high energies. 
In string theory, this
effect may occur below or above the fundamental string mass scale, 
depending on the string coupling constant $g_{\rm string}$. 
For example, black holes are expected to be produced at
energies of order $M_{\rm string}/g_{\rm string}^2$ \HorowitzNW, although some 
strong gravity effects may appear already at slightly lower energies
\MeadeSZ. Thus in weakly coupled
string theory with $g_{\rm string}<1$, black hole production and in general, 
the onset of strong gravity effects occur above the string mass scale. 
In this case, the lowest energy signals of strings at the LHC would be due 
to virtual Regge excitations produced in parton
collisions. The corresponding scattering amplitudes can be evaluated by using string
perturbation theory, with the dominant contributions originating from disk diagrams.
In this work, we discuss the disk amplitudes necessary for studying all $2\to 2$ 
scattering processes of gluons and quarks originating from D-brane intersections. 
% Some results  are new
% while other are known, nevertheless we believe that it is useful to collect all of 
% them in one place as a resource for the LHC string hunters.

String Regge resonances in models with low string scale are also discussed 
in \refs{\CullenEF,\AAB}, while KK graviton exchange, which
appears at the next order in perturbation theory, is discussed in
\refs{\CullenEF,\Dudas,\ChialvaGT}.

Let us now discuss the possible sizes of large  extra dimensions subject to the
experimental facts.
Cavendish type experiments test Newton's law up to a scale of millimeters. This provides
an upper bound on the large extra dimensions $R_j^\perp$ to be in the millimeter range.
On the other hand, QCD and electroweak scattering experiments give an upper bound
on the small extra dimensions $R_i^\parallel$ in the range of the electroweak scale
$M_{EW}^{-1}$.

A first look at the relations \Planckmass\ gives an estimate on the string
scale
$M_{\rm string}$ and the size of $d_\perp$ extra dimensions $R_j^\perp$.
For the $d_\parallel$ small directions to be of the order of the string scale
$M_{\rm string}$ and $d_\perp$ extra dimensions of size $R^\perp$ we
obtain\foot{The above values are computed for $g_{\rm string}\simeq g^2=\fc{1}{25}$,
\ie $\al=\fc{g^2}{4\pi}=0.003$. Furthermore, $E_R=\fc{hc}{R^\perp}$ and $1\
GeV^{-1}\sim 10^{-15}m$.} the numbers listed in Table~1.

\vskip0.1cm
{\vbox{\ninepoint{$$
\vbox{\offinterlineskip\tabskip=0pt
\halign{\strut\vrule#
%%%%%%%%%%%%%%%%%%
&~$#$~\hfil
&\vrule$#$
&~$#$~\hfil
&\vrule$#$
&~$#$~\hfil
&\vrule$#$
&~$#$~\hfil
&\vrule$#$
&~$#$~\hfil
&\vrule$#$
&~$#$~\hfil
&\vrule$#$
&~$#$~\hfil
&\vrule$#$\cr
\noalign{\hrule}
%%%%%%%%%%%%%%%%%%
&  &&d_\perp=1
&&d_\perp=2 && d_\perp=3  && d_\perp=4&& d_\perp=5&& d_\perp=6 &\cr
\noalign{\hrule}
&R^\perp\ [GeV^{-1}]&&1.6\cdot10^{26}&&4\cdot10^{11} && 5.4\cdot10^{6}&&2\cdot 10^4&&
693 &&74  &\cr
&R^\perp\ [m]&&1.6\cdot10^{11}&&4\cdot10^{-4} && 5.4\cdot10^{-9}&&2\cdot10^{-11}&&
7\cdot 10^{-13}&&7\cdot 10^{-14} &\cr
&E_R\ [MeV]&&7.7\cdot 10^{-24}   &&3\cdot 10^{-9}  &&2\cdot 10^{-4}&&0.06&&1&&16&\cr
\noalign{\hrule}}}$$
\vskip-6pt
\centerline{\noindent{\bf Table 1:}
{\it Size of $d_\perp$ large extra dimensions for a string scale of $M_{\rm
string}=1\ TeV$.}}
\vskip10pt}}}
\noindent
So, the case $d_\perp=1$ is ruled out experimentally.

\subsec{Exchanges of string Regge excitations and string contact interactions}

Due to the extended nature of strings, the world--sheet string amplitudes
are generically non--trivial functions of $\ap$ in addition to the
usual dependence on the kinematic invariants and degrees of freedom of the
external states.
In the effective field theory description this $\ap$--dependence gives rise to a
series of infinite many resonance channels\foot{In addition, there may be
additional resonance channels
due to the exchange of KK and winding states, as it is the case for four fermions 
amplitudes in four dimensions, \cf Section 5.} due to Regge excitations and/or
new contact interactions.

Generically, as we shall see in Section 5, tree--level string amplitudes involving
four gluons or amplitudes with two gluons and two fermions
are described by the Euler Beta function depending on the kinematic
invariants $s=(k_1+k_2)^2,\ t=(k_1-k_3)^2,\ u=(k_1-k_4)^2$, with $s+t+u=0$ and
$k_i$ the four external momenta.
The whole amplitudes $A(k_1,k_2,k_3,k_4;\ap)$ may be understood as an
infinite sum over $s$--channel poles with intermediate string states
$|k;n\rangle$ exchanged, \cf Figure 2.
\ifig\exchange{Exchange of string Regge excitations.}
{\epsfxsize=0.4\hsize\epsfbox{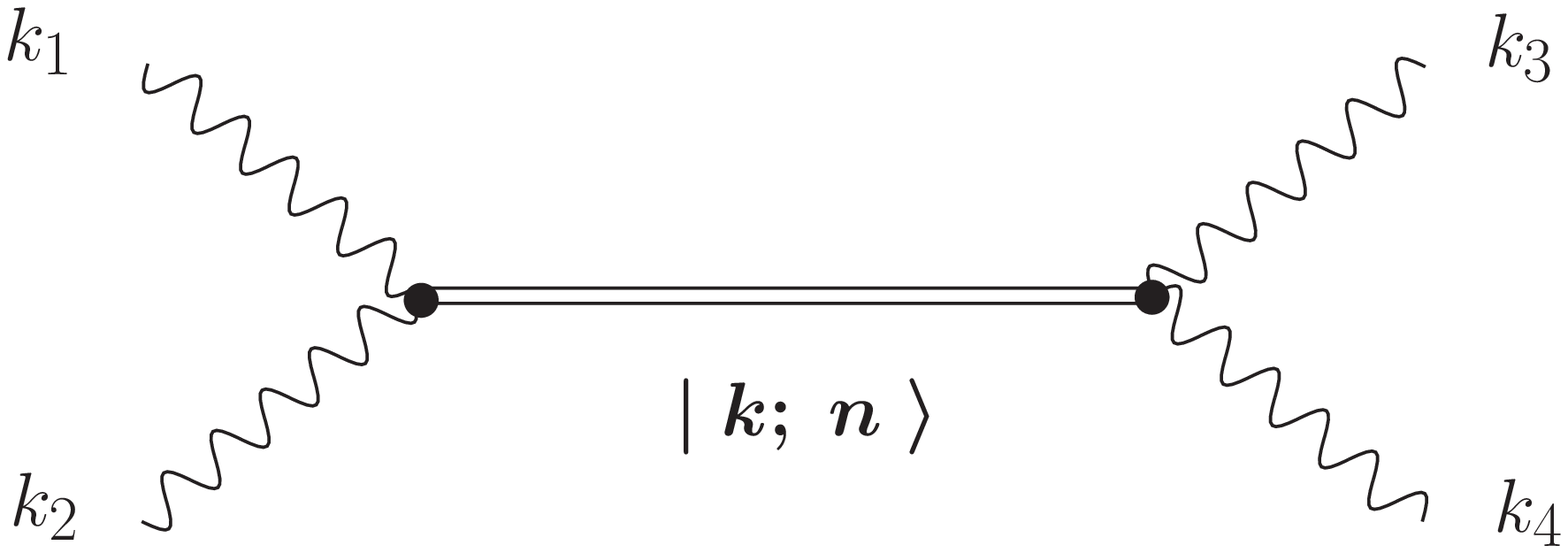}}
\noindent
After neglecting kinematical factors the string amplitude $A(k_1,k_2,k_3,k_4;\ap)$
assumes the form
\eqn\basic{
A(k_1,k_2,k_3,k_4;\ap)
\sim-\fc{\Gamma(-\ap s)\ \Gamma(1-\ap u)}{\Gamma(-\ap s-\ap u)}
=\sum_{n=0}^\infty\ \ \fc{\gamma(n)}{s-M_n^2}}
as an infinite sum over $s$--channel poles at the masses
\eqn\massSR{
M_n^2=M_{\rm string}^2\ n}
of the string Regge excitations.
In \basic\ the residues $\gamma(n)$  are determined by the three--point coupling of 
the intermediate states $|k;n\rangle$ to the external particles and given by
\eqn\DEG{
\gamma(n)={t\over n!}\fc{\Gamma(u\ap+ n)}{\Gamma(u\ap)}=\fc{t}{n!}\ \prod\limits_{j=1}^n
[u\alpha'-1+j]\sim (\ap\ u)^n\ \ \ ,}
with $n+1$ being the highest possible
spin of the state $|k;n\rangle$.

Another way of looking at the expression \basic\ appears when we express each
term in the sum as a power series expansion in $\ap$:
\eqn\Basic{\eqalign{
A(k_1,k_2,k_3,k_4;\ap)\ &\sim\ \fc{t}{s}\ -\ {\pi^2\over 6}\ tu\ \ap^2+\ldots\ .\cr
&\hskip0.3cm\underbrace{\hskip0.5cm}_{n=0}\hskip0.4cm
\underbrace{\hskip2.5cm}_{n\neq 0}}}
In this form \Basic\ the massless state $n=0$ gives rise to a field--theory
contribution ($\ap=0$), while at the order $\ap^2$
all massive states $n\neq 0$ sum up to a finite term.
The $n=0$ term in \Basic\ describes the field--theory contribution to
the diagram \exchange, \eg the exchange of a massless gluon.
On the other hand, the term at the order $\ap^2$ describes a new string contact
interaction as a result of summing up all heavy string states.
Expanding \Basic\ to higher orders in $\ap$ yields an infinite series of new
string contact interactions for the effective field theory.
{}For example, for a four gluon superstring amplitude the first string contact 
interaction is
given by  $\ap^2\ g_{Dp}^{-2}\ \tr F^4$, which represent a correction to YM theory:
\ifig\four{New string contact interaction.}
{\epsfxsize=0.35\hsize\epsfbox{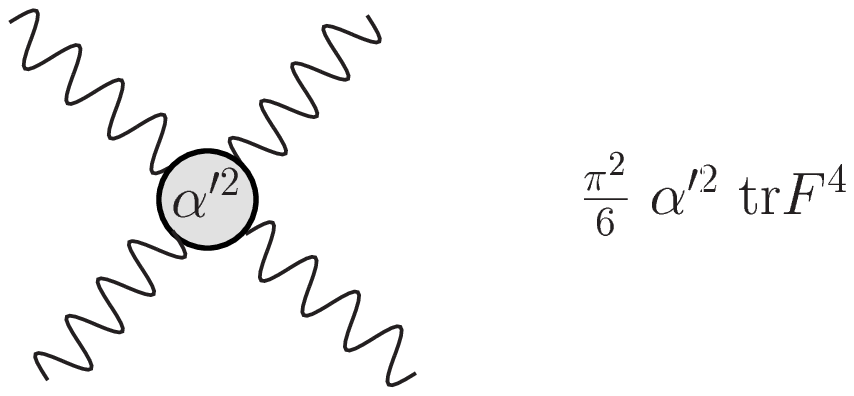}}
\noindent
While the first string correction for four--gluon scattering yields $\ap^2$ contact
terms, the scattering of four chiral fermions yields already a correction
at $\ap$, \cf Section 5.

\newsec{Standard Model from D-branes}

\subsec{Generalities}

We will consider type II
orientifolds\foot{Alternative
GUT constructions from F-theory have recently been discussed in \BeasleyKW.}
with several stacks of $Dp_a$-branes, each being wrapped
around individual compact homology ($p-3$)-cycles $\pi_a$ of the internal space.
Hence the effective open string gauge theories with
groups $G_a$ live in the $(p+1)$-dimensional subspaces ${\IR}^{1,3}\otimes
\pi_a$.

In order to incorporate non-Abelian gauge interactions and to obtain
massless fermions in non-trivial gauge representations, one has to introduce
D-branes in type II superstrings. Specifically there exist three classes of
four-dimensional models:

\vskip0.3cm
\noindent
{\sl (i) Type I compactifications with D9/D5 branes}

\vskip0.3cm
\noindent
This class of IIB models contain different stacks of D9-branes, which wrap 
the entire space ${\cal M}_6$, and which
also possess open string, magnetic,
Abelian gauge fields $F_{ab}$ on their world volumes (magnetized branes). 
This magnetic fields are
in fact required, if one wants to get chiral fermions from open strings.
%In other words, $F_{ab}$ corresponds to open string vector bundles, and this class of models
%is string dual to heterotic string compactifications.
Because of Ramond tadpole cancellation 
one also needs an orientifold 9-plane (O9-plane). 
In addition one can also include D5-branes and
corresponding O5-planes.
%Since the open string gauge fields $F_{ab}$ induce mixed boundary
%conditions on the D-branes, the internal compact space can be regarded as a non-commutative
%space.

\vskip0.3cm
\noindent
{\sl (ii) Type IIB compactifications with D7/D3 branes}

\vskip0.3cm
\noindent
Here we are dealing with different stacks of D7-branes, which wrap different internal 4-cycles,
which intersect each other. The D7-branes can also carry non-vanishing open string
gauge flux $F_{ab}$, which is needed for chiral fermions. 
In addition, one can also allow for D3-branes, which are located at different
point of ${\cal M}_6$. In order to cancel all Ramond tadpoles one needs in general O3- and O7-planes.
A specific class of chiral gauge models can be obtained by placing a stack of D3-branes at a singularity
of the internal space  ${\cal M}_6$.

\vskip0.3cm
\noindent
{\sl (iii) Type IIA compactifications with D6 branes}

\vskip0.3cm
\noindent
This class of models contains intersecting D6-branes, which are wrapped around 3-cycles
of ${\cal M}_6$. Now, orientifold O6-planes $\pi_{O6}$ are needed for Ramond tadpole
cancellation. In general, each stack of $D6_a$-branes, which is wrapped around the
cycle $\pi_a$, is accompanied by the the orientifold mirror stack, wrapped around the reflected
cycles $\pi'_a$.
The chiral massless spectrum is completely fixed
by the topological intersection numbers $I$ of the 3-cycles of the configuration,
\cf Table 2.

\vskip 0.8cm
\vbox{ \centerline{\vbox{
\hbox{\vbox{\offinterlineskip
\def\tablespace{height2pt&\omit&&\omit&&
\omit&\cr}
\def\tablerule{\tablespace\noalign{\hrule}\tablespace}

\hrule\halign{&\vrule#&\strut\hskip0.2cm \hfill #\hfill\hskip0.2cm\cr
& sector && representation && intersection number $I$ &\cr
\tablerule
& $a'\,a$ && $A_a$ && ${1\over 2}\left(\pi'_a\circ\pi_a+\pi_{O6}\circ\pi_a\right)$   &\cr
&  $a'\, a$ && $S_a$ && ${1\over 2}\left(\pi'_a\circ\pi_a-\pi_{O6}\circ\pi_a\right)$ &\cr
& $a\,b$ && $(\overline{N}_a,N_b)$ && $\pi_{a}\circ \pi_b$ &\cr
&  $a'\, b$ && $({N}_a,{N}_b)$ && $\pi'_{a}\circ \pi_b$ &\cr
}\hrule}}}}
\centerline{
\hbox{{\bf Table 2:}{\it ~~ Intersection of 3-cycles $\pi_a,\pi_b$, mirror cycles $\pi'_a$ and orientifold
plane $\pi_{O6}$}}}}
\vskip 0.5cm

In general some of the string $U(1)$'s are
anomalous and receive masses due to Green-Schwarz mechanism.
However, for intersecting brane worlds it may also happen that via
axionic couplings  some anomaly-free Abelian gauge groups become massive.
The condition that a linear combination $U(1)_Y=\sum_i c_i U(1)_i$
remains massless reads:
\eqn\green{    \sum_i  c_i\, N_i\, \left(\pi_i-\pi_i'\right)=0\ .}
In general, if the hypercharge is such a linear combination of $U(1)$'s,
$Q_Y=\sum_i c_i Q_i$, then the gauge coupling is given by
\eqn\hgauge{   {1\over \alpha_Y}=\sum_i {N_i\, c_i^2\over 2}\ 
                         {1\over \alpha_i}\ , }
where we have taken into account that the $U(1)$'s
are generically not canonically normalized (for all possible hyper charge
assignments in D-brane orientifolds see \AnastasopoulosDA).

In the following we will describe some local type IIA/IIB D-brane
configurations that lead to the SM
in a very economic way.

\subsec{Three stack D-brane models}

Here one starts with three stacks of D-branes with initial gauge symmetries:
\eqn\gaugein{\eqalign{ U(3)\times U(2) \times U(1)\times U(1)\ .}}
The (left-handed) SM spectrum is shown in Table 3.

\vskip 0.8cm
\vbox{
\centerline{\vbox{
\hbox{\vbox{\offinterlineskip
\def\tablespace{height2pt&\omit&&\omit&&\omit&&
\omit&\cr}
\def\tablerule{\tablespace\noalign{\hrule}\tablespace}

\hrule\halign{&\vrule#&\strut\hskip0.2cm\hfill #\hfill\hskip0.2cm\cr
\tablespace
& matter  && $SU(3)\times SU(2)\times U(1)^3$ && $U(1)_Y$ && $U(1)_{B-L}$ &\cr
\tablerule
& $ q$ && $({\bf 3},{\bf 2})_{(1,1,0)}$ &&  ${1\over 3}$ &&
${1\over 3}$ & \cr
\tablespace
&  $\bar u$ && $(\overline{\bf 3},{\bf 1})_{(2,0,0)}$ &&
      $-{4\over 3}$ &&  $-{1\over 3}$  & \cr
\tablespace
& $\bar d$ && $ (\overline{\bf 3},{\bf 1})_{(-1,0, 1)}$ &&
     ${2\over 3}$ && $-{1\over 3}$   & \cr
\tablerule
& $l$ && $({\bf 1},{\bf 2})_{(0,-1,1)}$ && ${-1}$ &&
         ${-1}$   & \cr
\tablespace
& $\bar e$ && $({\bf 1},{\bf 1})_{(0,2,0)}$ && ${2}$ &&
          ${1}$   & \cr
\tablespace
& $\bar\nu$ && $({\bf 1},{\bf 1})_{(0,0,-2)}$ && ${0}$ &&
                ${1}$    & \cr
\tablespace}\hrule}}}}
\centerline{
\hbox{{\bf Table 3:}{\it ~~ Left-handed fermions for the 3 stack
model.}}}
}
\vskip 0.5cm
\noindent
The hypercharge $Q_Y$ is given as the following  linear combination
of the three $U(1)'$s:
\eqn\hyperthreestack{Q_Y=-{2\over 3}\ Q_a+\h\ Q_b\ .}
Here one is forced to realize the left-handed $(\bar u,\bar c,\bar t)$-quarks
in the antisymmetric
representation of $U(3)$, which is the same as the anti-fundamental
representation $\overline 3$.
Note that  the three stack models with antisymmmetric  matter  are dual to
the D3-brane quivers at CY
singularities \refs{\BerensteinNK,\VerlindeJR,\MalyshevZZ}.
Alternative bottom-up constructions of the SM via D-branes can be found in 
\AntoniadisQM.

\subsec{Four stack D-brane models}

One of the most common ways to realize the SM is by considering four stacks of
D-branes.
%To be specific, let us consider D6-branes in type IIA orientifolds.
There are several simple ways to embed the SM
gauge group into products of unitary and symplectic gauge groups (see \report).
We will use as a prototype model four stacks of D-branes with
gauge symmetries:
\eqn\gaugein{U(3)_a\times U(2)_b \times U(1)_c\times U(1)_d\ .}
The intersection pattern of the four stacks of D6-branes can be
depicted as in Figure 4.
\ifig\figi{Intersection pattern of four stacks of D6-branes giving rise to the
MSSM.}
{\epsfxsize=0.7\hsize\epsfbox{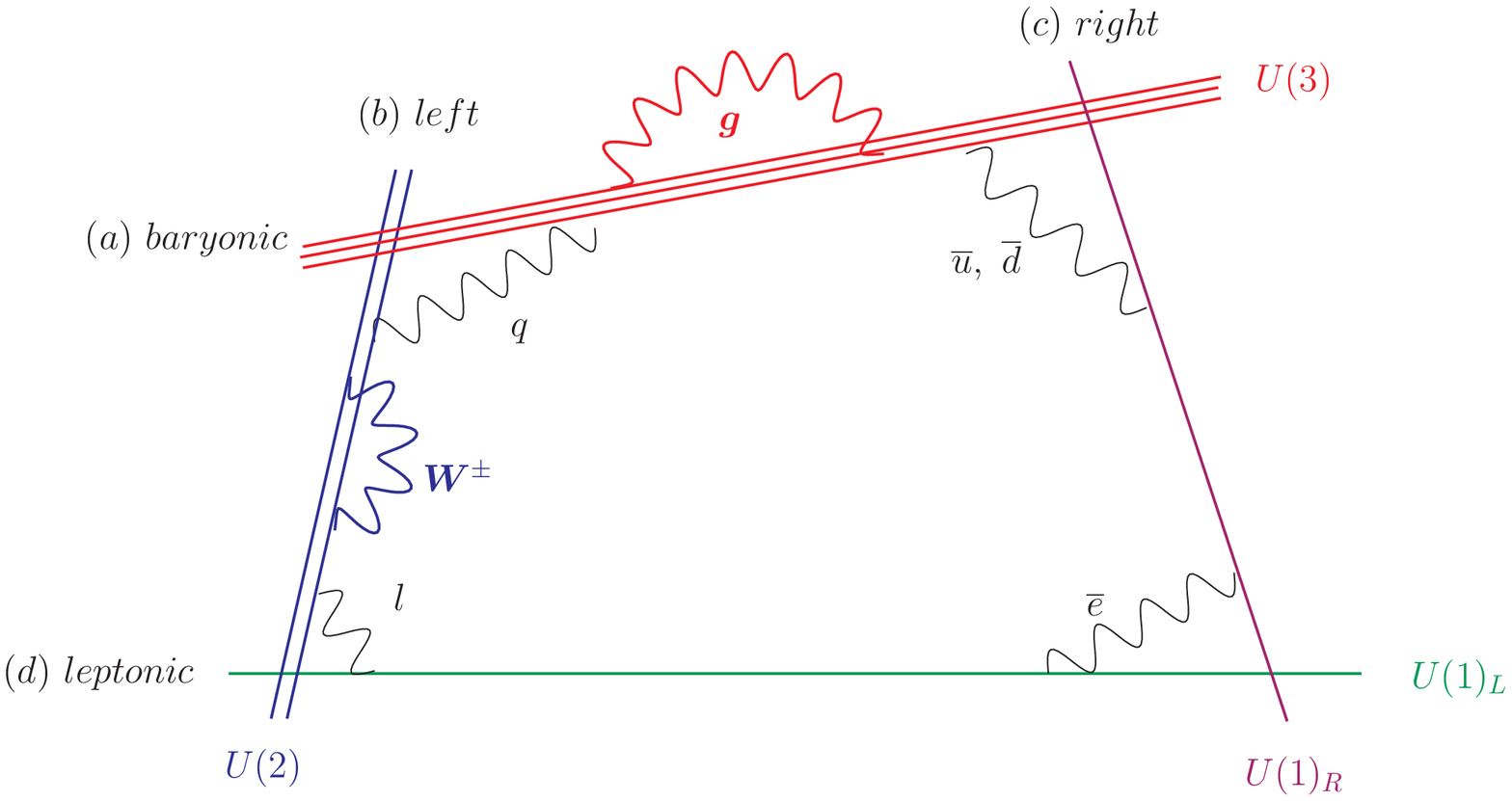}}
\noindent The chiral spectrum of the intersecting brane world
model should be identical to the chiral spectrum
of the SM particles.
In type IIA, this fixes uniquely the intersection numbers of the
3-cycles, $(\pi_a,\pi_b,\pi_c,\pi_d)$,
the four stacks of D6-branes are wrapped on.

There exist several ways to embed the hypercharge $Q_Y$ into the four $U(1)$
gauge symmetries.
The standard electroweak 
hypercharge $Q_Y^{(S)}$ is given as the following  linear combination
of three $U(1)'$s
\eqn\hyper{Q_Y^{(S)}={1\over 6}\ Q_a+\h\ Q_c+\h\ Q_d\ .}
Therefore, in this case the gauge coupling of the hypercharge
is given as
\eqn\hgaugebb{   {1\over \alpha_Y}={1\over 6}\ {1\over \alpha_a} +
                                {1\over 2}\ {1\over \alpha_c} +
                                {1\over 2}\ {1\over \alpha_d}\ .}

Now we turn to the particle content of our prototype model.
In compact orientifold compactifications each stack of D-branes is 
accompanied by a orientifold mirror
stack of D$'$-branes. In the next Section about the amplitudes, we will not
make a difference between
the the D-brane and the mirror D$'$-branes. Hence we will use in the
following the indices
$a,b,c,d$ collectively for the D-branes as well as for their mirror branes. Then
self-intersections among D-branes include intersections between D-
and D$'$-branes. Furthermore, for simplicity, we will
suppress from the spectrum those open string states which one also gets from
intersections between
D-branes and orientifold planes. With these restrictions
the left-handed fermion spectrum
for our prototype model is presented in Table~4.

\vskip 0.8cm
\vbox{ \centerline{\vbox{
\hbox{\vbox{\offinterlineskip
\def\tablespace{height2pt&\omit&&\omit&&\omit&&
\omit&\cr}
\def\tablerule{\tablespace\noalign{\hrule}\tablespace}

\hrule\halign{&\vrule#&\strut\hskip0.2cm \hfill #\hfill\hskip0.2cm\cr
&
particle && $U(3)_a\times U(2)_b \times U(1)_a\times U(1)_b\times U(1)_c \times U(1)_d$ && mult. &\cr\tablerule
&$q$ && $(\3,{\2})_{1,-1,0,0}+(\3,\2)_{1,1,0,0}$  & & $I_{ab}$   &\cr
\tablerule
&$\bar u$ &&  $(\ov{\3},1)_{-1,0,-1,0}+(\ov{\3},1)_{-1,0,0,-1}$  && $I_{ac}+I_{ad}$ &\cr\tablerule
&$\bar d$ &&  $(\ov{\3},1)_{-1,0,1,0}+(\ov{\3},1)_{-1,0,0,1}$ & & $I_{ac}+I_{ad}$ &\cr
&$\bar d'$ && $(\ov{\3}_A,1)_{2,0,0,0}$  &&  $\h I_{aa}$ &\cr \tablerule
&$l$ && $(1,\2)_{0,1,-1,0}+(1,\2)_{0,1,0,-1}$  &&         $I_{bc}+I_{bd}$       &           \cr
& &&+ $(1,{\2})_{0,-1,-1,0}+(1,{\2})_{0,-1,0,-1}$ &&      &\cr\tablerule
&$\bar e$ && $(\1,\1)_{0,0,2,0}$ && $\h I_{cc} $&\cr
&$\bar e'$ && $(\1,\1)_{0,0,0,2}$ && $\h I_{dd} $&\cr
&$\bar e''$ && $(\1,\1)_{0,0,1,1}$ && $I_{cd}$
&\cr
%\tablerule
}\hrule}}}}
\centerline{
\hbox{{\bf Table 4:}{\it ~~ Chiral spectrum for the four stack model with $Q_Y^{(S)}$}}}
%\multicolumn{3}{|c|}{$U(3)_a\times U(2)_b \times U(1)_a\times U(1)_b\times
%U(1)_c \times U(1)_d$ %%with $Q_Y^{(S)}$}&\cr
}
\vskip 0.5cm

\noindent
To derive three generations of quark and leptons, the intersection
number in Table 4 must satisfy certain phenomenological restrictions:
We must have $I_{ab}=3$. From the left-handed anti u-quarks, we get that
$I_{ac}=3$, and likewise for the two types of left-handed anti d-quarks, we infer that
$I_{ac}+I_{ad}+{1\over 2}I_{aa}=3$.
In the lepton sector we require  that $I_{bc}+I_{bd}=3$ and
${1\over 2}(I_{cc}+I_{dd})+I_{cd}=3$.

\newsec{Embedding of Standard Model D--branes into  large volume Calabi--Yau spaces}

Let us now discuss large extra dimensions in the context of
string
compactifications. In fact,  it is not
completely straightforward to construct SM-like D-brane models on CY spaces with large transverse dimensions.
In order to combine D-branes with SM particle content with the scenario of large extra dimensions, one
has to consider specific types of CY compactifications.
The three or four stacks of intersecting D-branes that give rise to the spectrum of the SM are
just local modules that have to be embedded into a global large volume CY-manifold in order to obtain a consistent
string compactification. For internal consistency several tadpole and stability conditions have to be
satisfied that depend on the details of the compactification, such as background fluxes etc. In this
work we will not aim to provide fully consistent orientifold compactifications with all tadpoles cancelled, since it is enough for us to
know the properties of the local SM D-brane modules for the
computation of the  scattering amplitudes among the SM open strings. However it is important to
emphasize that in order to allow for large volume compactification, the D-branes eventually cannot be wrapped
around  untwisted 3- or 4-cycles of a compact
torus or of toroidal orbifolds, but one has to consider twisted, blowing-up cycles of an orbifold or more general CY spaces
with blowing-up cycles.
The reason for this is that wrapping the three or four
stacks of D-branes around internal cycles of a six-torus or untwisted orbifold cycles, the volumes of these
cycles involve the toroidal radii. Therefore these volumes cannot be kept small
while making the overall volume of the six-torus very big. Hence,
the SM D-branes  must be wrapped around small cycles inside a blown up orbifold
or a CY manifold. Other
cycles have to become large, in order to get a CY space with large volume and a low string scale
$M_{\rm string}$.

The embedding of the local SM D-brane module into a large CY manifold is
depicted in Figure 5.
\ifig\figii{Embedding of the SM-branes into a large CY space.}
{\epsfxsize=0.6\hsize\epsfbox{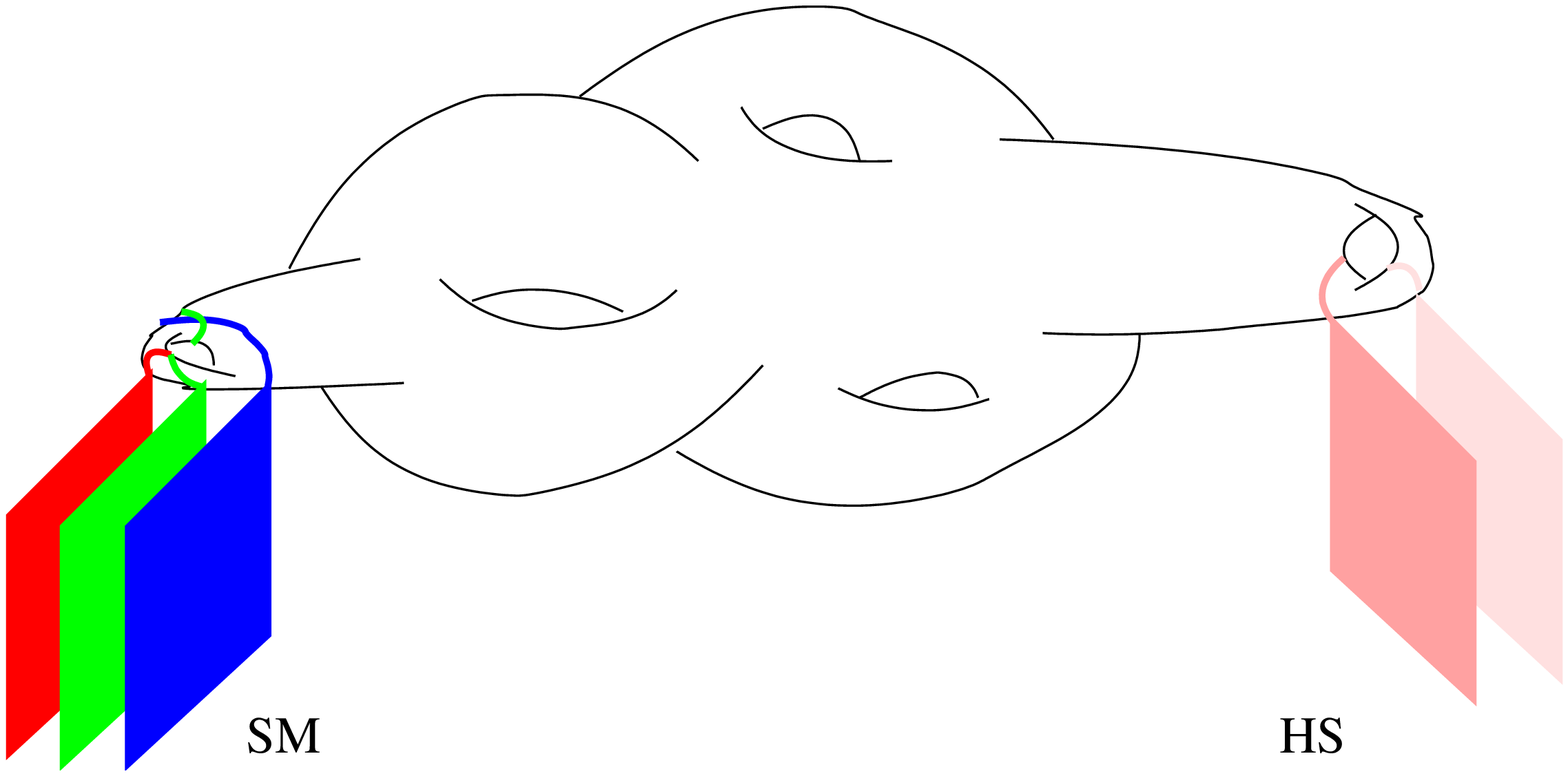}}
\noindent
At some other corner of the CY manifold there can be possibly other D-branes,
which do not
intersect the SM branes and build a hidden gauge sector of the theory.

In Section 5 we shall compute open string disk four--point amplitudes
involving SM matter fields. For those amplitudes
involving four gauge bosons or two gauge bosons and two matter fermions, the
amplitudes do not depend on the geometry of the underlying CY spaces.
On the other hand, the four--fermion amplitudes
depend on the internal CY geometry and topology. Concretely, the
four--fermion amplitudes in general
depend on the CY intersection numbers, and also on the rational
instanton numbers of the CY space.
However, to perform the open string CFT computations for the scattering amplitudes
of matter fields we shall assume that the SM D--branes are wrapped around
flat, toroidal like cycles. Therefore the four--fermion
amplitudes are functions of toroidal wrapping numbers.
Eventually switching from our toroidal-like
results to more general CY
expressions, some of the factors, which depend on the toroidal geometry, have
to be replaced
by geometrical or topological CY parameters.
However, the kinematical structure of the matter
field amplitudes is universal and not affected
by the underlying CY geometry. At any rate, as we shall argue at the end of
Section 5, for the case that the longitudinal brane directions are somewhat
greater than the string
scale $\Ms$ the four--fermion couplings depend only on the local
structure of the brane intersections, but not on the global CY geometry.

\subsec{Type IIB large volume compactifications with wrapped D7-branes}

In type IIB orientifolds we assume that the D7-branes are wrapped around
4-cycles inside a CY-orientifold.
The relation \vol\ for the volume $V_6$ applies only for
toroidal and orbifold compactifications. Therefore we shall generalize
the expressions \Planckmass\ for $M_{\rm Planck}$ to the case of large  volume CY
compactifications. In the string frame\foot{In the Einstein frame the
K\"ahler moduli $t_k$ are multiplied by the factor $e^{-\h\phi_{10}}$.
Therefore, in the Einstein frame the CY volume reads $V_6=\fc{1}{6}e^{-\fc{3}{2}\phi_{10}}\
\kappa_{ijk}\ t_it_jt_k$.}, the volume~$V_6$ of a CY space
$X$ is given by
\eqn\calabivolume{
V_6={1\over 3!}\int_XJ\wedge J\wedge J={1\over 6}\ \kappa_{ijk}\ t_it_jt_k\ ,}
with $t_i$ ($i=1,\dots , h^{1,1}$) the (real) K\"ahler moduli
in the string basis and $\kappa_{ijk}$ the triple
intersection numbers of $X$. The K\"ahler form $J$ is expanded
w.r.t. a base $\lbrace\hat D_i\rbrace$ of the cohomology
$H^{1,1}(X,{Z})$ as $J=\sum\limits_{i=1}^{h^{1,1}}{t}_i\
\hat D_i$. Without loss of generality we restrict
to orientifold projections with $h^{1,1}_-=0$, $h^{1,1}_+=h^{1,1}$.
On the other hand, the real parts of the physical K\"ahler moduli $T_i$ correspond to the
volumes of the CY homology four-cycles $D_k$ and
are computed from the relation:
\eqn\physT{
T_i={1\over 2}\ \int_{D_i}J\wedge J={\partial V_6\over\partial{t}_i}=\h\
\kappa_{ijk}\ t_jt_k\ .}
It follows that the volume $V_6$ of $X$
becomes a function of degree 3/2 in the K\"ahler moduli~$T_i$:
\eqn\voleinst{
V_6= {1\over 3!}\ \int_XJ\wedge J\wedge J={\cal O}(T_i^{3/2})\ .}

For D7--branes wrapped around the four-cycle $D_k$, the corresponding gauge 
coupling constant takes the form 
\eqn\gaugedseven{
g^{-2}_{D7_{k}}=(2\pi)^{-1}\ \alpha'^{-2}\ T_k\ ,}
in analogy\foot{On the other hands, for (space--time filling)
D3-branes the corresponding gauge coupling constant is given by:
\eqn\gaugedthree
{g^{-2}_{D3}=(2\pi)^{-1}e^{-\phi_{10}}\equiv S\ .}} to \eqq \gaugeAB.
In the case of magnetic F-fluxes on the D7-brane world--volume
the gauge couplings \gaugedseven\ 
receive an additional $S$-dependent contribution, \cf \LMRS.

Now we consider CY manifolds which allow for large
volume compactification. Here
we assume that a set of four-cycles $D^b_{\alpha}$ ($\alpha=1,\dots ,
h^{1,1}_b$) can be chosen arbitrarily large while
keeping the rest of the four-cycles $D^s_{\beta}$ ($\beta=1,\dots
,h^{1,1}-h^{1,1}_b$) small, \ie $T^b_{\alpha}\gg T^s_{\beta}$.
Since we want the gauge couplings of the SM gauge groups to have finite,
not too small values, we must assume that the SM gauge bosons originate from
D7-branes wrapped around the small 4-cycles $D^s_{\beta}$.
This splitting of the four-cycles into big and small cycles
is only possible, if the CY triple
intersection numbers form a specific pattern.
In addition, the Euler number of the CY space
must be negative, i.e. $h^{2,1}>h^{1,1}>1$.

For a simple class of CY spaces with this property the overall volume
$V_6$ is controlled by
one big four-cycle $T^b$.
In this case it has been shown \refs{\ConlonKI,\ConlonXV,\BergWT}  that one may indeed
find minima of the scalar potential, induced
by fluxes and radiative corrections in the K\"ahler potential,
that allow for $T^b\gg T^s_\beta$.
For these CY-spaces, the volume has to take the  form
\eqn\largevola{
V_6\sim  (T^b)^{3/2}-h(T^s_\beta)\ ,}
where $h$ is a homogeneous function of the small K\"ahler moduli $T^s_\beta$ of
degree 3/2.
E.g. one may consider the following more specific volume form:
\eqn\largevol{
V_6\sim (T^b)^{3/2}-\sum_{\beta=1}^{h^{1,1}-1}(T^s_\beta)^{3/2}\ .}
Looking from the geometrical point of view, these models have a "Swiss cheese"
like structures, with holes
inside the CY-space given by the small four-cycles.

The simplest example of a Swiss cheese example is the CY manifold
${\bf P}_{[1,1,1,6,9]}[18]$ with $h^{1,1}=2$.
In terms of the 2-cycles the volume is given by
\eqn\volpa{
V_6=6\ ({t}_1^{3}+{t}_2^{3})\ .}
According to \physT\ the corresponding 4-cycle volumes become:
\eqn\fourcyces{\eqalign{
T^b & ={\partial V_6\over\partial {t}_1}=18\ 
{t}_1^2\quad\Longleftrightarrow\quad {t}_1=
{\sqrt{T^b}\over 3\sqrt 2}\ ,\cr
T^s & ={\partial  V_6\over\partial {t}_2}=18\ 
{t}_2^2\quad\Longleftrightarrow\quad {t}_2=
-{\sqrt{T^s}\over 3\sqrt 2}\ .}}
Then the volume can be written  in terms of the 4-cycles as
\eqn\volp{
V_6={1\over 9\sqrt 2}\ \biggl[\ (T^b)^{3/2}-(T^s)^{3/2}\ \biggr]\ .}
Another interesting Swiss cheese example, the CY manifold ${\bf
P}_{[1,3,3,3,5]}[15]$ with $h^{1,1}=3$ has recently been discussed in 
\BlumenhagenSM\ in the context
of a phenomenologically attractive CY orientifold model where the small cycles
are wrapped by D7-branes.

However, in order to accommodate the SM with at least three or four stack of
wrapped D7-branes we need more small 4-cycles.
Therefore we assume that there exist
CY spaces which have a set of small, blowing-up
four-cycles that do not intersect the big cycles, i.e. the CY volume is of the form
\eqn\calabivolume{
V_6={t}_1^3+{1\over 6}
\kappa_{\beta_i\beta_j\beta_k}{t}_{\beta_i}{t}_{\beta_j}{t}_{\beta_k}\ .}
The SM D7-branes are wrapped around the four-cycles $D_{\beta_i}$ with volumes
$T_{\beta_i}$ that are kept small. At the same time
one is
allowed to choose   the four-cycle volume $T^b$ to be large.
Let us give one example of a hypothetical CY space
with
three distinct four-cycles  $D_\beta$
($\beta=1,2,3$), who still have toroidal-like  intersection numbers
$\kappa_{123}=1$, and whose intersections
with the large four-cycles are absent. The dual two-cycles locally form a
$T^2\times T^2\times T^2$ torus inside the CY space.
Its volume form  is assumed to be
\eqn\hypcal{
V_6={t}_1^3+{t}_2{t}_3{t}_4+\dots  \ ,}
where the dots stand for the contribution of other possible cycles.
The big four-cycle is just
\eqn\BIG{
T^b  ={\partial V_6\over\partial {t}_1}=3{t}_1^2\ ,}
and the three small four-cycles intersect in one point and are given as
\eqn\small{\eqalign{
T_1^s&={\partial V_6\over\partial {t}_2}={t}_3{t}_4, \quad T_2^s={\partial
V_6\over\partial {t}_3}={t}_2{t}_4,\quad
T_3^s={\partial  V_6\over\partial {t}_4}={t}_2{t}_3
\ ,\cr
{t}_2&=-\sqrt{T_2^sT_3^s\over T_1^s},\quad  {t}_3=-\sqrt{T_1^sT_3^s\over T_2^s},\quad
{t}_4=-\sqrt{T_1^sT_2^s\over T_3^s}\ .}}
In term of the four-cycle volumes, $ V$ is then given as
\eqn\hypcal{
V_6={1\over 3\sqrt 3}\ ({T}^b)^{3/2}-({T}_1^s{T}_2^sT_3^s)^{1/2}-\dots\ .}
Hence, this would-be CY has the form of a Swiss cheese with geometry,
where the intersecting four-cycle
holes cut themselves a local $T^2\times T^2\times T^2$ space out of the entire
CY manifold.
However we do not know if this kind of CY does exist.

Let us comment briefly on IIB large volume orientifolds with D5-branes, which
are wrapped around
CY 2-cycles. Again, the two have to be kept small, whereas the overall
volume of the CY-space
is very large. Models of this kind are e.g. possible on toroidal orbifolds,
where the D5-branes
are located at a singularity in a transversal two-dimensional space, as
discussed in \CremadesDH.

\subsec{Type IIA large volume compactifications with wrapped D6-branes}

Type IIA orientifolds with wrapped D6-branes can be obtained from the type IIB
compactifications via T-duality resp. via mirror transformations, which
basically exchange the role of the K\"ahler moduli
with the role of the complex structure moduli and vice versa, i.e going from
the type IIB CY space $X$ to its type IIA mirror space, denoted by $\tilde X$.
The volume of $\tilde X$ is still given by eqs. \vol\ resp. \calabivolume, now expressed in terms of properly
defined type IIA radii resp. IIA K\"ahler moduli $T_i$, which are the 2-cycle volumes
on $\tilde X$.
Moreover the orientifold O3/O7-planes in type IIB become O6-planes in type IIA, which are wrapped around
certain homology 3-cycles $\Pi_{O6}$ inside $\tilde X$. Similarly, the type IIB D3/D7-branes become D6-branes, wrapped
around homology 3-cycles $\Pi_a$, which are suitably embedded
into the large volume CY space $\tilde X$.
The $\Pi_a$  intersect each other at angles $\theta_{ab}$, and their intersection angles with
orientifold cycles $\Pi_{O6}$ are denoted by $\theta_a$.

The corresponding D6-brane gauge coupling constants
are proportional to the volumes of the wrapped 3-cycles, i.e.:
\eqn\dsix{
g^{-2}_{D6_a}=(2\pi)^{-1}\ \alpha'^{-2}\ {\rm Vol}(\Pi_a)\ .}
The volume of the cycle $\Pi_a$ is given in terms of the associated complex
structure moduli $U_a$ of $\tilde X$.
To accommodate type IIA orientifolds with low string scale and large overall volume,
the corresponding complex structure moduli $U^s_\beta$, around which the SM D6-branes are
wrapped, must be small
compared to the volume of $\tilde X$
to achieve finite values for the corresponding gauge coupling constants.
As in type IIB, the CY spaces $\tilde X$ must satisfy certain
restrictions for large volume
compactifications to be possible. In principle the structure of the allowed
IIA CY spaces can be
inferred from type IIB via mirror symmetry. E.g. one can wrap the D6-branes
around certain rigid
(twisted) 3-cycles of orbifold compactifications (see e.g. \BlumenhagenTN),
which can be kept small, whereas
the overall volume is made very large.

To perform the computation of the matter field scattering amplitudes,
as in type IIB  we assume
that the 3-cycles, which are are wrapped by the SM D6-branes,
are flat and have a kind of toroidal like intersection pattern.
Specifically, we assume that the SM sector is wrapped around 3-cycles inside
a local $T^2\times T^2\times T^2$,
and the D6-brane wrappings around the tree 2-tori are described by wrapping numbers
$(n^i_a,m^i_a)$ $(i=1,2,3)$, where the lengths $L^i_a$ of the wrapped
1-cycles in each $T^2$ is given by the following
equation:
\eqn\Wrapping{
L^i_a=\sqrt{(n^i_a)^2\ (R_i)^2+(m^i_a)^2\ (R_{i+1})^2}\ .}
Then the gauge coupling on  a D$6$--brane which is wrapped around a
$3$--cycle, is \refs{\phencvetic,\tye,\CIM}:
\eqn\gaugesix{
g_{D6_{a}}^{-2}=(2\pi)^{-1}\ \ap^{-3/2}\ e^{-\phi_{10}}\
\prod_{i=1}^{3} L^i_a\ .}
Here, the $3$--cycle $\Pi_a$ is assumed to be a direct product of
three $1$--cycles with wrapping numbers $(n^i,m^i)$  w.r.t. a pair of two internal
directions\foot{In type IIB orientifolds, the gauge coupling of a D7-brane,
wrapped around the 4-cycle $T^{2,j}\times T^{2,k}$ with wrapping numbers
$m^j$, $m^k$ and magnetic fluxes
$f^j$, $f^k$ is (\cf \gaugedseven)
\eqn\gaugeflux{
g^{-2}_{D7_i}=(2\pi)^{-1}\ \alpha'^{-2}\ |m^jm^k|\ \re(T_j-f^jf^kS)\ .}}.
In terms of the corresponding three complex structure moduli $U_i$ of the $T^2$'s
this equation becomes
\eqn\gaugesixa{
g_{D6_{a}}^{-2}=(2\pi)^{-1}e^{-\phi_4}\prod_{i=1}^3\
{|n^i_a-m^i_aU_i|\over\sqrt{\im(U_i)}}\ .}
Finally, the intersection angles of the D6-branes with the O6-planes along the
three $y_i$ directions can be expressed as
\eqn\anglesint{
\tan(\theta^i_a)={m^i_aR_{i+1}\over n^i_aR_i}\ ,}
and the D6-brane intersection angles are simply given as $\theta_{ab}^i=
\theta_b^i-\theta_a^i$.
More details about the effective gauge couplings,
and also about matter field metrics of these
kind of intersecting D-brane models can be found in \refs{\LMRS,\LustFI}.

\newsec{Four--point string amplitudes of gauge and matter Standard Model fields}

In this Section we compute the four-particle amplitudes relevant to LHC physics at the leading order of string perturbation theory, with the string disk world-sheet incorporating the propagation of virtual Regge string excitations at the tree level of effective field theory\foot{The SM amplitudes reappear in the formal 
$\alpha'\to 0~\ (\Ms\to\infty)$ limit.}. With the protons colliding at LHC, there are always two incident partons, gluons or quarks, while the two outgoing particles are partons fragmenting into jets, electroweak gauge bosons or leptons produced via the Drell--Yan mechanism. In  all these processes, the baryonic stack of branes plays a special role. We will call it stack $a$. Note that in addition to gluons $g$ in the adjoint representation of $SU(N_a)=SU(3)$ color group, this stack gives rise to a color singlet gauge boson $A$ coupled to the baryon number. This boson combines with gauge bosons associated to other stacks to form the vector boson coupled to electroweak hypercharge. We will not enter into details of the mixing mechanism because they are model-dependent. Thus
we simply consider $A$ as one of the particles possibly produced in parton collisions.
Starting from the amplitudes involving $A$ and gauge bosons associated to different stacks,
one can easily obtain
the physical amplitudes describing the production of photons, $Z^0$, or hypothetical $Z'$s
in the framework of specific models. Since four-point disk amplitudes can involve as many as four different stacks, it is very important to establish a transparent notation. In this Section, we are still using the string (hep-th) conventions, with the metric signature $(-++\,+)$ and some kinematic invariants defined in the string units. In the next Section, we will make transition to the conventions used in experimental literature (hep-ex).

\subsec{Notation and Conventions}
\item{$\bullet$} $a,~b,~c,~d$: Dp-brane stacks. Associated gauge groups are\hfill\break $G_a=SU(3)\times U(1)_a,~ G_b=SU(2)\times U(1)_b,~G_c=U(1)_c,~G_d=U(1)_d$\hfill\break
gauge couplings: $g_{Dp_a}\equiv g,~ g_{Dp_b},~
g_{Dp_c},~  g_{Dp_d}$
\item{$\bullet$} Summary of index notation\hfill\break
indices $a,~\alpha$ $(\alpha=1,2,3)$ (possibly with subscripts): are associated to stack $a$\hfill\break
indices $b,~\beta$ $(\beta=1,2)$ (possibly with subscripts): are associated to stack $b$\hfill\break
indices $c,~\gamma$ $(\gamma=1)$ (possibly with subscripts): are associated to stack $c$\hfill\break
indices $d,~\delta$ $(\delta=1)$ (possibly with subscripts): are associated to stack $d$
\item{$\bullet$} gauge bosons:\hfill\break $(A^a,A^0)=(g^{a}, A)$:~  indices $a$ labels adjoint representation of $G_a$\hfill\break
$B^{b}$:~indices $b$ labels adjoint representation of $G_b$\hfill\break\vdots
\item{$\bullet$} quarks (compare with Table 4):\hfill\break
$q^{\alpha}_{\beta}:~$ left-handed quarks\hfill\break
$ u^{\alpha}_{\gamma},~ u^{\alpha}_{\delta},~d^{\alpha}_{\gamma}~,d^{\alpha}_{\delta}:~$ right-handed quarks\hfill\break
$\bar q_{\alpha}^{\beta}~$ right-handed antiquarks\hfill\break
$\bar u_{\alpha}^{\gamma},~\bar u_{\alpha}^{\delta},~\bar d_{\alpha}^{\gamma},~\bar d_{\alpha}^{\delta}:~$ left-handed antiquarks\hfill\break
$G_a$: superscripts $\alpha$ label fundamental rep. ${\bf N}_a={\bf 3}$, subscripts $\alpha$ label rep. ${\bf \bar N}_a={\bf\bar 3}$ \hfill\break
$G_b$: superscripts $\,\beta$ label rep. ${\bf N}_b={\bf 2}$, subscripts $\beta$ label ${\bf \bar N}_b={\bf\bar 2}$ \hfill\break
\vdots
\item{$\bullet$} leptons (compare with Table 4)\hfill\break
$l^{\gamma}_{\beta},~l^{\delta}_{\beta}:~$ left-handed leptons\hfill\break
$ e^{\gamma}_{\delta},~e^{\gamma}_{\gamma},~e^{\delta}_{\delta}:~$ right-handed leptons\hfill\break
$\bar l_{\gamma}^{\beta},~\bar l_{\delta}^{\beta}:~$ right-handed antileptons\hfill\break
$\bar e_{\gamma}^{\delta},~\bar e_{\gamma}^{\gamma},~\bar e_{\delta}^{\delta}:~$ left-handed antileptons
\item{$\bullet$} Chan-Paton factors (all traces in the fundamental representation): gauge bosons\hfill\break
gluons $A^a=g^a$:~ $[T^a]^{\alpha_1}_{\alpha_2}$~ generators of $SU(N_a)=SU(3)$\hfill\break
$A^0=A$ boson:~~ $\,[T^0]^{\alpha_1}_{\alpha_2}={1\over\sqrt{2N}}\delta^{\alpha_1}_{\alpha_2}$\hfill\break
$B^b$ bosons: $[T^{\,b}]^{\beta_1}_{\beta_2}$~ generators of $G_b$\hfill\break\vdots\hfill\break
The generators are normalized according to ${\rm Tr}(T^{a_1}T^{a_2})={1\over 2}\delta^{a_1a_2}$.
\item{$\bullet$} Chan-Paton factors: quarks and leptons\hfill\break
left-handed quarks:~~~ $[T^{\alpha_1}_{\,\beta_1}]_{\alpha_2}^{\beta_2}=\delta^{\alpha_1}_{\alpha_2}
\delta^{\beta_2}_{\beta_1}$\hfill\break
right-handed quarks:~ $[T^{\alpha_1}_{\,\gamma_1}]_{\alpha_2}^{\gamma_2}=\delta^{\alpha_1}_{\alpha_2}
\delta^{\gamma_2}_{\gamma_1}$\hfill\break\vdots
\item{$\bullet$} Wave functions: fermions\hfill\break
left-handed helicity:~ $\;u^{-\lambda}(k_i)\equiv u^{\lambda}_i$\hfill\break
right-handed helicity: $\bar u^+_{\dot\lambda}(k_i)~ \,\equiv\bar u_{i\dot\lambda}$\hfill\break
$k_{\mu}\sigma^{\mu}_{\lambda\dot\lambda}=u_{\lambda}(k)\bar u_{\dot\lambda}(k)$\hfill\break
helicity notation:~ $\langle ij\rangle=u^{\lambda}_iu_{j\lambda}\equiv u_iu_j
\qquad [ij]=\bar u_{i\dot
\lambda}\bar u^{\dot\lambda}_j\equiv\bar u_i\bar u_j$
\hfill\break important property: $|\langle ij\rangle|=|[ij]|=\sqrt{-2k_ik_j}~,\qquad
\;\langle ij\rangle[ij]=2k_ik_j$
\item{$\bullet$} Wave functions: vector bosons (with reference momentum $r$)\hfill\break
left-handed polarization:~~ $\displaystyle \;\xi^{-\mu }(k_i,r)={u^{\lambda}(i)\sigma^{\mu}_{\lambda\dot\lambda}\bar u^{\dot\lambda}(r)\over\sqrt{2} \,[r\,i]}$\hfill\break
right-handed polarization:~ $\displaystyle\xi^{+\mu }(k_i,r)={\bar u_{\dot\lambda}(i)\bar\sigma^{\mu\dot\lambda\lambda} u_{\lambda}(r)\over\sqrt{2} \langle i\,r\rangle}$
\item{$\bullet$} Four-particle kinematics (all momenta incoming)\hfill\break
$k_1+k_2+k_3+k_4=0$\br
$\hat s=\alpha'(k_1+k_2)^2=2\alpha' k_1k_2$\hfill\break
$\hat t=\,\alpha'(k_1+k_3)^2=2\alpha' k_1k_3$\hfill\break
$\hat u=\alpha'(k_1+k_4)^2=2\alpha' k_1k_4$\hfill\break
$\hat s+\hat t+\hat u=0$\br
In this Sections, carets are dropped; standard Mandelstam variables will be reintroduced in the next Section.

\vskip0.2cm
For the correct normalization of the various fields we recall
the low--energy effective ($N=1$ SUSY) action of the gauge and matter sectors, which
reads up to the two derivative level:
\eqn\LOW{
\Lc=-\Tr\sum_{r=a,b,\dots}\fc{1}{2 g^{2}_{Dp_r}}
(F^r_{\mu\nu}\ F^{r\mu\nu}+
4i\bar\lambda^r\slashchar{D}\lambda^r)-
\Tr\sum_{r\cap  s} (D_{\mu}^{\scriptscriptstyle r\cap s}\bar\phi D^{\mu}_{\scriptscriptstyle r\cap s}\phi+
i\bar \psi\slashchar{D}^{\scriptscriptstyle r\cap s} \psi)+\ldots,}
where $F^r_{\mu\nu}$ is the field strength of $A^r_{\mu}$ and $\lambda^r$ is its
partner gaugino. The first sum runs over all stacks of
D-branes, while the second over their intersections. All traces are in the fundamental representations.
The gauge covariant derivatives of matter fermions $\psi^{\alpha}_{\beta}$ associated to the intersection of $a$ and $b$ are given by:
\eqn\gauge{
(D^{\scriptscriptstyle a\cap b}_{\mu} \psi)^{\alpha}_{\beta}=\p_\mu \psi^{\alpha}_{\beta}+i[T^a]^{\alpha}_{\alpha'} A_\mu^a \psi^{\alpha'}_{\beta}-i [T^b]^{\beta'}_{\beta} A_\mu^b \psi^{\alpha}_{\beta'} ,}
and  similar  expressions for scalars $\phi^{\alpha}_{\beta}$.
Note that all matter fields are canonically normalized in Eq. \LOW, {\it i.e}.\ the moduli-dependent metrics have been absorbed by appropriate field redefinitions.

\subsec{Four--point string amplitudes and open string vertex operators}

Let $\Phi^i$, $i=1,2,3,4$, represent gauge bosons, quarks of leptons of the SM realized on three or more stacks of intersecting D-branes. The corresponding string vertex operators $V_{\Phi^i}$ are constructed from the fields of the underlying superconformal field theory (SCFT) and contain explicit (group-theoretical) Chan-Paton factors. In order to obtain the scattering amplitudes, the vertices are inserted at the boundary of a disk world-sheet, and the following SCFT correlation function is evaluated:
\eqn\Vier{
\Mc(\Phi^{1},\Phi^{2},\Phi^{3},\Phi^{4})=\sum_{\pi\in S_4/\IZ_2}
V_{CKG}^{-1}\int\limits_{\Ic_\pi}
\lf(\prod_{k=1}^4 dz_k\ri)\ \vev{V_{\Phi^{1}}(z_1)\ V_{\Phi^{2}}(z_2)\
V_{\Phi^{3}}(z_3)\ V_{\Phi^{4}}(z_4)}\ .}
Here, the sum runs over all six cyclic inequivalent orderings $\pi$ of the four vertex
operators along the boundary of the disk. Each permutation $\pi$ gives rise to
an integration region
$\Ic_\pi=\{z\in \IR\ | z_{\pi(1)}<z_{\pi(2)}<z_{\pi(3)}<z_{\pi(4)}\}$. The group-theoretical factor is determined by the trace of the product of individual Chan-Paton factors,
ordered in the same way as the vertex positions. The disk boundary contains four segments which may be associated to as many as four different stacks of D-branes, since each vertex of a field originating from a D-brane intersection connects two stacks. Thus the Chan-Paton factor may actually contain as many a four traces, all in the fundamental representations of gauge groups associated to the respective stacks. However, purely partonic amplitudes for the scattering of quarks and gluons involve no more than three stacks.

In order to cancel the total background ghost charge of $-2$ on the disk,
the vertices in the
correlator \Vier\ have to be chosen in the appropriate ghost picture and the picture ``numbers'' must add to $-2$.
Furthermore, in Eq. \Vier, the factor $V_{CKG}$
accounts for the volume of the
conformal Killing group of the disk after choosing the conformal gauge.
It will be canceled by fixing three vertex positions and introducing the
respective $c$--ghost correlator.
Because of the $PSL(2,\IR)$ invariance on the disk, we can fix three positions
of the vertex operators. Depending on the ordering $\Ic_\pi$
of the vertex operator positions we obtain six partial amplitudes.
The first set of three partial amplitudes may be obtained by the choice
\eqn\choice{
z_1=0\ \ \ ,\ \ \ z_3=1\ \ \ ,\ \ \ z_4=\infty\ ,}
while for the second set we choose:
\eqn\choicei{
z_1=1\ \ \ ,\ \ \ z_3=0\ \ \ ,\ \ \ z_4=\infty\ .}
The two choices imply the ghost factor $\vev{c(z_1)c(z_2)c(z_3)}=z_{13}z_{14}z_{34}$.
The remaining vertex position $z_2$ takes arbitrary values along the
boundary of the disk. After performing all Wick contractions in \Vier\ the
correlators become basic \refs{\STiii,\STiv} and
generically for each partial amplitude the integral \Vier\
may be reduced to the Euler Beta function:
\eqn\veneziano{
B(s,u)=\int_0^1 x^{s-1}\ (1-x)^{u-1}=\fc{\Gamma(s)\
\Gamma(u)}{\Gamma(s+u)}=\fc{1}{s}+\fc{1}{u}-{\pi^2\over 6}\ (s+u)+\Oc(\ap^2)\ .}
Although we are mainly interested in the amplitudes involving the particles of the SM, 
we give below, for completeness,
the vertex operators $V_\Phi$ for the full $N=1$ SUSY  multiplets.

\vskip0.5cm
\noindent
{\sl (i) Gauge vector multiplet:}\br

\noindent
The gauge boson vertex operator in the $(-1)$-ghost picture reads
\eqn\fieldsiii{\hskip -3.7cm
V_{A^a}^{(-1)}(z,\xi,k) ~=~ g_{A} [T^a]^{\alpha_1}_{\alpha_2}\ e^{-\phi(z)}\ \xi^\mu\ 
\psi_\mu(z)\ e^{ik_\rho X^\rho(z)}\ ,}
while in the zero--ghost picture we have:
\eqn\gaugevertexzero{
V_{A^a}^{(0)}(z,\xi,k)=\fc{g_{A}}{(2\ap)^{1/2}} [T^a]^{\alpha_1}_{\alpha_2}\ \xi_\mu\
[\ i\p X^\mu(z)+2\ap\ (k\psi)\ \psi^\mu(z)\ ]\ e^{ik_\rho X^\rho(z)}\ .}
where $\xi^\mu$ is the polarization vector. The vertex must be inserted on 
the segment of disk boundary on stack $a$, with the indices $\alpha_1$ and $\alpha_2$ 
describing the two string ends.

{}For our purposes, the most important property of the gluon vertex operators \fieldsiii\
and \gaugevertexzero\ is
that they do not depend on the internal (CY) part of the SCFT. They depend only on the 
SCFT fields describing string coordinates $X^\mu$ in four dimensions, and on their 
world--sheet superpartners $\psi^\mu$. Although the construction of these vertices 
utilizes SCFT, their form is universal to all compactifications 
and remains unaffected by eventual SUSY breaking in the bulk or by D--brane 
configurations. This is the reason why the results for $N$--gluon disk amplitudes 
\refs{\STi,\STii,\STiii,\STiv} are completely universal and hold even if SUSY
is broken in four dimensions.

In case of $N=1$ supersymmetry,
the gaugino vertex operators, in the $(-1/2)$-ghost picture, are
\eqn\fieldsii{\eqalign{
V_{\lambda^{a,I}}^{(-1/2)}(z,u,k)
&=g_\lambda\ [T^a]^{\alpha_1}_{\alpha_2}\ e^{-\phi(z)/2}\ u^{\lambda} S_{\lambda}(z)\ 
\Si^I(z)\ e^{ik_\rho X^\rho(z)}\ ,\cr
V_{\ov\lambda^{a,I}}^{(-1/2)}(z,\bar u,k)
&=g_\lambda\ [T^a]^{\alpha_1}_{\alpha_2}\ e^{-\phi(z)/2}\ \ov u_{\dot\lambda} 
S^{\dot\lambda}(z)\ \ov\Si^I(z)\
e^{ik_\rho X^\rho(z)}\ .}}
where $S_{\lambda}$ and $S^{\dot\lambda}$ are the  world-sheet  spin fields associated
to the negative and positive helicity fermions, respectively.
The index $I$ labeling gaugino species may range from 1 to 4,
depending on the amount of supersymmetries on the D--brane world--volume,
while the associated world-sheet fields $\Si^I$ of conformal dimension $3/8$
belong to the Ramond sector of SCFT
\refs{\banksi,\banksii,\flt}.
In the case of extended supersymmetry on the D--brane world--volume
we also have scalars $\phi^{a,i}$ in the adjoint representation of the gauge group.
Their vertex operators take the form:
\eqn\ffieldsiii{\eqalign{
V_{\phi^{a,i}}^{(-1)}(z,k)
&=g_\phi[T^a]^{\alpha_1}_{\alpha_2}\ e^{-\phi(z)}\ \Psi^i \ e^{ik_\rho X^\rho(z)}\ ,\cr
V_{\ov\phi^{a,i}}^{(-1)}(z,k)
&=g_\phi[T^a]^{\alpha_1}_{\alpha_2}\ e^{-\phi(z)}\ \ov\Psi^i \ e^{ik_\rho X^\rho(z)}\ .}}

For this multiplet, the open string  couplings are:
\eqn\opscoupling{
g_\lambda=(2\ap)^{1/2}\ap^{1/4}\
g_{Dp_a}\ \ \ ,\ \ \
g_A=(2\ap)^{1/2}\ g_{Dp_a}\ \ \ ,\ \ \
g_\phi=(2\ap)^{1/2}\ g_{Dp_a}\ .}
These prefactors, together with the universal 
factor\foot{See \JOE\ for the derivation of this factor.}
\eqn\diskfactor{C_{D_2}={1\over g_{Dp_a}^2\,\ap^2}\ ,}
which must be inserted in all disk amplitudes with the boundary on a single 
stack $a$ of D-branes, ensure agreement of the string computations with
the effective action \LOW.
Indeed, with these normalizations the three--gluon superstring disk amplitude is:
\eqn\basicc{
\eqalign{
\Mc[A^{a_1}(\xi_1,k_1)A^{a_2}(\xi_2,k_2)A^{a_3}(\xi_3,k_3)]&=g_{Dp_a}\
\Tr(T^{a_1}[T^{a_2},T^{a_3}])\cr
&\times
\lf[(\xi_1\xi_2)(\xi_3k_{12})+(\xi_1\xi_3)(\xi_2k_{31})+
(\xi_2\xi_3)(\xi_1k_{23})\ri]\ .}}
Furthermore, the string coupling of one gauge boson \fieldsiii\
to two gauginos \fieldsii\ is:
\eqn\basicci{
\Mc[\lambda^{a_1}(k_1,u_1)\bar\lambda^{a_2}(k_2,\bar u_2)A^{a_3}(\xi,k_3)]=g_{Dp_a}\
(u_1^{\lambda}\si_{\lambda\dot\lambda}^\mu\bar u_2^{\dot\lambda})\ \xi_\mu\times
\Tr(T^{a_1}[T^{a_2},T^{a_3}]) \ ,          }
in agreement\foot{The results \basicc\ and \basicci\
can be matched directly with the interaction vertices of Eq. \LOW\
by an additional rescaling
$V_A\ra g_{Dp_a}^{-1} V_A~,~V_\lambda\ra g_{Dp_a}^{-1} V_\lambda$.} with Eq. \LOW.

\vskip0.5cm
\noindent
{\sl (ii) Matter multiplet:}\br

\noindent

The chiral fermion vertex operators of the quarks and leptons are:
\eqn\chiralfermion{\eqalign{
V^{(-1/2)}_{\psi^{\alpha}_{\beta}}(z,u,k)&=g_\psi [T^{\alpha}_{\beta}]_{\alpha_1}^{\beta_1}e^{-\phi(z)/2}\ u^{\lambda}
S_{\lambda}(z)\ \Xi^{\scriptscriptstyle a\cap b}(z)\ e^{ik_\rho X^\rho(z)}\ ,\cr
V^{(-1/2)}_{\bar\psi^{\beta}_{\alpha}}(z,\bar u,k)&=g_\psi [T_{\alpha}^{\beta}]^{\alpha_1}_{\beta_1}
e^{-\phi(z)/2}\ \bar u_{\dot\lambda}
S^{\dot\lambda}(z)\ \ov\Xi^{\scriptscriptstyle a\cap b}(z)\ e^{ik_\rho X^\rho(z)}\ .}}
These vertices connect two segments of disk boundary, associated to stacks $a$ and $b$, with the indices $\alpha_1$ and $\beta_1$ representing the string ends on the respective stacks.
The internal field $\Xi^{\scriptscriptstyle a\cap b}$ of conformal dimension $3/8$
is the fermionic boundary
changing operator.
In the intersecting D-brane models, the intersections are characterized by angles $\theta_{ba}$. Then
$\Xi^{\scriptscriptstyle a\cap b}$ can be expressed
in terms of bosonic and fermionic twist fields $\sigma$ and $s$:
\eqn\explicit{
\Xi^{\scriptscriptstyle a\cap b}=\prod_{j=1}^3\sigma_{\theta^j_{ba}}\ s_{\theta^j_{ba}}\ \ \ ,\ \ \
\ov\Xi^{\scriptscriptstyle a\cap b}=\prod_{j=1}^3\sigma_{-\theta_{ba}^j}\ s_{-\theta_{ba}^j}\ .}
The spin fields
\eqn\spinfield{
s_{\theta^j}=e^{i(\theta^j-\h)H^j}\ \ \ ,\ \ \
s_{-\theta^j}=e^{-i(\theta^j-\h)H^j}}
have conformal dimension $h_s=\h(\theta^j-\h)^2$ and twist the internal part
of the Ramond ground state spinor.
The field $\sigma_{\theta}$ has conformal dimension
$h_\sigma=\h\theta^j(1-\theta^j)$
and produces discontinuities in the boundary conditions of the
internal complex bosonic Neveu--Schwarz coordinates $Z^j$.

In case of $N=1$ supersymmetry,
the vertex operators of chiral matter scalars
originating from strings stretching between stacks $a$ and $b$ are
\eqn\chiralsc{\eqalign{
V^{(-1)}_{\phi^{\alpha}_{\beta}}(z,k)=g_\phi[T^{\alpha}_{\beta}]_{\alpha_1}^{\beta_1}\ e^{-\phi(z)}\ \Pi^{\scriptscriptstyle a\cap b}(z)\ e^{ik_\rho X^\rho(z)}\ ,\cr
V^{(-1)}_{\bar\phi^{\beta}_{\alpha}}  (z,k)=g_\phi[T_{\alpha}^{\beta}]^{\alpha_1}_{\beta_1}\ e^{-\phi(z)}\
\ov\Pi^{\scriptscriptstyle a\cap b}(z)\ e^{ik_\rho X^\rho(z)}\ .}}
where $\Pi^{\scriptscriptstyle a\cap b}$ is the scalar boundary changing operator of conformal
dimension $1/2$.
Again, for the D-branes $a$ and $b$ intersecting at angles $\theta_{ba}^j$, an
explicit representation can be given in terms of bosonic and fermionic twist fields
\eqn\explicit{
\Pi^{\scriptscriptstyle a\cap b}=\prod_{j=1}^3\sigma_{\theta_{ba}^j}\ s_{\theta_{ba}^j}\ \ \ ,\ \ \
\ov\Pi^{\scriptscriptstyle a\cap b}=\prod_{j=1}^3  \sigma_{-\theta_{ba}^j}\  s_{-\theta_{ba}^j}}
The spin fields
\eqn\spinfield{
s_{\theta^j}=e^{i\theta^j H^j}\ \ \ ,\ \ \ s_{-\theta^j}=e^{-i \theta^j H^j}}
have conformal dimension $h_s=\h(\theta^j)^2$ and twist the internal part
of the Neveu--Schwarz  ground state.

For the chiral multiplet, the open string couplings are:
\eqn\opscouplings{
g_\phi=(2\ap)^{1/2}\,e^{\phi_{10}/2}\quad,\quad
g_\psi= (2\ap)^{1/2}\ap^{1/4}\,e^{\phi_{10}/2} .}
These prefactors, together with the universal factor\foot{See \JOE\ for the 
derivation of this factor.}
\eqn\diskfactor{\tilde C_{D_2}={e^{-\phi_{10}}\over \ap^2}\ ,}
which must be inserted in the presence of operators changing disk boundary, ensure 
agreement of the string computations with
the effective action \LOW. Indeed, the coupling of fermions \chiralfermion\ to the 
gauge boson \fieldsiii\ is
\eqn\bbasici{
\Mc[\psi^{\alpha_1}_{\beta_1}(k_1,u_1)\bar\psi_{\alpha_2}^{\beta_2}(k_2,\bar u_2)
A^a(\xi,k_3)]=g_{Dp_a}\
(u_1^{\lambda}\si_{\lambda\dot\lambda}^\mu\bar u_2^{\dot\lambda})\ \xi_\mu\times
[T^a]^{\alpha_1}_{\alpha_2}\delta^{\beta_2}_{\beta_1}\, ,}
in agreement with Eq. \LOW.

\subsec{Four gluon amplitudes}

Four-gluon amplitudes have been known for many years \JSCH.
The corresponding string disk  diagram is shown in  Figure 6:
\ifig\intert{\ Gauge boson vertex operators
along the boundary of the disk world--sheet.}
{\epsfxsize=0.3\hsize\epsfbox{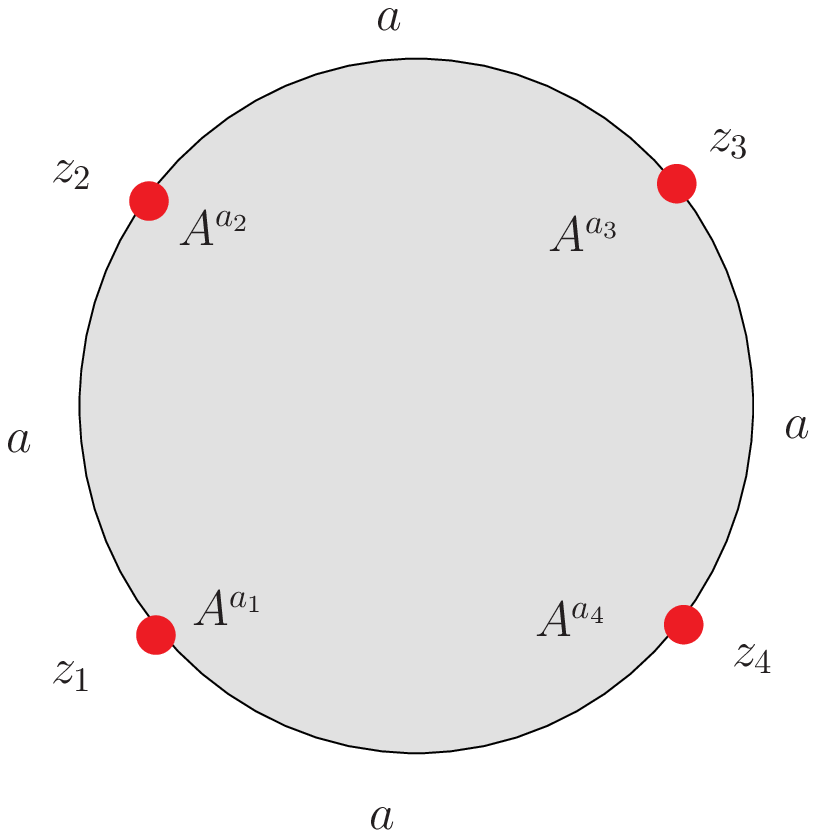}}
\noindent
The complete amplitude can be generated from the maximally helicity violating 
MHV amplitudes \refs{\STi,\STii}. Usually only one amplitude is written 
explicitly -- a {\it partial\/} amplitude associated to one specific Chan-Paton factor. 
The full expression is necessary, however, for collider applications.
Let us start from the partial amplitude \refs{\STi,\STii}
\eqn\ampl{
{\cal
   M}_P(g^-_1,g^-_2, g^+_3, g^+_4) ~=~ 4\, g^2\, {\rm Tr}
 \, (\, T^{a_1}T^{a_2}T^{a_3}T^{a_4})\ {\langle 12\rangle^4\over
   \langle 12\rangle\langle 23\rangle\langle 34\rangle\langle
   41\rangle}\ \hat V(k_1,k_2,k_3,k_4)\ ,}
where $\pm$ refer to polarizations. The Veneziano formfactor is given by
\eqn\formf{
\hat V(k_1,k_2,k_3,k_4)=\hat V(s,t,u)= {su\over (s+u)
   }B(s,u)\ ,}
Its low-energy
 expansion reads
\eqn\vexp{\hat
V(s,t,u)\approx 1-{\pi^2\over 6}\  s\,
     u+\zeta(3)\ s\,tu+\dots}
It is convenient to introduce:
\eqn\vexpdef{
\hat V_t =\hat V(s,t,u) ~,\qquad \hat V_s=\hat V(  t\leftrightarrow  s) ~,\qquad \hat V_u=\hat V(  t\leftrightarrow  u) ~.}
The color
factor can be written as
\eqn\colf{\eqalign{{\rm
   Tr}(T^{a_1}T^{a_2}T^{a_3}T^{a_4}) ~=~ & d^{a_1a_2a_3a_4}+{i\over 2}(d^{a_1a_4n}f^{a_2a_3n}-
d^{a_2a_3n}f^{a_1a_4n})\cr & +{1\over 12}( f^{a_1a_4n}f^{a_2a_3n}-
f^{a_1a_2n}f^{a_3a_4n}        ) }}
where the totally symmetric symbols
\eqn\dtrace{d^{a_1a_2a_3}={\rm STr}(T^{a_1}T^{a_2}T^{a_3})~,
\quad d^{a_1a_2a_3a_4} ={\rm STr}(T^{a_1}T^{a_2}T^{a_3}T^{a_4}) }
are the symmetrized traces \groupf\
while $f^{a_1a_2a_3}$ is the totally antisymmetric structure constant.

The full MHV amplitude can be obtained \refs{\STi,\STii} by summing
the partial amplitudes \ampl\ with the indices permuted in the
following way: \eqn\afull{ {\cal M}(g^-_1,g^-_2,g^+_3,g^+_4)
 =4\,g^{2}\langle 12\rangle^4 \sum_{\sigma\in S_4/Z_4 } { {\rm Tr} \, (\,
   T^{a_{1_{\sigma}}}T^{a_{2_{\sigma}}}T^{a_{3_{\sigma}}}T^{a_{4_{\sigma}}})\
   \hat V(k_{1_{\sigma}},k_{2_{\sigma}},k_{3_{\sigma}},k_{4_{\sigma}})\over\langle
   1_{\sigma}2_{\sigma} \rangle\langle
   2_{\sigma}3_{\sigma}\rangle\langle 3_{\sigma}4_{\sigma}\rangle \langle
   4_{\sigma}1_{\sigma}\rangle }\ .}
where $S_4$ is the set of all permutations of
$\{1,2,3,4\}$ while $Z_4$ is the subset of cyclic permutations.
As a result, the imaginary part of
the color factor \colf\ cancels and one obtains
\eqn\mhva{\eqalign{
{\cal
   M}(g^-_1,g^-_2,&\ g^+_3, g^+_4) ~=~ 8\, g^2\langle
 12\rangle^4\,\times\cr &
\;\,\,\lf\{\  {\hat V_t\over\langle 12\rangle\langle
     23\rangle\langle 34\rangle\langle
     41\rangle}\ \lf[\ d^{a_1a_2a_3a_4}+{1\over 12}\ ( f^{a_1a_4n}f^{a_2a_3n}-
f^{a_1a_2n}f^{a_3a_4n} )\ \ri]\ri. \cr
& +{\hat V_s\over\langle 14\rangle\langle
     42\rangle\langle 23\rangle\langle
     31\rangle}\ \lf[\ d^{a_1a_2a_3a_4}+{1\over 12}\ ( f^{a_2a_4n}f^{a_3a_1n}-
f^{a_2a_3n}f^{a_1a_4n} )\ \ri]\cr
&\lf. +{\hat V_u\over\langle 13\rangle\langle
     34\rangle\langle 42\rangle\langle
     21\rangle}\ \lf[\ d^{a_1a_2a_3a_4}+{1\over 12}\ ( f^{a_3a_4n}f^{a_1a_2n}-
f^{a_3a_1n}f^{a_2a_4n} )\ \ri]\ \ri\}\ ,}}
All
non-vanishing amplitudes can be obtained from the above expression by
appropriate crossing operations.

\subsec{Two gauge bosons and two fermions}

We consider the following correlation function:
\eqn\setup{
\vev{V_{A^{x}}^{(0)}(z_1,\xi_1,k_1)\ V_{A^{y}}^{(-1)}(z_2,\xi_2,k_2)\
V^{(-1/2)}_{\psi^{\alpha_3}_{\beta_3}}(z_3,u_3,k_3)
\ V^{(-1/2)}_{\bar\psi^{\beta_4}_{\alpha_4} }(z_4,\bar u_4,k_4)}\ .}
The fact that fermions originate from the same pair of stacks, say $a$ and $b$
is forced
upon us by the conservation of twist charges, in a similar way as their
opposite helicities are
forced  by the internal charge conservation.
It follows that both gauge bosons must be associated either to one of these stacks, say
$(x,y)=(a_1,a_2)$, or one of them is associated to $a$ while the other to $b$,
say $(x,y)=(a,b)$. The corresponding disk diagrams are shown in  Figure~7.
\iifig\interz{Ordering of gauge boson and fermion vertex operators,}{for
$(x,y)=(a_1,a_2)$ (left diagram) and $(x,y)=(a,b)$ (right diagram).}
{\epsfxsize=0.7\hsize\epsfbox{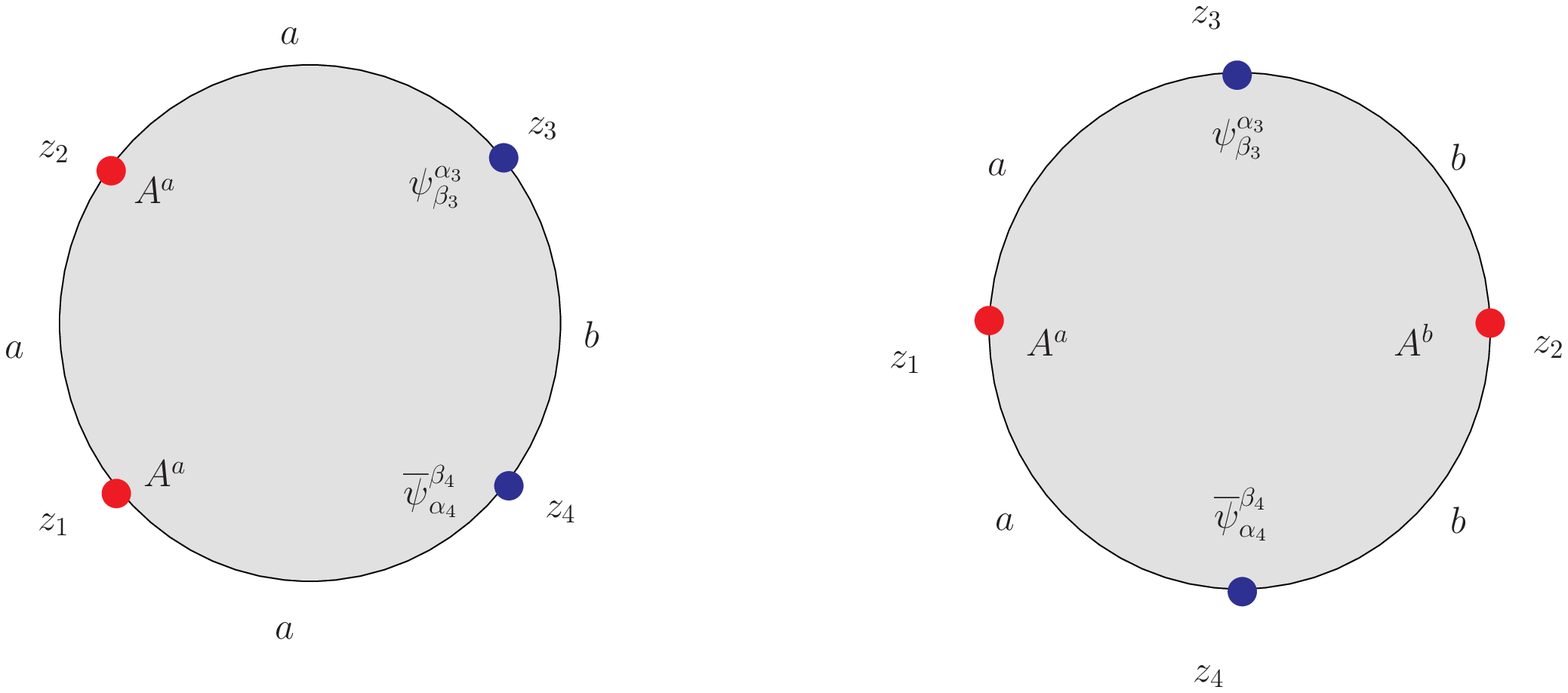}}
\noindent
With the position $z_4=\infty$ as in \choice\ and \choicei, the correlator
\setup\ becomes
\eqn\final{\eqalign{
2\ap\ g_{Dp_{x}}g_{Dp_y}&\
\lf\{\fc{1}{z_{12}}\ \lf[k_{1\rho}\ (\xi_1\xi_2)-\xi_{1\rho}\
(\xi_2k_1)+\xi_{2\rho}\ (\xi_1k_2)+\xi_{2\rho}\ (\xi_1k_3)\
\fc{z_{12}}{z_{13}}\ri]\ (u_3\sigma^{\rho} \ov u_4)\ri.\cr
&+\lf.\h\ \fc{1}{z_{13}}\ k_{1\lambda}\ \xi_{1\mu}\ \xi_{2\rho}\
(u_3\sigma^\lambda\ov\sigma^{\mu}\sigma^\rho \ov u_4)\ri\}\
\fc{z_{13}}{z_{23}}\ |z_{12}|^{  s}\ |z_{23}|^{  u}\ }}
times the Chan Paton factor which is determined by the relative position of
$z_2$ with respect  to $z_1$ and $z_3$.
If $(x,y)=(a_1,a_2)$, there are two allowed orderings of vertex positions: as
in \choice, with $z_2<0$ or $0<z_2<1$, see the left side of \interz. Then
\eqn\finalpartial{\eqalign{
\Mc[A^{a_1}&(\xi_1,k_1)A^{a_2}(\xi_2,k_2)\psi^{\alpha_3}_{\beta_3}(k_3,u_3)
\bar\psi^{\beta_4}_{\alpha_4}(k_4,\bar u_4)]=-2\ \ap\ g_{Dp_a}^2\ \Kc\cr
&\times\lf[\Tr(T^{a_1}T^{a_2}T^{\alpha_3}_{\beta_3}
T^{\beta_4}_{\alpha_4})\ B(s,u)+\Tr(T^{a_2}T^{a_1}T^{\alpha_3}_{\beta_3}
T^{\beta_4}_{\alpha_4})\ \fc{t}{u}\ B(s,t)\ri]\ ,}}
where the kinematic factor:
\eqn\kin{\eqalign{
\Kc&=\lf\{\lf[k_{1\rho}\ (\xi_1\xi_2)-\xi_{1\rho}\ (\xi_2k_1)+\xi_{2\rho}\
(\xi_1k_2)-\fc{s}{t}\ \xi_{2\rho}\ (\xi_1k_3)\ri]\
(u_3\sigma^{\rho} \ov u_4)\ri.\cr
&\hskip4cm\lf.-\h\fc{s}{t}\ k_{1\lambda}\ \xi_{1\mu}\
\xi_{2\rho}\ (u_3\sigma^\lambda\ov\sigma^{\mu}\sigma^\rho \ov u_4)\ri\}\ .}}
On the other hand, if $(x,y)=(a,b)$, then there is only one allowed ordering,
as in \choice, with $z_2>1$, see the right side of \interz, and we obtain
\eqn\ffinalpartial{\eqalign{
\Mc[A^{a}(\xi_1,k_1)&A^{b}(\xi_2,k_2)\psi^{\alpha_3}_{\beta_3}(k_3,u_3)
\bar\psi^{\beta_4}_{\alpha_4}(k_4,\bar u_4)]=-2\ \ap\ g_{Dp_a}\ g_{Dp_b}\ \Kc\cr
&\times\Tr(T^{a}T^{\alpha_3}_{\beta_3}T^{b}
T^{\beta_4}_{\alpha_4})\ \fc{t}{s}\ B(t,u)\ .}}

The amplitudes \finalpartial\ and \ffinalpartial\ may be written as sums over
infinite many $s$--channel poles at the masses \massSR\
in exactly the same way as \basic. On the other hand,
for the lowest string correction of \eqqs \finalpartial\ and \ffinalpartial,
which gives rise to a string contact interaction in lines
of Subsection 2.3, we use the expansion of the Euler Beta function \veneziano:
\eqn\Contact{\eqalign{
\Mc[A^{a_1}(\xi_1,k_1)A^{a_2}(\xi_2,k_2)&\lf.\psi^{\alpha_3}_{\beta_3}(k_3,u_3)
\bar\psi^{\beta_4}_{\alpha_4}(k_4,\bar u_4)]\ \ri|_{\ap^2}=-2\ \ap\ g_{Dp_a}^2\ {\pi^2\over
6}\ t\ \Kc\cr
&\times\lf[\Tr(T^{a_1}T^{a_2}T^{\alpha_3}_{\beta_3}
T^{\beta_4}_{\alpha_4})+\Tr(T^{a_2}T^{a_1}T^{\alpha_3}_{\beta_3}
T^{\beta_4}_{\alpha_4})\ri]\ ,\cr
\Mc[A^{a}(\xi_1,k_1)A^{b}(\xi_2,k_2)&\lf.\psi^{\alpha_3}_{\beta_3}(k_3,u_3)
\bar\psi^{\beta_4}_{\alpha_4}(k_4,\bar u_4)]\ \ri|_{\ap^2}=-2\ \ap\ g_{Dp_a}\ g_{Dp_b}\
{\pi^2\over 6}\ t\ \Kc\cr
&\times \Tr(T^{a}T^{\alpha_3}_{\beta_3}T^{b}
T^{\beta_4}_{\alpha_4})\ .}}
Hence, the first string contact interaction appears at the order $\Ms^{-4}$ as
in the case of four--gluon scattering (\cf \four).

In order to write the kinematic factor \kin\ more explicitly, we choose $k_2$ as the
reference
momentum for the polarization vector $\xi_1$ and $k_1$ as the reference momentum for
$\xi_2$. Then it is easy to see that the kinematic factor vanishes if both
gauge bosons have the same helicity, while for the opposite helicities
$\xi_1^{\pm}\xi_2^{\mp}=0$ and most of terms the r.h.s.\ of \eqq \kin\ vanish except for
\eqn\epsk{\xi_{2\rho}^+ (\xi_1^-k_3)
(u_3\sigma^{\rho} \bar u_4)={1\over \alpha' s}\ \langle 13\rangle^2[23][24]~,\quad
\xi_{2\rho}^- (\xi_1^+k_3)
(u_3\sigma^{\rho} \bar u_4)={1\over \alpha' s}\ \langle 23\rangle^2[13][14]\ .}
After combining the kinematic and Chan-Paton factors, and renaming external
fields, \eqqs \finalpartial\ and \ffinalpartial\ become
\eqn\aggff{
\Mc(g^{-}_1,g^{+}_2,q_3^-,\bar q_4^+)~=~
2\, g^2\, \delta^{\beta_4}_{\beta_3}\,{\langle 13\rangle^2\over \langle
23\rangle\langle 24\rangle}\
\Big[(T^{a_1}T^{a_2})^{\alpha_3}_{\alpha_4}\ {t\over s}\ \hat V_t+
(T^{a_2}T^{a_1})^{\alpha_3}_{\alpha_4}\ {u\over s}\ \hat V_u\Big]\ ,}
\eqn\aggfg{
\Mc(g_1^-, B_2^+,q_3^-,\bar q_4^+)~=~
2\,  g_{Dp_b}\ g\ {\langle 13\rangle^2\over \langle 23\rangle\langle
24\rangle}\
(T^{a})^{\alpha_3}_{\alpha_4}(T^{b})^{\beta_4}_{\beta_3}\,\hat V_s\ .}
All other helicity amplitudes can be obtained from the above expressions by
appropriate crossing operations. The low energy expansions of the above
amplitudes can be obtained by using \eqq \veneziano.
The leading term agrees with the well-known QCD result.

\subsec{Four chiral fermions}

Let us first discuss what are the possible four-point disk amplitudes among four
fermion fields.
Of course the amplitudes are constrained by the $SU(3)\times SU(2)\times
U(1)_Y$ gauge invariance
of the SM. In addition,
the allowed disk scattering amplitudes are constrained by the
conservation of the additional $U(1)$ charges of the matter fields.
Recall that the these $U(1)$'s are part of the original $U(N)$ gauge symmetries of the
different stack of D-branes,
and are in general massive due to the generalized Green-Schwarz mechanism. Only the SM
hypercharge stays as
a massless, anomaly free linear combination.
Nevertheless, the massive $U(1)$'s act as global symmetries
and provide selection rules that constrain the allowed tree level
couplings. Note that space-time
instantons, i.e. wrapped Euclidean D-branes, may violate the conservation
of the global $U(1)$
symmetries, and can hence lead to new processes, which we do not discuss here.

There are two classes of fermion disk amplitudes. The first class contains the 
amplitudes describing the processes that 
occur in the SM by the exchange of massless particles. They have
poles in the associated exchange channels.
Exchange of heavy Regge excitations of mass
$M_{\rm string}$  provide additional string corrections to this kind of amplitudes.
Therefore the SM background must be subtracted in order to see the new signal  from
string theory. 
Second, there are four fermion disk amplitudes, which do 
not exhibit any massless poles. Hence there are only the 
string contact interactions due to the exchange of Regge like
excitations. Therefore these amplitudes do not possess SM, tree--level
background.
In the following we discuss  the possible amplitudes in the prototype
model of Section 4, with particle content as given in Table 4.
We indicate which amplitudes occur already in the SM due to the tree level exchange
of SM gauge bosons, and which can occur only in the D-brane model under investigation.
At this stage we would like to stress that the computations of the four fermion
disk amplitudes are  performed for intersecting D6-branes wrapped around
flat 3-cycles of a six-dimensional torus. It follows that the explicit expressions,
which we shall present in the following, may contain also contributions from exchange 
of massive states (scalar fields),
whose masses depend on the intersection angles of the 3-cycles. In addition there are
contributions from  fields that
are due to the extended supersymmetries on the D6-branes on the torus, e.g. moduli fields
that describe the positions of the D6-branes on the tori. All these model dependent 
fields may
in general appear in intermediate channels of the four fermion amplitudes. 

%%%%%%%%%%%%%%%%%%%%%%%%%%%%%%%%%%%%%%%%%%%%%%%%%%%%%%%%%%%%%%%%%%%%%%%%%%%%%%%%

\vskip0.3cm
\noindent
(i) All four fermions at the same intersection point (one angle).
\vskip0.3cm
\noindent
The simplest case is the scattering of two pairs of fermion/anti--fermion fields
which are located at the same intersection point of two stacks, $a$ and $b$,
\cf Figure 8.
\iifig\exi{\ Two intersecting stacks $a$ and $b$}{
and two pairs of conjugate fermions.}
{\epsfxsize=0.35\hsize\epsfbox{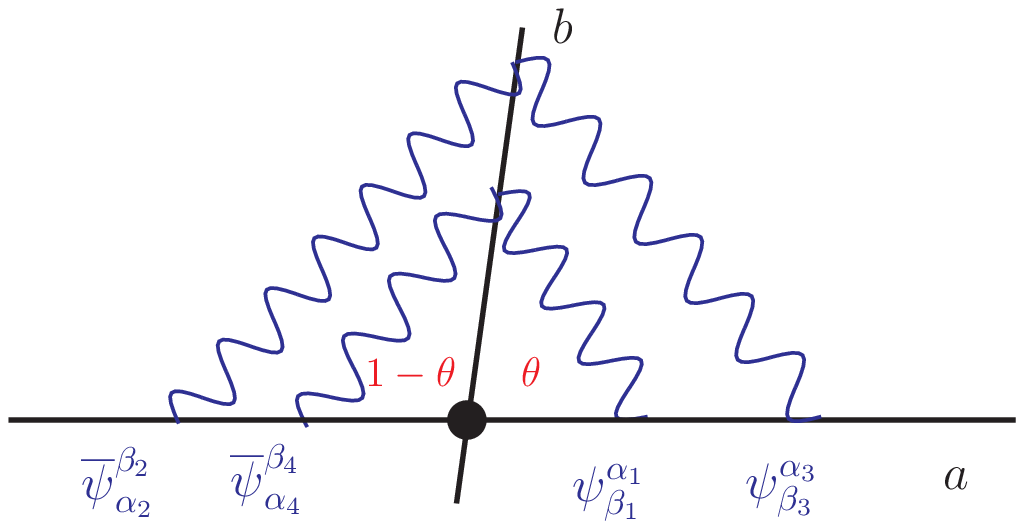}}
\noindent
The amplitude is
\eqn\fermionone{
\vev{\psi_{\beta_1}^{\alpha_1}\bar \psi_{\alpha_2}^{\beta_2}
\psi_{\beta_3}^{\alpha_3}\bar \psi_{\alpha_4}^{\beta_4}}}
and exhibits poles due to the exchange of massless
gauge bosons from the stack $a$ and stack $b$.
In addition, Regge excitations, KK and winding states are exchanged.
An example for this amplitude is the process
\eqn\quarksa{q^-\bar q^+\to q^-\bar q^+\qquad
\big(q_{L} q_{R}^c\to q_{L} q_{R}^c\big) }
which receives contributions from the exchange of gluons, photons,
$W,Z$-bosons, as well as the Regge and
KK excitations.

\vskip0.3cm
\noindent
(ii) Two pairs of conjugate fermions at two different intersection points (two angles)

\vskip0.3cm
\noindent
Now we are considering three stacks of D-branes,  $a$, $b$ and $c$
with two intersection angles $\theta:=\theta_b-\theta_a$ and
$\nu:=\theta_c-\theta_a$, \cf  Figure 9.
\iifig\exii{\ Three intersecting stacks $a$, $b$ and $c$}{
and two pairs of conjugate fermions.}
{\epsfxsize=0.35\hsize\epsfbox{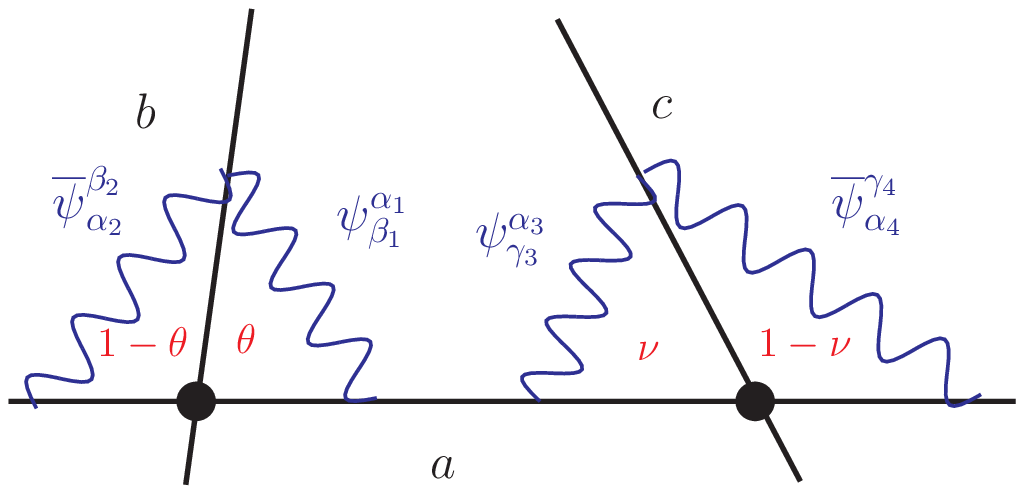}}
\noindent
We have open strings spanned between $a$ and $b$, as well as open strings
stretched between $a$ and $c$\foot{In case of
multiple intersections, namely considering different families,
stack $b$ and $c$ can be the same, but the fermions may be associated to
different intersections.}.
Hence we consider the following  amplitude:
\eqn\fermiontwo{\vev{\psi_{\beta_1}^{\alpha_1}\bar \psi_{\alpha_2}^{\beta_2}
\psi_{\gamma_3}^{\alpha_3}\bar \psi_{\alpha_4}^{\gamma_4}}\ .}
This amplitude exhibits massless poles due to the exchange of the common massless
gauge boson from stack $a$ as well from its Regge excitations, KK and
windings states thereof.
An example for this amplitude is the process
\eqn\quarksaa{
q^-\bar q^+\to q^+\bar q^-\qquad \big(q_{L} q_{R}^c\to q_{R} q_{L}^c\big)\ .}

\vskip0.3cm
\noindent
(iii) Four different fermions at four intersection points (three angles):

\vskip0.3cm
\noindent
Finally we are considering four stacks of D-branes,
$a,b,c$ and $d$ with all four chiral fermions originating from different
intersections, \cf  Figure 10.
\iifig\exiii{\ Four intersecting stacks $a,b,c$ and $d$}{
and fermions at four different intersections.}
{\epsfxsize=0.35\hsize\epsfbox{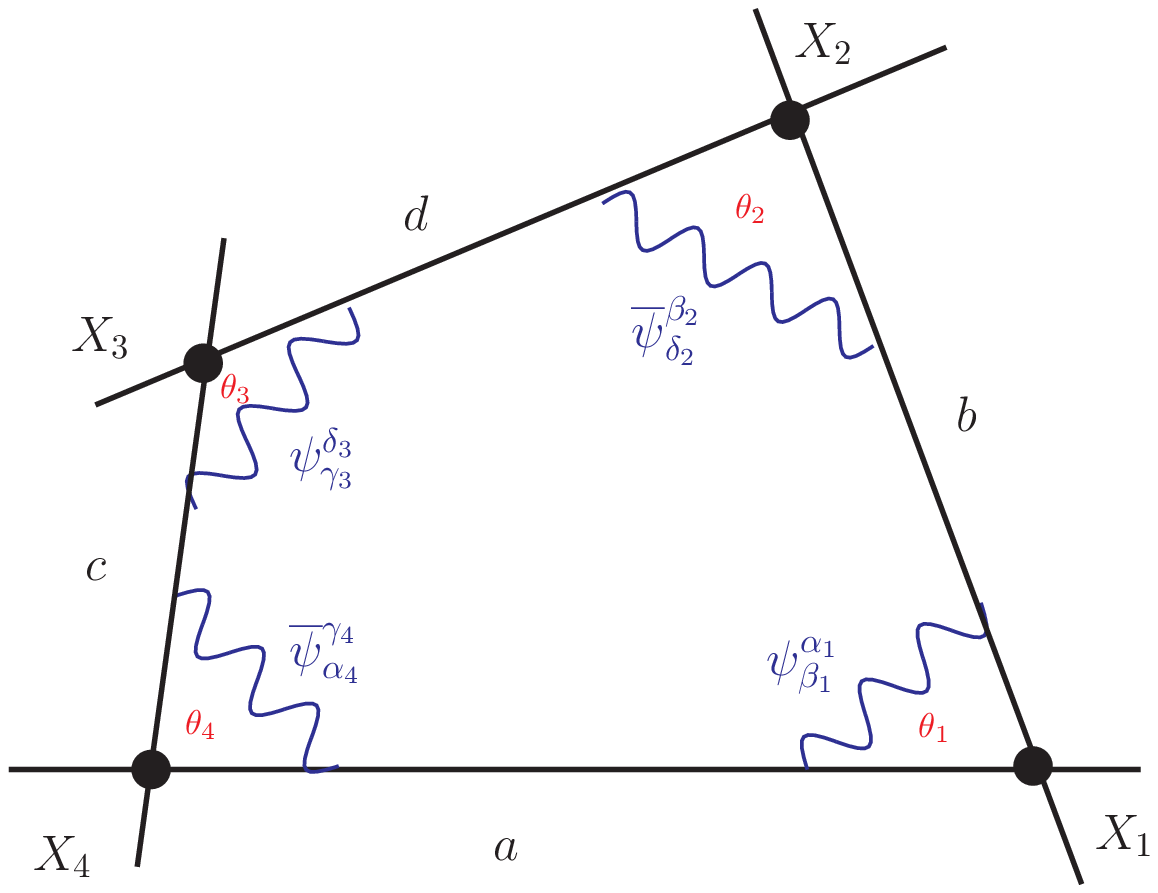}}
\noindent
The corresponding amplitude is of the form
\eqn\fourfermion{\vev{\psi_{\beta_1}^{\alpha_1}\bar \psi^{\beta_2}_{\delta_2}
\psi_{\gamma_3}^{\delta_3}\bar \psi^{\gamma_4}_{\alpha_4}}\ .}
No massless gauge bosons are exchanged in this amplitude.
However, there may be the exchange of the SM Higgs field as well as some exotic
states with masses of the order of the string scale.
An example of this kind of amplitude is the following contribution to the
Drell--Yan process:
\eqn\quarkleptonc{q^-\bar q^-\to \mu^-\bar \mu^-\qquad
\big(q_{L} q_{L}^c\to \mu_{L} \mu_{L}^c\big)\ .}
Note that purely hadronic $2\to 2$ parton scattering processes with quarks and
antiquarks belonging to the same family involve no more than three stacks
because all partons share the QCD D-brane stack.
\br

In order to compute the four--fermion amplitudes, we evaluate
\eqq \Vier\ with the following correlator:
\eqn\setupferm{
\vev{V_{\psi^{\alpha_1}_{\beta_1}}^{(-1/2)}(z_1,u_1,k_1)\
V_{\ov\psi^{\beta_2}_{\delta_2}}^{(-1/2)}(z_2,\ov u_2,k_2)\
V^{(-1/2)}_{\psi^{\delta_3}_{\gamma_3}}(z_3,u_3,k_3) \
V^{(-1/2)}_{\ov\psi^{\gamma_4}_{\alpha_4}}(z_4,\ov u_4,k_4)}}
involving two chiral $\psi^{\alpha_1}_{\beta_1}, \psi^{\delta_3}_{\gamma_3}$
and two anti--chiral $\ov\psi^{\beta_2}_{\delta_2},
\ov\psi^{\gamma_4}_{\alpha_4}$ matter fermions.
In the type IIA picture the latter are represented by open strings stretched
between two intersecting stacks of branes  with
the vertex operator\foot{Vertex operators for the case if one brane is on top
of an orientifold plane have been constructed in \Richter.} \chiralfermion.
Although the following discussion is carried out for intersecting D6--branes
it may be translated into other setups.
We shall present the explicit expressions for the four--fermion
string amplitudes
in the case of the prototype four stack model introduced in Section 4.

The most general case may involve four different $D$--branes $a,b,c$ and $d$,
which intersect at
the four points $X_i$ with the angles $\th_i$, $i=1,\ldots,4$, respectively,
\cf Figure 11.
In addition we have the relation $\theta_1+\theta_2+\theta_3+\theta_4=2$.
\ifig\inter{\ Four D-brane stacks and chiral fermions
in the target space and on the disk world--sheet.}
{\epsfxsize=0.85\hsize\epsfbox{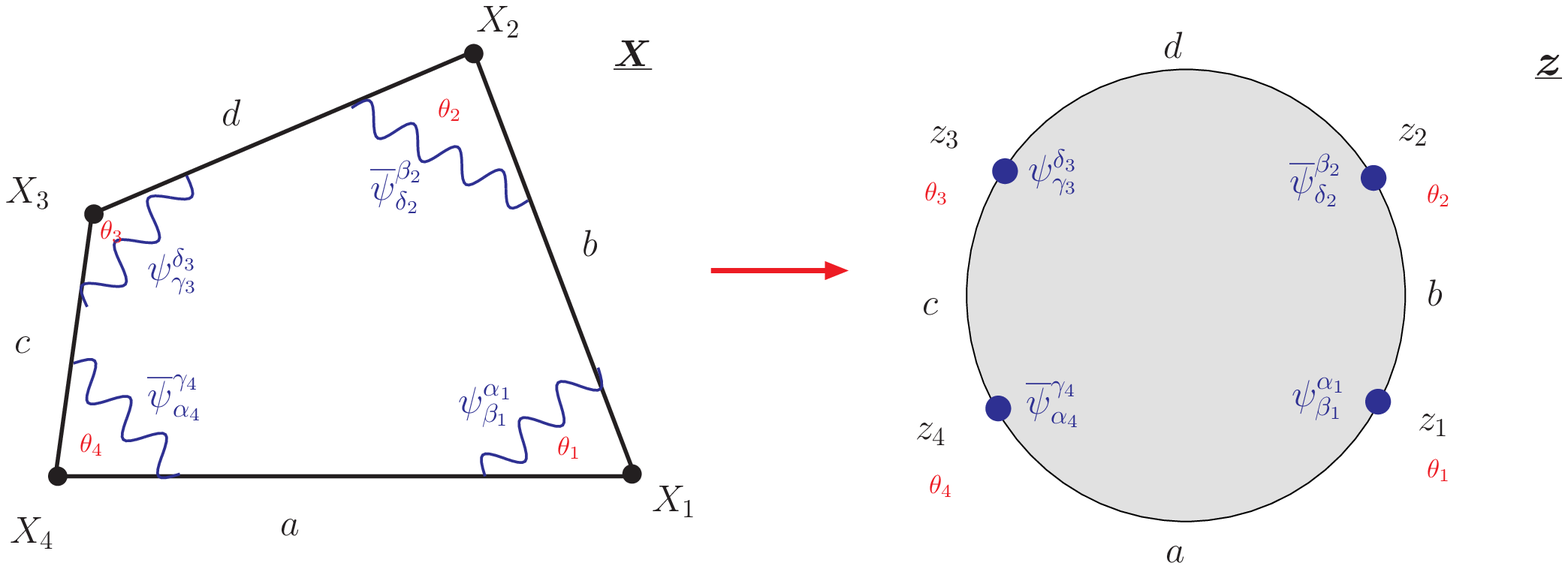}}
\noindent
The explicit expression of \setupferm\ depends on the number of different
branes $a,\ldots,d$ involved and the location of the four intersection points
$X_1,\ldots,X_4$.
On the world--sheet of the disk the ordering $X_i$ of adjacent D$6$--branes is
translated into a related ordering of vertex operator positions $z_i$ through
the map $z_i\longrightarrow X_i:=X(z_i)$.
Due to the chirality and twist properties of the four fermions
the dual resonance channels of the full amplitude are very restricted.
Further details depend on the specific configuration of intersections
and will be discussed below.

In what follows we need to define the straight line distance between two different
brane intersection points:
\eqn\whatf{
\delta^j_a:=X^j_1-X^j_4\ \ ,\ \ \delta^j_b=X^j_2-X^j_1\ \ ,\ \
\delta^j_c=X^j_4-X^j_3\ \ ,\ \ \delta^j_d=X^j_3-X^j_2\ .}
For a given brane $a$ the latter are decomposed into
\eqn\whatff{
\delta^j_a=\eps_a^j\ L^j_a+d_a^j\ \ \ ,\ \ \ \eps_a^j\in\IR}
with the longitudinal direction $L_a^j$ along the brane $a$ and
transverse component $d_a^j$, \cf  Figure 12.
\ifig\configz{\ Two branes $a$ and $b$ and the distance $\delta_a$.}
{\epsfxsize=0.5\hsize\epsfbox{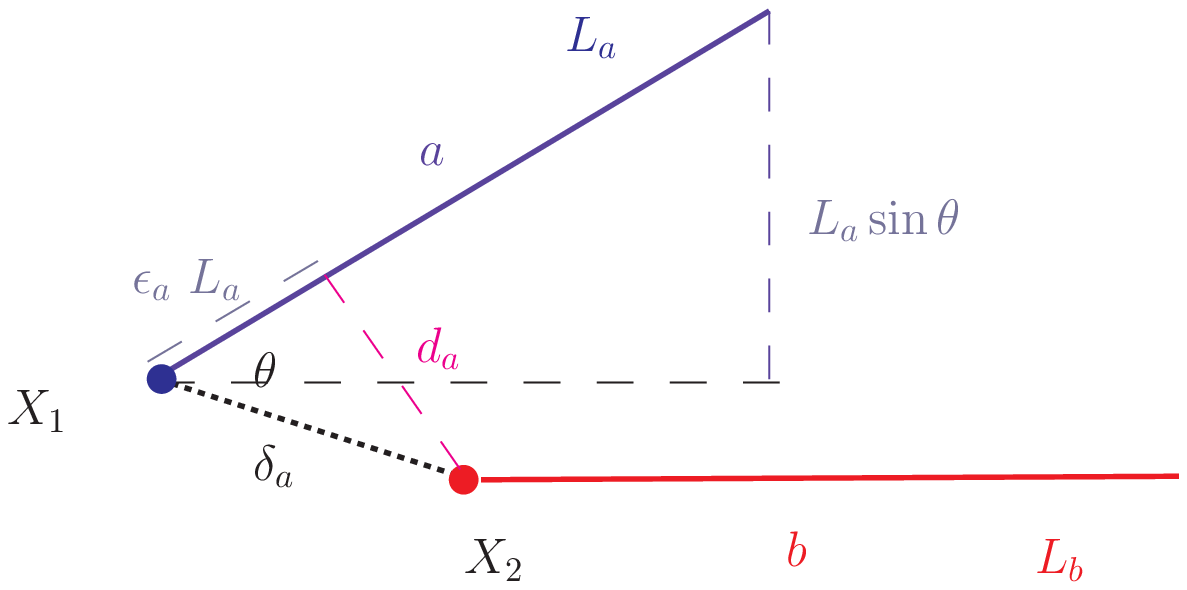}}
\noindent
The distance of two equivalent intersections from one pair $(a,b)$
of intersecting branes is given by a multiple of the lattice vectors $L_a^j$
or $L_b^j$, \ie $d^j_a=0$ and $\eps^j_a\in \IZ$.

Depending on the angles $\th_i$ for particular brane configurations
the amplitude \setupferm\ may furnish massless gauge boson exchange.
In those cases massless gauge bosons, internal KK states w.r.t. the longitudinal
brane directions and winding states w.r.t. their orthogonal directions are exchanged.
The masses of internal KK states w.r.t. the compact direction $L_b^j$ of brane $b$
and winding states w.r.t
the direction between brane $b$ and $a$ related to the gauge boson from stack
$b$ are given by:
\eqn\KKmassb{
\ap\ (m^j_{ab})^2=\fc{\sin^2(\pi\theta^j)}{\ap}\ \lf|p^j_aL_a^j+\delta^j_a\ri|^2+
\fc{\ap\ ({\tilde p}^j_b)^2}{|L^j_b|^2}\ \ \ ,\ \ \ p_a,\tilde p_b\in\IZ^6\ .}

In the following we discuss the three cases introduced before
\eqn\CASE{\eqalign{
(i)&\ \ \ \ \th_1=\th_3=\th,\ \th_2=\th_4=1-\th\ ,\cr
(ii)&\ \ \ \ \th_1=\th,\ \th_2=1-\th,\ \th_3=\nu,\ \th_4=1-\nu\ ,\cr
(iii)&\ \ \ \ \th_1,\ \th_2,\ \th_3,\ \th_4=2-\th_1-\th_2-\th_3\ ,}}
separately. Furthermore we must specify the intersections classes $f_i$ the four
intersection points  $X_i$ are related to.
In the following let us describe these cases in more detail.\br

\noindent
$\underline{(i)\ \rm{\it One\ angle}\ \th:}$\br
\noindent
In case $(i)$ we may either have $(1)$ chiral fermions stemming from
intersections $f_i$ of a single pair of branes $a,b$.
For this case we have $d^j=0$, \ie $c\simeq b,\ d\simeq a$. In the second case $(2)$
we consider chiral fermions stemming from
intersections $f_i,\tilde f_j$ of two pairs of branes $a,b$ and $c,d$,
which are mutually shifted
by some distance $d^j$ orthogonal to the brane directions $L^j$.\br
$(i.1)$ The pair of two intersecting branes $a$ and $b$ has $I_{ab}$
intersection points $f_i$ and all four points $X_i$ are assumed to be elements thereof.
A generic case with $I_{ab}=3$ is depicted in Figure 13 with the three
different intersection points drawn in black, red and blue, respectively.
We must further specify the class of intersection points $f_i$ involved, which
yields to the subcases $(i.1a)$ and $(i.1b)$.\br
\iifig\config{Two  possible polygons between two pairs of conjugate
intersection points}
{of an intersecting brane pair $(a,b)$ with intersection number $I_{ab}=3$.}
{\epsfxsize=0.55\hsize\epsfbox{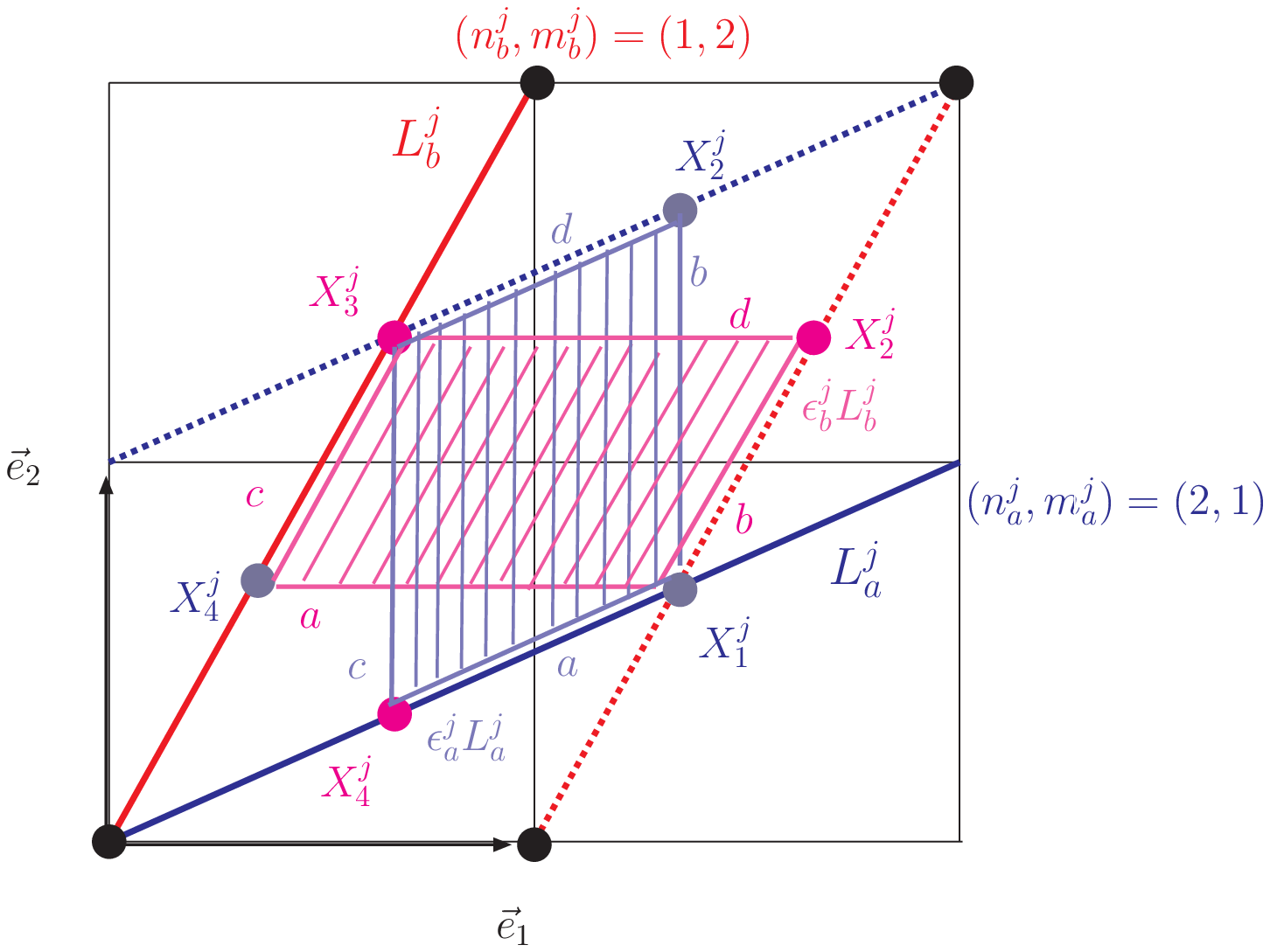}}

\noindent
$(i.1a)$ All four chiral fermions all located at the same intersection $f$.
In that case all four intersection points $X_i$
differ a by an integer lattice shift, \ie $X_i=f+\IZ\ L$ and hence
$\eps^j\in\IZ\ ,\ d^j=0$.
In \config\ the class $f$ may be \eg the set of four black dots,
which span one polygon. \br

\noindent
$(i.1b)$ A pair of two chiral fermions from the same
intersection $f_i$ and an other pair from an other intersection $f_j$.
In \config\ the intersections $f_i,f_j$ may be \eg the two blue and two
red dots, respectively.
Obviously, the points from the same set of intersections are separated by
a lattice vector, \ie $\eps^j\in\IZ,\ d^j=0$.
However, the length between points from different intersections is smaller.
It is given by the intersection number $I_{ab}$, \ie $\eps^j\in
\fc{\IZ}{I_{ab}},\ d^j=0$.
Generically, there are two different configurations for the polygon spanned by
the four points $X_i$. In \config,  these two polygons are
drawn in red and blue, respectively:
$\eps_a^j,\eps_d^j\in\IZ\ ,\ \eps_b^j,\eps_c^j\in\IQ$ or
$\eps_b^j,\eps_c^j\in\IZ\ ,\ \eps_a^j,\eps_d^j\in\IQ$ and $d^j=0$.\br

\noindent
$(i.2)$ The case $d^j\neq 0$ describes two intersecting D--brane pairs $(a,b)$ and
$(c,d)$ with the intersection angle $\th$. This situation is similar as in
$(i.1)$ except, that intersection points $\tilde f_i$ of one D--brane pair
are mutually shifted by some distance $d^j$ orthogonal
to the brane directions $L^j_a,L^j_b$.
The intersection numbers do not change, \ie  $I_{ab}=I_{cd}$. \br

\noindent
$\underline{(ii)\ \rm{\it Two\ angles}\ \th,\nu:}$\br
\noindent
For this case we consider two different intersecting D--brane pairs $(a,b)$ and
$(c,d)$ with the intersection angles $\th,\nu$ respectively, \cf \inter.
In this case, there is no relation between the set of intersections of $(a,b)$
and $(c,d)$ as in the previous case $(i.2)$.
All four chiral fermions may originate from different intersections $f_i$, \cf \inter.
One pair of fermions is related to the
twist--antitwist pair $(\th,1-\th)$ at the intersections $(X_1,X_2)$ and the second pair
of fermions is related to the twist--antitwist pair $(\nu,1-\nu)$ from the intersections
$(X_3,X_4)$. Hence only one polygon configuration is possible.\br

\noindent
$\underline{(iii)\ \rm{\it Three\ angles}\ \th_1,\th_2,\th_3:}$\br
The most general case of \setupferm\ involves four fermions from four different
intersections $X_i$. For that case we have three arbitrary angles $\th_1,\th_2,\th_3$ and
$\th_4=2-\th_1-\th_2-\th_3$. This is the situation depicted in \exiii.\br

%%%%%%%%%%%%%%%%%%%%%%%%%%%%%%%%%%%%%%%%%%%%%%%%%%%%%%%%%%%%%%%%%%%%%%%%%%%%%
\br
\noindent
{\sl (i) Four-fermion amplitudes involving two twist--antitwist pairs $(\th,1-\th)$}\br

\noindent
We consider the case of two pairs $(a,b)$ and $(c,d)\simeq (b',a')$ of D--branes
intersecting at the angle $\th$.
In \setupferm\ we have two pairs of twist--antitwist fields $(\th,1-\th)$
from intersections $f_i$ and $f_j$, respectively.
One pair of fermions is related to a twist--antitwist
pair $(\th,1-\th)$  at the points $X_1,X_2$ related to the intersection
$f_i$ (or $X_2,X_3$ related to the intersection $f_j$)
and a second fermion pair to an other  twist--antitwist
pair $(\th,1-\th)$ at the points $X_3,X_4$ related to the intersection
$f_j$ (or $X_1,X_4$ related to the intersection $f_i$).

The explicit expression of \setupferm\ for the case $(i)$
has been computed\foot{Four--fermion interactions in D--brane models have also
been discussed in \refs{\Narain,\Benakli}.}
in \refs{\Cvetic,\Abeli,\KW} and extended in \Chemtob
\eqn\fourferm{\eqalign{
\Mc[\psi^{\alpha_1}_{\beta_1}(u_1,k_1)\ \ov\psi^{\beta_2}_{\alpha'_2}(\ov u_2,k_2)\
\psi^{\alpha_3'}_{\beta'_3}(u_3,k_3)\  &\ov\psi^{\beta_4'}_{\alpha_4}(\ov u_4,k_4)]
=4\pi\ \ap\ e^{\phi_{10}}\
(u_{1L}u_{3L})\ (\ov u_{2R}\ov u_{4R})\cr
&\times \int_0^1dx\ x^{s-1}\ (1-x)^{u-1}\
\lf(\prod_{i=1}^3 I_i(x)^{-1/2}\ri)\cr
&\hskip-5.25cm\times\lf\{\Tr(T^{\alpha_1}_{\beta_1}T^{\beta_2}_{\alpha'_2}
T^{\alpha_3'}_{\beta'_3}T^{\beta'_4}_{\alpha_4})
\sum_{p_b,p_d\in\IZ^6} e^{-S^{bd}_{\rm inst.}(x)}+
\Tr(T^{\alpha_1}_{\beta_1}T^{\beta'_4}_{\alpha_4}T^{\alpha_3'}_{\beta'_3}
T^{\beta_2}_{\alpha'_2})\sum_{p_a,p_c\in\IZ^6} e^{-S_{\rm inst.}^{ac}(x)}\ri\}\ ,}}
with the disk instanton action
\eqn\diskaction{
S^{bd}_{\rm inst.}(x)=\fc{\pi}{\ap}\ \sum_{j=1}^3\sin(\pi\theta^j)\
\lf[\ \lf|p^j_bL_b^j+\delta^j_b\ri|^2\ \tau_j(x)+
\lf|p^j_dL^j_d+\delta^j_d\ri|^2\ \tau_j(1-x)\ \ri]\ ,}
and the combinations of hypergeometric functions:
\eqn\hypergeom{
\tau_j(x)=\fc{F_j(1-x)}{F_j(x)}\ \ \ ,\ \ \
I_j(x)=\fc{F_j(x)\ F_j(1-x)}{\sin(\pi\theta^j)}\ \ \ \ ,\ \ \
F_j(x)=\FF{2}{1}\lf[\th^j,1-\th^j,1;x\ri]\ .}
The first term of \fourferm\ accounts for the polygon with the
twist--antitwist pairs at $X_1,X_2$ and $X_3,X_4$, while the second term
describes the polygon with the twist--antitwist pairs at $X_1,X_4$ and $X_2,X_3$.
In \fourferm\ the spinor products $(u_{1L}u_{3L})\ (\ov u_{2R}\ov u_{4R})$ arise from
contracting the space--time spin fields of the fermion vertex operators
\chiralfermion. The following identity is useful for extracting the
gauge boson exchange channels:
\eqn\Peskin{
(u_{1L}u_{3L})\ (\ov u_{2R}\ov u_{4R})=\h\ (u_{1L}\sigma^\mu \ov u_{2R})\
(u_{3L}\sigma_\mu \ov u_{4R})=-
\h\ (u_{1L}\sigma^\mu \ov u_{4R})\ (u_{3L}\sigma_\mu\ov u_{2R})\ .}
The normalization of \fourferm\ simply arises from our convention
\opscouplings\ and \diskfactor, \ie $g_\psi^4\tilde C_{D_2}=4\ap e^{\phi_{10}}
g_{10}^2\equiv 4\ap e^{\phi_{10}}$.

With the limits
\eqn\limes{
F(x)\ra 1\ \ \ ,\ \ \ F(1-x)\ra\fc{\sin(\pi\theta)}{\pi}\
\ln\lf(\fc{\delta}{x}\ri)\ \ \ ,\ \ \ \ln\delta(\theta)=2\psi(1)-\psi(\theta)-
\psi(1-\th)}
as $x\ra 0$ we may extract from \fourferm\ the $s$--channel  pole contribution
\eqn\fourferms{\eqalign{
\Mc\longrightarrow&\ \ap\ (u_{1L}\si^\mu \ov u_{2R})\ (u_{3L} \si_\mu \ov u_{4R})\cr
&\times\lf\{\
\Tr(T^{\alpha_1}_{\beta_1}T^{\beta_2}_{\alpha'_2}
T^{\alpha_3'}_{\beta'_3}T^{\beta'_4}_{\alpha_4})\ g_{D6_{a'}}^2\
\sum_{p_b,{\tilde p}_d\in \IZ^6}
\fc{\prod\limits_{j=1}^3\delta(\th^j)^{-\ap(m^j_{bd})^2}\ e^{2\pi i {\tilde p}^j_d
\fc{\delta^j_d}{|L^j_d|}}}{s+\ap\sum\limits_{j=1}^3 (m^j_{bd})^2}\ri.\cr
&\lf.+\Tr(T^{\alpha_1}_{\beta_1}T^{\beta'_4}_{\alpha_4}
T^{\alpha_3'}_{\beta'_3}T^{\beta_2}_{\alpha'_2})\
g_{D6_{b'}}^2\ \sum_{p_a,{\tilde p}_c\in \IZ^6}
\fc{\prod\limits_{j=1}^3\delta(\th^j)^{-\ap(m^j_{ac})^2}\ e^{2\pi i {\tilde p}^j_c
\fc{\delta^j_c}{|L_c^j|}}}{s+\ap\sum\limits_{j=1}^3 (m^j_{ac})^2}\ \ri\}\ ,}}
with the gauge couplings
\eqn\gaugeb{
g^{-2}_{D6_a}=(2\pi)^{-1}\ \ap^{-3/2}\ e^{-\phi_{10}}\ \prod\limits_{j=1}^3 |L_a^j|}
introduced in \gaugesix\ and the masses \KKmassb\ of internal KK and 
winding states. In deriving \fourferms\ a Poisson resummation on the integers $p_d$
and $p_c$ is involved.
On the other hand, the $u$--channel of \fourferm\ gives rise to:
\eqn\fourfermt{\eqalign{
\Mc\longrightarrow&\ -\ap\ (u_{1L}\si^\mu \ov u_{4R})\ (u_{3L}\sigma_\mu \ov u_{2R})\cr
&\times\lf\{\
\Tr(T^{\alpha_1}_{\beta_1}T^{\beta_2}_{\alpha'_2}
T^{\alpha_3'}_{\beta'_3}T^{\beta'_4}_{\alpha_4})\ g_{D6_b}^2\
\sum_{p_d,{\tilde p}_b\in\IZ^6}
\fc{\prod\limits_{j=1}^3\delta(\th^j)^{-\ap(m^j_{db})^2}\ e^{2\pi i {\tilde p}^j_b
\fc{\delta^j_b}{|L_b^j|}}}{u+\ap\sum\limits_{j=1}^3 (m^j_{db})^2}\ri.\cr
&\lf.+\Tr(T^{\alpha_1}_{\beta_1}T^{\beta'_4}_{\alpha_4}
T^{\alpha_3'}_{\beta'_3}T^{\beta_2}_{\alpha'_2})\
g_{D6_{a}}^2\ \sum_{p_c,{\tilde p}_a\in\IZ^6}
\fc{\prod\limits_{j=1}^3\delta(\th^j)^{-\ap(m^j_{ca})^2}\ e^{2\pi i {\tilde p}^j_a
\fc{\delta^j_a}{|L_a^j|}}}{u+\ap\sum\limits_{j=1}^3 (m^j_{ca})^2}\ \ri\}\ .}}
In deriving \fourferms\ a Poisson resummation on the integers $p_b$
and $p_c$ is involved.
In the case of $d^j=0$ with the identifications $b'=c\simeq b$ and $a'=d\simeq a$,
we have $m_{ca}^j,m_{bd}^j\equiv m^j_{ba}$ and $m_{ac}^j,m_{db}^j\equiv m^j_{ab}$.

Finally let us discuss the special subcases of $(i)$, introduced above:\br
\noindent
$(i.1)$ The pairs of two branes $a,b$ and $d,c$ are identical, \ie
$d\simeq a,\ c\simeq b$, \cf Figure 14.
\ifig\intero{Four D-brane stacks and chiral fermions
in the target space and on the disk world--sheet.}
{\epsfxsize=0.85\hsize\epsfbox{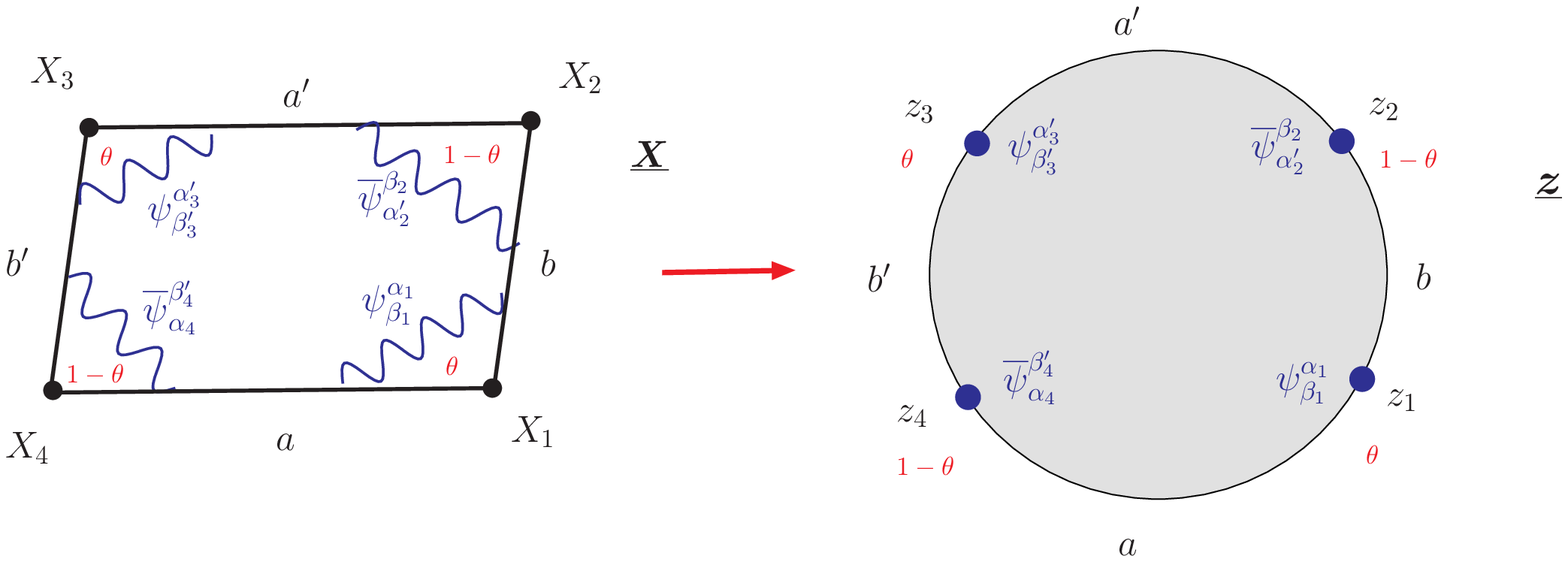}}
\noindent
$(i.1a)$ For all intersections $X_i$ related to {\it one} intersection $f$ we have
$\delta^j=0$. From \KKmassb\ we see, that
massless gauge boson exchange from stack $a$ or stack $b$ appears
both in the $s$--  and $u$--channel.
The case under consideration corresponds \eg to a scattering process
of four quarks from one family, \eg the scattering of  $u,\bar u$--quarks (\cf
\exi).
For this case we introduce the function
\eqn\tomi{
V_{abab}(s,u)=2\pi\ \ap\ e^{\phi_{10}}\int_0^1dx\ x^{s-1}\ (1-x)^{u-1}\
\lf(\prod_{i=1}^3 I_i(x)^{-1/2}\ri)\ 
\sum_{p_a,p_b\in\IZ^6} e^{-S^{ba}_{\rm inst.}(x)}\ ,}
which will become relevant in the next Section. The instanton action becomes in that case:
\eqn\diskactionn{
S^{ba}_{\rm inst.}(x)=\fc{\pi}{\ap}\ \sum_{j=1}^3\sin(\pi\theta^j)\
\lf[\ |p^j_bL_b^j|^2\ \tau_j(x)+|p^j_aL^j_a|^2\ \tau_j(1-x)\ \ri]\ .}
According to \fourferms\ and \fourfermt\ in the limit $s,t,u\ra0$
the function $V_{abab}(s,u)$ exhibits the behaviour:
\eqn\behav{
V_{abab}(s,u)=\ap\ \lf(\ \fc{g_{D6_a}^2}{s}+\fc{g_{D6_b}^2}{u}\ \ri)+\ldots\ .}\br
$(i.1b)$ In the case of two distinct intersections $f_i,f_j$ we have:
$\delta_b^j\equiv 0,\ \delta_d^j\sim\fc{L_d^j}{I_{db}^j}\simeq\fc{L_a^j}{I_{ab}^j}$ and
$\delta_a^j\equiv 0,\ \delta_c^j\sim\fc{L_c^j}{I_{ac}^j}\simeq\fc{L_b^j}{I_{ab}^j}$.
Hence, there are massless gauge boson exchanges from stack $a$ or stack $b$
only in the $s$--channel \fourferms. However, no massless gauge boson
contribution stems from the $u$--channel \fourfermt.
The case under consideration corresponds \eg to a scattering process
of a pair of quarks associated
to two different families, \eg the scattering of  $u,\bar u$--quarks
with $c,\bar c$--quarks.

\noindent
$(i.2)$ In the case of $d^j\neq 0$ the two polygon
contributions are  parameterized by:
$\delta_a^j =d_a^j,\ \delta_c^j\sim\fc{L_c^j}{I_{ac}^j}$ and
$\delta_b^j=d_b^j,\ \delta_d^j\sim\fc{L_d^j}{I_{db}^j}$.
For this case the mass \KKmassb\ is always non--zero and no massless gauge bosons
are exchanged neither in the $s$--channel \fourferms\ nor in the $u$--channel
\fourfermt.

\br\noindent
{\sl (ii) Four-fermion amplitudes involving two twist--antitwist pairs
$(\th,1-\th)$ and $(\nu,1-\nu)$}\br

\noindent
Here we consider the generic case of four different stacks of D$6$--branes
$a,b,c$ and $d$ with two pairs of twist--antitwist fields,
$(\th,1-\th)$ and $(\nu,1-\nu)$. We have two
intersection angles $\th$ and $\nu$ referring to the pairs $(a,b)$ and
$(c,d)$, respectively, \cf Figure 15.
\ifig\inters{\ Four D-brane stacks and chiral fermions
in the target space and on the disk world--sheet.}
{\epsfxsize=0.85\hsize\epsfbox{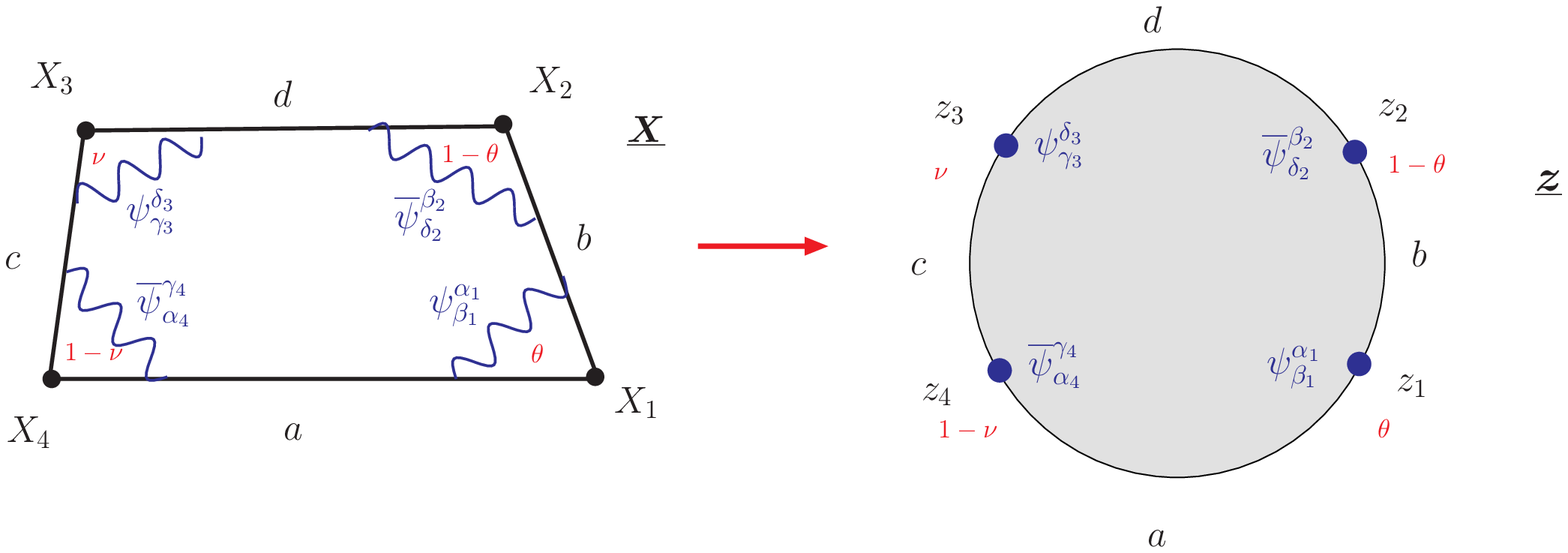}}
\noindent
In the previous case $\th=\nu$ we have encountered two possible polygon contributions
as a matter of how the two twist--antitwist pairs are paired: $(X_1,X_2)$ and
$(X_3,X_4)$ or $(X_1,X_4)$ and  $(X_2,X_3)$, respectively.
Obviously, for $\th\neq\nu$ we have
only one polygon from the twist--antitwist pairs $(X_1,X_2)$ and
$(X_3,X_4)$.

The explicit expression of \setupferm\ for $\th\neq\nu$ has been computed
in \refs{\Cvetic,\Abelii}
\eqn\Fourferm{\eqalign{
\Mc[\psi^{\alpha_1}_{\beta_1}(u_1,k_1)&\
\ov\psi^{\beta_2}_{\delta_2}(\ov u_2,k_2)\
\psi^{\delta_3}_{\gamma_3}(u_3,k_3)\
\ov\psi^{\gamma_4}_{\alpha_4}(\ov u_4,k_4)]=
4\pi\ \ap\ e^{\phi_{10}}\cr
&\times \lf[\Tr(T^{\alpha_1}_{\beta_1} T^{\beta_2}_{\delta_2}\
T^{\delta_3}_{\gamma_3}T^{\gamma_4}_{\alpha_4})+
\Tr(T^{\alpha_1}_{\beta_1} T^{\gamma_4}_{\alpha_4}\
T^{\delta_3}_{\gamma_3}T^{\beta_2}_{\delta_2})\ri] (u_{1L}u_{3L})\ 
(\ov u_{2R}\ov u_{4R})\ \cr
&\times \int_0^1dx\ x^{s-1}\ (1-x)^{u-1}\
\lf(\prod_{i=1}^3 I_i(x)^{-1/2}\ri)\  \sum_{p_b,p_d\in\IZ^6}
e^{-S_{\rm inst.}(x)}\ ,}}
with the disk instanton action
\eqn\Diskaction{
S_{\rm inst.}(x)=\fc{\pi}{\ap}\
\sum_{j=1}^3\sin(\pi\theta^j)\
\lf\{\ \lf|v^j_b\ri|^2\ \lf(\tau_j(x)+\fc{\beta^j}{2}\ri)+
\lf|v_d^j+\fc{\beta^j}{2}\ v_b^j\ri|^2\ 
\fc{1}{\tau_j(x)+\fc{\beta^j}{2}}\ \ri\}\ ,}
with $v_r=p_rL_r+\delta_r$,\ 
$\beta^j=-\fc{\sin[\pi(\th^j-\nu^j)]}{\sin(\pi\nu^j)}$ and the function:
\eqn\Hypergeom{
2\pi\ I_j(x)=B(1-\th^j,\nu^j)\ F_{1;j}(1-x)\ K_{2;j}(x)+B(\th^j,1-\nu^j)\
F_{2;j}(1-x)\ K_{1;j}(x)\ .}
Above we have introduced 
the hypergeometric functions $K_{1;j}(x)=\FF{2}{1}[1-\th^j,\nu^j,1;x]$,\
$K_{2;j}(x)=\FF{2}{1}[\th^j,1-\nu^j,1;x]$,\
$F_{1;j}(x)=\FF{2}{1}[1-\th^j,\nu^j;1-\th^j+\nu^j;x]$, and
$F_{2;j}(x)=\FF{2}{1}[1-\nu^j,\th^j,1+\th-\nu^j;x],$
and the Euler Beta function \veneziano.
Finally, we have
\eqn\TAU{
\tau_j(x)=\fc{1}{\pi}\ B(1-\th^j,\nu^j)\ \sin(\pi\th^j)\
\fc{F_{1;j}(1-x)}{K_{1;j}(x)}\ ,}
and the relation:
\eqn\relation{
I_j(x)=\fc{1}{\sin(\pi\theta^j)}\
\lf(\tau_j(x)+\fc{\beta^j}{2}\ri)\ K_{1;j}(x)\ K_{2;j}(x)\ .}
As result of respecting the global
monodromy conditions we have the relations: 
$\beta^jv^j_b=v_a^j-v_d^j$, \ie $v_d^j+\fc{\beta^j}{2} v_b^j=\h(v_a^j+v_d^j)$
and $\sin(\pi\nu^j)\ |v^j_c|=\sin(\pi\lambda^j)\ |v_b^j|$.
The function \Hypergeom, which determines the quantum part of the amplitude
\Fourferm, is the square root of the relevant closed string piece \doubref\BKM\EJSS.
For $\th^j=\nu^j$ we have $\beta^j=0$ and the functions
$\tau_j(x)$ and $I_j(x)$ reduce to the expressions \hypergeom.
With this information it is straightforward to show, that \Fourferm\ boils down
to the first term of \fourferm\ as $\nu^j\ra\th^j$.

Due to the chirality and twist properties of the four fermions
the amplitude \Fourferm\ furnishes massless gauge boson exchange through
the $s$--channel only.
On the other hand, for $\th^j\neq \nu^j$
the limit $x\ra 1$ does not imply massless gauge boson exchange and factorizes
onto Yukawa couplings. This property is discussed in more detail for case $(iii)$.

In the following let us consider the case $d=a,\ \delta=\alpha$
and $\delta_c^j,\delta^j_b=0$, which corresponds to \exii.
Only in the $s$--channel a massless gauge boson exchange occurs.
To extract from \Fourferm\ the $s$--channel  pole contribution we need the limit
\eqn\limes{\tau_j(x)\ra\fc{1}{\pi}\ \sin(\pi\th^j)\
\ln\lf(\fc{\delta(1-\th^j,\nu^j)}{x}\ri)\ ,}
as $x\ra 0$, with $\ln\delta(\theta^j,\nu^j)=2\psi(1)-\psi(\th^j)-\psi(\nu^j)$.
For the limit $x\ra0$ we obtain
\eqn\Fourferms{\eqalign{
\Mc&\longrightarrow\
\ap\ \lf\{\Tr(T^{\alpha_1}_{\beta_1} T^{\beta_2}_{\alpha_2}\
T^{\alpha_3}_{\gamma_3}T^{\gamma_4}_{\alpha_4})+
\Tr(T^{\beta_2}_{\alpha_2}T^{\alpha_1}_{\beta_1} T^{\gamma_4}_{\alpha_4}\
T^{\alpha_3}_{\gamma_3})\ri\}\
(u_{1L}\gamma_\mu \ov u_{2R})\ (u_{3L} \gamma^\mu \ov u_{4R})\cr
&\times g_{D6_a}^2\ \sum_{p_b,{\tilde p}_a\in\IZ^6}
\fc{\prod\limits_{j=1}^3\delta(1-\th^j,\nu^j)^{-\ap(m^j_{ba})^2}\
e^{2\pi i {\tilde p}^j_a\fc{\delta^j_a}{|L^j_a|}}\
e^{\pi i\lf(\fc{1}{1-e^{2\pi i\th^j}}-\fc{1}{1-e^{2\pi i\nu^j}}\ri)\ \ap(m_{ba}^j)^2}}
{s+\ap\sum\limits_{j=1}^3 (m^j_{ba})^2}\ ,}}
with the gauge coupling \gaugeb\ and the mass $m^j_{ba}$ of KK and winding states
given in \eqq \KKmassb.
In deriving \fourferms\ a Poisson resummation on the integer $p_a$ is involved.
Again, for $\nu^j=\th^j$ the limit \Fourferms\
reduces to the corresponding expression of \fourferms.
The case under consideration describes \eg the process
$q^-\bar q^+\to q^+\bar q^-\ \big(q_{L} q_{R}^c\to q_{R} q_{L}^c\big)$,
\cf \quarksaa. To accommodate the
different choices of helicities of the four fermions we introduce the function
\eqn\tomii{
V_{abac}^{(n_1,n_2)}(s,u)=2\pi\ \ap\ e^{\phi_{10}}\int_0^1dx\
x^{s-1-\fc{n_1}{2}}\ (1-x)^{u-1-\fc{n_2}{2}}\
\lf(\prod_{i=1}^3 I_i(x)^{-1/2}\ri)
\sum_{p_a,p_b\in\IZ^6} e^{-S^{ba}_{\rm inst.}(x)}\ ,}
with the following assignment of helicity configurations
\eqn\WITH{\eqalign{
V_{abac}(s,u)&:=V_{abac}^{(0,0)}(s,u)\ \ \ {\rm for}\ \ \
(13)_L(24)_R\ \ \ {\rm and}\ \ \
(24)_L(13)_R\ ,\cr
V'_{abac}(s,u)&:=V_{abac}^{(0,1)}(s,u)\ \ \ {\rm for}\ \ \
(14)_L(23)_R\ \ \ {\rm and}\ \ \
(23)_L(14)_R\ ,\cr
V''_{abac}(s,u)&:=V_{abac}^{(1,0)}(s,u)\ \ \ {\rm for}\ \ \
(12)_L(34)_R\ \ \ {\rm and}\ \ \
(34)_L(12)_R\ ,}}
and the common instanton part:
\eqn\Diskactionn{
S^{ba}_{\rm inst.}(x)=\fc{\pi}{\ap}\
\sum_{j=1}^3\sin(\pi\theta^j)\
\lf\{\ |p^j_bL_b^j|^2\ \lf(\tau_j(x)+\fc{\beta^j}{2}\ri)+
|p^j_aL^j_a+\delta^j_a|^2\ \fc{1}{\tau_j(x)+\fc{\beta^j}{2}}\ \ri\}\ .}
According to \Fourferms\ in the limit $s,t,u\ra0$
the functions $V_{abac}(s,u)$ and $V'_{abac}(s,u)$  furnish massless gauge
boson exchange:
\eqn\behavi{
V_{abac}(s,u)=\ap\ \fc{g_{D6_a}^2}{s}+\ldots\ \ \ ,\ \ \
V'_{abac}(s,u)=\ap\ \fc{g_{D6_a}^2}{s}+\ldots\ .}
On the other hand, the function $V_{abac}''(s,u)$ does not imply
massless gauge boson exchange and behaves as:
\eqn\behavii{
V''_{abac}(s,u)=\ap\ \fc{g_{D6_a}^2}{s-\h}+\ldots\ .}
This case is discussed in more detail for case $(iii)$.\br

\noindent
{\sl (iii) Four-fermion amplitudes involving fermions from four different
intersections}\br

\noindent
The most general case, depicted in \inter, involves four fermions from four
different
intersections $X_i$ with the four angles $\th_i$ and $\th_4=2-\th_1-\th_2-\th_3$.
The explicit expression of \setupferm\ for this case has been computed
in \Abelii
\eqn\FFourferm{\eqalign{
\Mc[\psi^{\alpha_1}_{\beta_1}(u_1,k_1)&
\ov\psi^{\beta_2}_{\delta_2}(\ov u_2,k_2)
\psi^{\delta_3}_{\gamma_3}(u_3,k_3)\
\ov\psi^{\gamma_4}_{\alpha_4}(\ov u_4,k_4)]=4\pi\ \ap\ e^{\phi_{10}}\
\lf(\prod_{i=1}^3\fc{\sin(\pi\th^i_2)\ \sin(\pi\th^i_4)}
{\sin(\pi\th^i_1)\ \sin(\pi\th^i_3)}\ri)^{1/4}\cr
&\times \lf[\Tr(T^{\alpha_1}_{\beta_1} T^{\beta_2}_{\delta_2}\
T^{\delta_3}_{\gamma_3}T^{\gamma_4}_{\alpha_4})+
\Tr(T^{\alpha_1}_{\beta_1} T^{\gamma_4}_{\alpha_4}\
T^{\delta_3}_{\gamma_3}T^{\beta_2}_{\delta_2})\ri] (u_{1L}u_{3L})\
(\ov u_{2R}\ov u_{4R})\ \cr
&\times \int_0^1dx\ x^{s-1}\ (1-x)^{u-1}\
\lf(\prod_{i=1}^3 I_i(x)^{-1/2}\ri)\ \sum_{p_b,p_d\in\IZ^6}
e^{-S_{\rm inst.}(x)}\ ,}}
with the disk instanton action
\eqn\DDiskaction{
S_{\rm inst.}(x)=\fc{\pi}{\ap}\ \sum_{j=1}^3\sin(\pi\theta_2^j)\
\fc{\lf|v^j_b\ \tau_j-v^j_d\ri|^2+\gamma^j\tilde\gamma^j\
\lf|v^j_b\ (\beta^j+\tau_j)+v_d^j\ (1+\alpha^j\ \tau_j)\ri|^2}
{\beta^j+2\ \tau_j+\alpha^j\ \tau_j^2}\ ,}
with $v_r=p_rL_r+\delta_r$ and the quantum contribution:
\eqn\HHypergeom{\eqalign{
2\pi\ I_j(x)&=\fc{\Gamma(1-\th^j_1)\ \Gamma(\th_3^j)}{\Gamma(\th_2^j+\th_3^j)\
\Gamma(\th_3^j+\th_4^j)}\ F_{1;j}(1-x)\ K_{2;j}(x)\cr
&+\fc{\Gamma(\th^j_1)\
\Gamma(1-\th_3^j)}{\Gamma(\th_1^j+\th_2^j)\ \Gamma(\th_1^j+\th_4^j)}\
F_{2;j}(1-x)\ K_{1;j}(x)\ .}}
Above we have introduced
the hypergeometric functions
$K_{1;j}(x)=\FF{2}{1}[\th_2^j,1-\th_4^j,\th_1^j+\th_2^j;x]$,\
$K_{2;j}(x)=\FF{2}{1}[\th_4^j,1-\th_2^j,\th_3^j+\th_4^j;x]$,\
$F_{1;j}(x)=\FF{2}{1}[\th_2^j,1-\th_4^j;\th^j_2+\th^j_3;x]$, and
$F_{2;j}(x)=\FF{2}{1}[\th_4^j,1-\th_2^j,\th^j_1+\th^j_4;x],$\
$\beta^j=-\fc{\sin[\pi(\th_2^j+\th_3^j)]}{\sin(\pi\th^j_3)},\ \alpha^j=
-\fc{\sin[\pi(\th_1^j+\th_2^j)]}{\sin(\pi\th^j_1)}$ and
$\gamma^j=\fc{\Gamma(1-\th^j_2)\ \Gamma(1-\th^j_4)}{\Gamma(\th^j_1)\
\Gamma(\th^j_3)},\ \tilde\gamma^j=\fc{\Gamma(\th^j_2)\ \Gamma(\th^j_4)}
{\Gamma(1-\th^j_1)\ \Gamma(1-\th^j_3)}$.
Finally, we have
\eqn\TTAU{
\tau_j(x)=\fc{B(\th_2^j,\th_3^j)}{B(\th_1^j,\th_2^j)}\
\fc{F_{1;j}(1-x)}{K_{1;j}(x)}\ .}
Note, that the function \HHypergeom\ and the classical action \DDiskaction\
have crossing symmetry under the combined manipulations $x\leftrightarrow1-x$ and
$\th_1^j\leftrightarrow\th_3^j$.
For $\th_1^j=\th^j, \th_2^j=1-\th^j,\ \th_3^j=\nu^j$ and $\th_4^j=1-\nu^j$
we have $\alpha^j=0,\ \gamma^j,\tilde\gamma^j=1$ and the functions
$I_j(x)$ and $\tau_j(x)$ reduce to the expressions \Hypergeom\ and \TAU,
respectively.
With this information it is straightforward to show, that \FFourferm\ boils down
to \Fourferm\ in this limit.
Furthermore, for $\th_1^j=\th^j_3=\th^j$ and $\th_2^j=\th_4^j=1-\th^j$
we have $\alpha^j,\beta^j=0,\ \gamma^j,\tilde\gamma^j=1$ and the
functions  $I_j(x)$ and $\tau_j(x)$ reduce to the expressions \hypergeom.

The amplitude \FFourferm\ does not furnish massless gauge boson exchange limits.
On the other hand, it factorizes onto Yukawa couplings.
In the following we investigate the helicity configuration
$(12)_L(34)_R$ and $(34)_L(12)_R$, which
corresponds to the function
\eqn\tomiii{\eqalign{
V_{abcd}(s,u)&=2\pi\ \ap\ e^{\phi_{10}}\
\lf(\prod_{i=1}^3\fc{\sin(\pi\th^i_2)\ \sin(\pi\th^i_4)}
{\sin(\pi\th^i_1)\ \sin(\pi\th^i_3)}\ri)^{1/4}\cr
&\times\int_0^1dx\ x^{s-3/2}\ (1-x)^{u-1}\
\lf(\prod_{i=1}^3 I_i(x)^{-1/2}\ri)\ \sum_{p_b,p_d\in\IZ^6}
e^{-S_{\rm inst.}(x)}\ ,}}
with the instanton action \DDiskaction.
For $x\ra0$ and $0<\th_1^j+\th_2^j<1$ the integral \tomiii\
gives rise to the $s$--channel limit
\eqn\tchannel{
V_{abcd}(s,u)\lra\ap\ e^{\phi_{10}}\ Y(\th_1,\th_2)\ Y(1-\th_3,1-\th_4)\ 
\fc{1}{s+\ap\ m_H^2}\ ,}
with the intermediate mass
\eqn\Higgs{
\ap\ m_H^2=1-\h\ \sum_{j=1}^3(\th^j_1+\th_2^j)\ ,}
and the Yukawa couplings \doubref\Cvetic\FBM:
\eqn\YUKA{
Y(\th_a,\th_b)=(4\pi)^{1/2}\prod_{j=1}^3\lf(2\pi\fc{\Gamma(1-\th_a^j)\ \Gamma(1-\th_b^j)\
\Gamma(\th_a^j+\th_b^j)}{\Gamma(\th_a^j)\
\Gamma(\th_b^j)\ \Gamma(1-\th_a^j-\th_b^j)}\ri)^{1/4} \sum_{v_{ba}^j} e^{-\fc{\pi}{\ap}\
\fc{\sin(\pi\th_a^j)\
\sin(\pi\th_b^j)}{\sin[\pi(\th_a^j+\th_b^j)]} |v_{ba}^j|^2}.}
Hence, in the limit $x\ra0$ (heavy) string states with mass \Higgs\ are exchanged.
These states may represent the SM Higgs field as well as some
exotic states. The latter may give possible stringy signatures at the LHC  \Berenstein.

The (relevant) four--point fermion amplitudes $V(s,u)$, whose explicit form is given
in \eqqs \tomi, \tomii\ and \tomiii, receive world--sheet disk instanton
corrections from holomorphic mappings of the string world--sheet into the
polygon spanned by the four intersection points $X_i$, respectively.
The three--point couplings \YUKA\ are derived from the latter by appropriate
factorization and the relevant polygon splits into two triangles, \cf \eqq  \tchannel.

The amplitudes  \tomi, \tomii\ and \tomiii\ give rise to string corrections to
the contact four fermion interaction. The first correction appears at the order $\ap$.
For $(n_1,n_2)=(0,0)$ corresponding to the helicity configurations
$(13)_L(24)_R$ or $(24)_L(13)_R$ the latter are extracted by setting $s,u=0$ and may be
summarized in the expression:
\eqn\FAP{
\lf.V^{(0,0)}_{abcd}(s,u)\ri|_{\ap}=2\pi\ \ap\ e^{\phi_{10}}
\lf(\prod_{i=1}^3\fc{\sin(\pi\th^i_2)\ \sin(\pi\th^i_4)}
{\sin(\pi\th^i_1)\ \sin(\pi\th^i_3)}\ri)^{\fc{1}{4}}
\int_0^1dx\ \fc{\prod\limits_{i=1}^3 I_i(x)^{-\h}}{x\ (1-x)}\ 
\sum_{p_b,p_d\in\IZ^6}e^{-S_{\rm inst.}(x)}.}
The three cases $V_{abab}$, $V_{abac}^{(0,0)}$ and $V^{(0,0)}_{abcd}$, which follow the
classification \CASE, may be obtained from \FAP\ by inserting the
corresponding 
instanton actions \diskactionn, \Diskactionn\ and \DDiskaction, respectively.
Note, that in \FAP\ the contributions of the poles $x\ra0,1$ cancel
due to the relation $\fc{1}{x(1-x)}=\fc{1}{x}+\fc{1}{1-x}$.
To conclude, in contrast to gluon scattering, the first string contact interaction
appears already at the order $\ap$.

There is yet an other way of writing the expressions \tomi\ and \tomii\
following after a Poisson resummation on $p_a$
\eqn\FAPP{\eqalign{
V_{abac}^{(n_1,n_2)}(s,u)&=\ap\ g_{D6_a}^2\ \int_0^1dx\ x^{s-1-\fc{n_1}{2}}\ 
(1-x)^{u-1-\fc{n_2}{2}}\
\lf(\prod_{i=1}^3 K_{1;i}(x)K_{2;i}(x)\ri)^{-1/2}\cr 
&\times \sum_{p_b,\tilde p_a\in\IZ^6}\prod_{j=1}^3 
e^{2\pi i {\tilde p}^j_a\fc{\delta^j_a}{|L^j_a|}}\
e^{\pi i\lf(\fc{1}{1-e^{2\pi i\th^j}}-\fc{1}{1-e^{2\pi i\nu^j}}\ri)\ap(m_{ba}^j)^2}\ 
e^{-\pi\fc{\tau_j(x)}{\sin(\pi\theta^j)}\ \ap (m_{ba}^j)2},}}
with the mass $m_{ba}$ of KK and winding modes, given in \eqq \KKmassb.
If the longitudinal brane directions are somewhat greater than the string
scale $\Ms$ the world--sheet instanton corrections are suppressed and the
exponential sum in  \FAPP\ may be ignored.
In that case the four--fermion couplings are insensitive to how the D6--branes
are wrapped around the compact space and they depend only on the local
structure of the brane intersections encoded in the intersection angles $\th_i^j$.
In other words, the quantum part of \tomi, \tomii\ and \tomiii,
given by the function $I_j(x)$,
depends only on the angles $\th_i^j$ and the 
string scale $\Ms$ and is not sensitive to
the scales of the internal space.
In that case the four--fermion couplings may be written as sum over
$s$--channel poles in lines of \basic. The massive intermediate states exchanged are
twisted states with masses of the order of $\Ms$.

%%%%%%%%%%%%%%%%%%%%%%%%%%%%%%%%%%%%%%%%%%%%%%%%%%%%%%%%%%%%%%%%%%%%%%%%%%%%%
\newsec{From Amplitudes to Parton Cross Sections}

The purpose of this Section is to present the squared moduli of
disk amplitudes derived in the previous Section, averaged over helicities and colors of the incident partons and summed over helicities and colors of the outgoing particles. In order to respect
the notation and conventions used in experimental literature, which are rooted in the classic exposition of Bj\"orken and Drell, several steps have to be accomplished. The first one is to revert to the $(+--\,-)$ metric signature.
Furthermore, appropriate crossing operations have to be performed on the amplitudes,
to ensure that the incident particles are always number 1 and 2 while the outgoing are number 3 and 4. They carry the initial four-momenta $k_1, k_2$ and final four-momenta $k_3, k_4$, respectively, that satisfy the conservation law
\eqn\law{k_1+k_2=k_3+k_4\ .}
A generic process written as $ef\to g\,h$ has the momenta assigned as $e(k_1)f(k_2)\to
g(k_3)\,h(k_4)$.
The kinematic invariants (Mandelstam variables) are defined in the standard way:
\eqn\mandel{s=(k_1+k_2)^2=2k_1k_2~,\quad t=(k_1-k_3)^2=-2k_1k_3~,\quad u=(k_1-k_4)^2=-2k_1k_4\ .}
They are constrained by
\eqn\stu{s+t+u=0~,\qquad (s>0,~ t<0,~ u<0)\ .}

Since in the previous Section, we implicitly used string mass units $M_{\rm string}\equiv M$ for the Mandelstam variables $\hat s,~ \hat t,~ \hat u$, and a metric of opposite signature, we need to redefine the universal string formfactor:
\eqn\redef{\hat V ~\to~ V(s,t,u)={su\over tM^2}\ B(-s/M^2,-u/M^2)=
{\Gamma(1-s/M^2)\Gamma(1-u/M^2)\over\Gamma(1+t/M^2)}}
and similarly
\eqn\vexpdef{
V_t =V(s,t,u) ~,\qquad V_s=V(t\leftrightarrow s) ~,\qquad V_u=V(t\leftrightarrow u) ~.}
Now the low-energy
 expansions read
\eqn\vexp{
V(s,t,u)\approx 1-{\pi^2\over 6}\ M^{-4}\ s\,
   u-\zeta(3)\ M^{-6}\ s\, t\, u+\dots}
Similarly, we introduce the following functions describing the four-fermion amplitudes:
\eqn\otha{F(s,u) =\,{s\over g_a^2}\ V_{abab}(-s/M^2,-u/M^2)\equiv\, F_{su}}
\eqn\othb{G(s,u) =\,{s\over g_a^2}\ V_{abac}(-s/M^2,-u/M^2)\equiv\, G_{su}}
\eqn\othc{G'(s,u) ={s\over g_a^2}\ V'_{abac}(-s/M^2,-u/M^2)\equiv\, G'_{su}}
where $V_{abab}$, $V_{abac}$ and $V'_{abac}$ are defined in Eqs. \tomi\ and \WITH.
The above functions depend on details of compactifications, therefore they are model-dependent.
Note that the above redefinitions single out the QCD coupling $g_a$ because it is the strongest coupling.
Thus the results presented below coincide with the QCD predictions in the limit $M\to\infty$, {\it i.e}.\ $V=F=G=G'=1$. The effects due to electro-weak forces can also be extracted, although with more care in taking this limit.

There are two basic operations performed when squaring the amplitudes and summing over colors and polarizations. First, the moduli squared of helicity amplitudes  containing some spinor products (twistors) are expressed in terms of Mandelstam variables. This involves a repeated use of the following identities:\foot{It is worth mentioning that in the step of inverting the momenta from incoming to outgoing ones,
$k_3\to -k_3,~ k_4\to -k_4 ~\Longrightarrow~ 2k_1k_3\to - 2k_1k_3=t~,~ 2k_1k_4\to - 2k_1k_4=u$.}
\eqn\spinid{\langle ij\rangle^*= [ji]~,\quad  \langle ij\rangle[ji] =2k_ik_j}
\eqn\cons{\sum_{n\neq i,j}\langle in\rangle[nj]=0 \quad {\rm(momentum~conservation)}\ ,}
\eqn\shout{\langle ij\rangle\langle mn\rangle =\langle im\rangle\langle jn\rangle -\langle in\rangle\langle jm\rangle~ \quad {\rm(Schouten's~identity)}\ .}
The second operation is the summation over color indices.
It depends on the representations of external particles, therefore we include it case by case in the following
discussion of all parton scattering processes.

\subsec{$gg\to gg,\ gg\to gA,\ gg\to AA$}
The starting expression is Eq. \mhva\ that holds for $SU(N)$ gluons and $U(1)$ vector bosons $A$ coupled to the baryon number.
In order to obtain the cross section for the (unpolarized) partonic
subprocess $gg\to gg$, we take the squared moduli of individual
amplitudes, sum over final polarizations and colors, and average over
initial polarizations and colors.
The following formulae are useful for summing over $SU(N)$ colors:
\eqn\dsq{\eqalign{
\sum_{a_1,a_2,a_3}d^{a_1a_2a_3}d^{a_1a_2a_3}&  ={(N^2-1)(N^2-4)\over 16 N}\cr
\sum_{a_1,a_2,a_3,a_4}d^{a_1a_2a_3a_4}d^{a_1a_2a_3a_4}& ={(N^2-1)(N^4-6N^2+18)\over 96 N^2}\cr
\sum_{a_1,a_2}f^{i_1a_1a_2}f^{i_2a_1a_2}&  = N\ \delta^{i_1i_2}\cr
\sum_{a_1,a_2,a_3}f^{i_1a_1a_2}f^{i_2a_2a_3}f^{i_3a_3a_1}& = {N\over 2}f^{i_1i_2i_3}
}}
As an example, the modulus square of
the amplitude \afull, {\it summed\/} over initial and final colors, is:
\eqn\msqu{\eqalign{|{\cal
   M}(g^-_1,g^-_2,&\ g^+_3, g^+_4)|^2 =~  g^4(N^2-1)\,s^4\,\cr &
\times \Bigg[2N^2\left({V^2_t\over s^2u^2}+ {V^2_s\over t^2u^2}
+{V^2_u\over s^2t^2}\right)
+{4(-N^2+3)\over N^2}\left({V_t\over s\,u}
+ {V_s\over tu}+{V_u\over s\,t}\right)^{\! 2}\Bigg]}}
The modulus squared of the $gg\to gg$ amplitude, summed over final polarizations and colors,
and averaged over all $4(N^2-1)^2$ possible initial polarization/color
configurations, reads
\eqn\mhvav{\eqalign{|{\cal M}(gg\to
 gg)|^2&=  g^4\bigg({1\over s^2}+{1\over t^2}+{1\over u^2}\bigg)\,\cr 
&\times\Bigl[C(N)\!\left(\, s^2V^2_s+ t^2V^2_t
+u^2V^2_u\,\right) + D(N)\!\left(\, sV_s
+ tV_t+uV_u\,\right)^{2}\;\Bigr],}}
where
\eqn\abbs{C(N)={2N^2\over N^2-1}~,\qquad\qquad D(N)={4(-N^2+3)\over N^2(N^2-1)}~.}
Note that $D(N)$ is of order ${\cal O}(1/N^2)$ with respect to $C(N)$. Furthermore, the corresponding kinematic factor is suppressed in the low-energy limit at the rate ${
\cal O}(M^{-8})$ with respect to the leading QCD contribution which emerges from the first term (with $C(N)$ factor), {\it c.f}.\ Eqs. \stu\ and \vexp. Thus the second term in Eq. \mhvav\ is suppressed in both large $N$ and in low-energy limits.

The $U(1)$ gauge bosons $A$ can be produced by gluon fusion, $gg\to gA$ and $gg\to AA$, the processes that appear at the disk level as a result of tree level couplings of gauge bosons to massive Regge excitations \refs{\AnchordoquiDA,\AnchordoquiAC}.
It is convenient to relax the normalization constraint on the $U(1)$ generator:
\eqn\phgen{T^{a_4}=Q_AI\!\!I_N}
where $I\!\!I_N$ is the $N{\times}N$ identity matrix and $Q_A$ is an arbitrary charge.
In our conventions, the standard normalization corresponds to $Q_A=1/\sqrt{2N}$.
Then
\eqn\phd{d^{a_1a_2a_3a_4}=Q_A \,d^{a_1a_2a_3}}
and all non-Abelian structure constants drop out from Eq. \mhva.
The corresponding helici\-ty amplitudes can be obtained from the four-gluon
amplitudes
by the respective replacement of the color factors in Eqs. \mhva\ {\it etc}.
In this way, the averaged squared amplitudes become
\eqn\mhvph{|{\cal M}(gg\to
 gA)|^2~=~4g^4{(N^2-4)Q_A^2\over N(N^2-1)}\,\bigg({1\over s^2}+{1\over t^2}+{1\over u^2}\bigg)\left(\, sV_s
+ tV_t+uV_u\,\right)^{2}}
%where
%\eqn\cbbs{C(N)={4(N^2-4)\over N(N^2-1)}.}
%Similarly,
\eqn\mmhvph{|{\cal M}(gg\to
 AA)|^2~=~16g^4{Q_A^4\over N^2-1}\bigg({1\over s^2}+{1\over t^2}+{1\over u^2}\bigg)\left(\, sV_s
+ tV_t+uV_u\,\right)^{2}.}
In the low-energy limit, the Abelian gauge boson production rates \mhvph\ and \mmhvph\ are of order
${\cal O}(M^{-8})$ compared to the gluon production.
However, they can be larger than QCD rates in the string resonance region \refs{\AnchordoquiDA,\AnchordoquiAC}.

\subsec{$gg\to q\bar q,\ gq\to gq,\ gq\to qA,\ gq\to qB,\ q\bar q\to gg,\
q\bar q\to gA,\ q\bar q\to gB$}

All non-vanishing helicity amplitudes involving two quarks and two gauge
bosons can be obtained by appropriate crossing operations
from Eqs. \aggff\ and \aggfg.
The squared moduli of these amplitudes, {\it summed\/} over initial and final gauge indices, read:
\eqn\mffsqu{\eqalign{|{\cal M}(g^-_1, g^+_2, q^-_3,\bar q^+_4 )|^2=4\,g_a^4N_b\Bigg[ &\sum_{a_1,a_2}\Tr(T^{a_1}T^{a_1}T^{a_2}T^{a_2}){t\over us^2}(tV_t+uV_u)^2\cr & -
{1\over 2}\!\sum_{a_1,a_2,i}f^{a_1a_2i}f^{a_1a_2i}{t^2\over s^2}V_tV_u\Bigg],
}}
\eqn\mfbsq{|{\cal M}(g^-_1, B^+_2, q^-_3,\bar q^+_4 )|^2=4\,g_a^2g_b^2
\sum_{a,b}\Tr(T^{a}T^{a})\Tr(T^{b}T^{b})\,{t\over u}V_s^2.}
The above expression are written in a form suitable for non-Abelian as well as Abelian gauge bosons. In the latter case, the second term drops out from Eq. \mffsqu.

As an example, consider the gluon fusion  $gg\to q\bar q$. In this case, the following identity is used for summing over the color indices:
\eqn\cf{\sum_a T^aT^a={N^2-1\over 2N}I\!\!I_N\ .}
This process takes place entirely on the QCD stack $a$ while stack $b$ is a spectator and its only effect is to supply the overall factor $N_b$ in Eq. \mffsqu. Note that summing over quark helicities requires some attention because left- and right-handed quarks originate from different stacks. We will handle this by adding contributions from both stacks, with the net result of doubling the square of the chiral amplitude \mffsqu\ and replacing $N_b$ by the number of flavors $N_f$.\foot{Note that summing over vector boson helicities amounts to adding $(t\leftrightarrow u)$ terms to the r.h.s.\ of Eqs. \mffsqu\ and \mfbsq.} We will also apply this  procedure to other channels of the same reaction.
The squared modulus of the corresponding amplitude, summed over final polarizations and colors, and averaged over
all $4(N^2-1)^2$ initial polarization/color configurations, is
\eqn\fmsq{|{\cal M}(gg\to q\bar q)|^2=g^4{N_f\over 2N}\Bigg[ {t^2+u^2\over uts^2}(tV_t+uV_u)^2
-{2N^2\over (N^2-1)}{t^2+u^2\over s^2}V_tV_u\Bigg]\ ,}
%where
%\eqn\efss{E(N)={1\over 2N_a}~,\qquad F(N)=-{N_a\over N_a^2-1}\ .}

The hadroproduction of $B$ vector bosons from a non-QCD stack $b$ involves at least one incoming quark or antiquark.
We average over $N_b$ species of each of them.
We also sum over all $N_b^2{-}1$ $SU(N_b)$ $B$-bosons;
depending on the model, we can always add the $U(1)_b$ boson by hand.
Since $B$-bosons couple to chiral quarks, we do {\it not\/} add initial quarks
of opposite helicity because they are coupled to other stacks. Thus in order
to average the $B$ production rates over incident helicities, we simply divide
by the number of available initial helicity configuration.
All  amplitudes obtained by using Eqs. \mffsqu\ and \mfbsq\ are collected in Tables 5-8.

\subsec{$qq\to qq,\ q\bar q\to q\bar q$}

These amplitudes are more complicated for several reasons. Their computations are sensitive to left-right asymmetry of the SM, {\it i.e}.\ to the fact that different helicity states come in different gauge group representations, originating from strings stretching between distinct stacks of D-branes. Furthermore, by construction,  the intermediate channels of quark scattering processes include all $N^2$ gauge bosons of each $U(N)$, therefore $SU(N)$ gauge bosons, as well as their string and KK excitations, must be separated ``by hand'' from their $U(1)$ counterparts. Whenever this problem is encountered, we will implement the following identity on the group factors:
\eqn\chpat{\delta^{{\alpha}_1}_{\alpha_2}\delta^{\alpha_3}_{\alpha_4}
=2\sum_a
(T^{a})^{\alpha_1}_{\alpha_4}
(T^{a})^{\alpha_3}_{\alpha_2}+2\,Q^2_A\delta^{\alpha_1}_{\alpha_4}
\delta^{\alpha_3}_{\alpha_2} \ ,}
where the sum is over all $SU(N)$ generators of $N$-color QCD and $Q_A=1/\sqrt{2N}$.
Note that due to Fierz identity,
\eqn\fierzz{\langle 13\rangle [24]={1\over 2}\sum_{\mu}(u_1\sigma_{\mu}\bar u_2)(\bar u_4\bar\sigma^{\mu}u_3)=-{1\over 2}
\sum_{\mu}(u_1\sigma_{\mu}\bar u_4)(\bar u_2\bar\sigma^{\mu}u_3),}
the factor $\langle 13\rangle[24]$ can be interpreted as arising from the exchange of intermediate vector bosons in either $s$ or $u$ channels, depending on the nature of  accompanying kinematic singularity.
%Up to complex conjugation and crossing, there are six independent helicity amplitudes:
%$\Mc[q^\mp,\bar q^\pm,q^\mp,\bar q^\pm]$, $\Mc[q^\mp,\bar q^\pm,q^\pm,\bar q^\mp]$ and
%$\Mc[q^\mp,\bar q^\mp,q^\pm,\bar q^\pm]$.

If the amplitude involves left-handed quarks and right-handed antiquarks only, $q^-$ and $\bar q^+$, respectively, then all fermions come from one intersection, say of stack $a$ and stack $b$. The corresponding amplitude reads:
\eqn\gfff{\eqalign{\Mc[\,  q^{\alpha_1}_{\beta_1}(&1-)\, \bar q^{\beta_2}_{\alpha_2}(2+)
q^{\alpha_3}_{\beta_3}(3-)\,\bar q^{\beta_4}_{\alpha_4}(4+)
\,] ~=~ 4\,g_a^2\langle 13\rangle[24]\cr  \times &\Bigg({F_{su}\over s}
\sum_{a,b}\Big[
(T^{a})^{\alpha_1}_{\alpha_2}
(T^{a})^{\alpha_3}_{\alpha_4}+Q^2_{A}\delta^{\alpha_1}_{\alpha_2}
\delta^{\alpha_3}_{\alpha_4}\Big]{\times}
\Big[
(T^{b})^{\beta_4}_{\beta_1}
(T^{b})^{\beta_2}_{\beta_3}+Q^2_{B}\delta^{\beta_4}_{\beta_1}
\delta^{\beta_2}_{\beta_3}\Big]
\cr &+
{F_{us}\over u}
\sum_{a,b}\Big[
(T^{a})^{\alpha_1}_{\alpha_4}
(T^{a})^{\alpha_3}_{\alpha_2}+Q^2_{A}\delta^{\alpha_1}_{\alpha_4}
\delta^{\alpha_3}_{\alpha_2}\Big]{\times}
\Big[
(T^{b})^{\beta_2}_{\beta_1}
(T^{b})^{\beta_4}_{\beta_3}+Q^2_{B}\delta^{\beta_2}_{\beta_1}
\delta^{\beta_4}_{\beta_3}\Big]
\Bigg),
}}
where the functions $F_{su}=F(s,u)$ and $F_{us}=F(u,s)$ are defined in Eq. \otha.
The most important difference between the above amplitude and the amplitudes involving gauge bosons is that its  intermediate channels include not only massless particles and their string (Regge) excitations, but also KK excitations and winding modes associated to the extra dimensions spanned by intersecting D-branes. Even in the limit $M\to \infty$, the function $F(s,u)$ contains, in addition to the poles due to intermediate gauge bosons, an infinite number of poles associated to such massive particles. In fact, $F(s,u)$ encompasses the effects of gauge bosons from both stacks $a$ and $b$, as reflected by the residues of its massless poles,
\eqn\bres{(s\to 0,~u\to 0,~t\to 0):~ F(s,u)\approx 1+{g_b^2\,s\over g_a^2\,u},}
and of all their excitations.

In order to explain how Eqs. \chpat\ and \fierzz\ are useful for the interpretation of
kinematic poles, let us extract from the amplitude \gfff\ the singularities associated to intermediate gluons, coming from the limit $g_b\to 0$ in Eq. \bres, in which the strength of other interactions is negligible:
\eqn\bbres{(s\to 0,~u\to 0,~t\to 0):~ F_{su}\approx F_{us}\approx 1\ .}
To be precise, it is the $M\to\infty$ (string zero slope) limit, with the additional assumption that $g_b\ll g_a=g$.
Then stack $b$ is a spectator, therefore we should use Eq. \chpat\ to revert the factors involving $G_b$ generators back to their original form. Furthermore, we rewrite the kinematic factor by using Eq. \fierzz\ in order to exhibit a $s$-channel vector exchange in the first term of Eq. \gfff\ and a $u$-channel vector boson exchange in the second term. As a result, the amplitude becomes
\eqn\ggfff{\eqalign{\Mc[\,  q^{\alpha_1}_{\beta_1}(&1-)\, \bar q^{\beta_2}_{\alpha_2}(2+)
q^{\alpha_3}_{\beta_3}(3-)\,\bar q^{\beta_4}_{\alpha_4}(4+)
\,] ~\to~ \cr &
{g^2\over s}\sum_{\mu,a}(u_1\sigma_{\mu}\bar u_2)(\bar u_4\bar\sigma^{\mu}u_3)
\Big[
(T^{a})^{\alpha_1}_{\alpha_2}
(T^{a})^{\alpha_3}_{\alpha_4}+Q^2_{A}\delta^{\alpha_1}_{\alpha_2}
\delta^{\alpha_3}_{\alpha_4}\Big]\,
\delta^{\beta_2}_{\beta_1}
\delta^{\beta_4}_{\beta_3}\cr -&\,{g^2\over u}\sum_{\mu,a}(u_1\sigma_{\mu}\bar u_4)(\bar u_2\bar\sigma^{\mu}u_3)
\Big[
(T^{a})^{\alpha_1}_{\alpha_4}
(T^{a})^{\alpha_3}_{\alpha_2}+Q^2_{A}\delta^{\alpha_1}_{\alpha_4}
\delta^{\alpha_3}_{\alpha_2}\Big]\,
\delta^{\beta_4}_{\beta_1}
\delta^{\beta_2}_{\beta_3}
}}
which does indeed reproduce the well-known QCD result after setting $Q_A=0$, {\it i.e.\/} subtracting the unwanted contribution of the  $U(1)$ gauge boson $A$.

The squared modulus of the amplitude \gfff, {\it summed\/} over initial and final gauge indices, is
\eqn\mfff{\eqalign{|{\cal M}( q^-_1, \bar q^+_2,q^-_3,\bar q^+_4)|^2 ~=~ &\ g^4K(N_a)K(N_b) \bigg({t^2F_{su}^2\over s^2}+{t^2F_{us}^2\over u^2}\bigg)\cr &+
2\,g^4\,L(N_a)L(N_b)\,{t^2 \over su}\,F_{su}F_{us} }}
where
\eqn\kks{\eqalign{K(N)&=N^2-1+4Q^4N^2\cr
L(N)&={(1-N^2)\over N}+4Q^2(N^2-1)+4Q^4N}}
The charges $Q$ should be adjusted in certain regions of parameter space
and/or kinematic limits. For instance, in the QCD limit \bbres, with $g_b\ll
g$, the appropriate choice is $Q_{A}=0$ [$U(1)$ component eliminated] 
and $Q_{B}=1/\sqrt{2N_b}$ (stack $b$ treated as spectator). Note that $N_a=N$
for $N$-color QCD and $N_b=2$ for one [electroweak $SU(2)$] quark doublet. Thus
\eqn\qcdlim{K(N_a)K(N_b)=(N^2-1)N_b^2  ~,\qquad L(N_a)L(N_b)=-{(N^2-1)\over N}N_b ~,}
and Eq. \mfff\ becomes
\eqn\mfqcd{|{\cal M}( q^-_1, \bar q^+_2,q^-_3,\bar q^+_4)|^2 
\buildrel\hbox{\fiverm QCD}\over\longrightarrow g^4\bigg[(N^2-1)N^2_b 
\Big( {t^2\over s^2}+{t^2\over u^2}\Big)-2{(N^2-1)\over N}N_b\,{t^2 \over su }\bigg],}
which could be obtained directly from Eq. \ggfff. In some cases however, it is
more appropriate to consider the cases of identical and different flavors
[$SU(2)$ doublet components] separately, instead of summing over them. Then
Eqs. \gfff\ and \mfff\ can be easily disentangled into the contributions  
describing $(u\bar u u \bar u),~(d\bar d d\bar d),~(u\bar u u\bar u)$ 
and $(d\bar d d\bar d)$.

There are five more helicity configuration that remain to be included in
unpolarized cross sections. The amplitude with the helicity assignments
reversed with respect to \gfff\ $(+\leftrightarrow -)$ is very similar because
it involves right-handed quarks and left-handed antiquarks originating from
one intersection (of the QCD stack $a$ with one of $U(1)$ stacks, $c$ or $d$), 
provided that all quarks are of the same flavor. For each flavor, one obtains
the same result as on the r.h.s.\ of Eq. \mfff, with $N_a=N$ and $N_b=1$,
although the function $F$ is associated now to a different 
intersection\foot{It is defined as in Eq. \otha, but starting from $V_{acac}$
or $V_{adad}$.}. 
If the flavors are different, then the amplitude falls into the category
discussed below, because it couples the QCD stack to two other stacks and the
disk boundary connects three different stacks. Then the function $F$ has to be 
replaced by $G$ defined in Eq. \othb.

The four remaining helicity configurations fall into one class. They involve  
$SU(2)$ doublets and singlets at the same time, therefore they mix the QCD
stack with two other stacks: $SU(2)$ stack $b$ and one of $U(1)$ stacks, say
$c$. The corresponding helicity amplitudes contain massless poles in only one
channel, due to intermediate gluons and the $A$-boson. They are:
\eqn\gabc{\eqalign{\Mc[\,  q^{\alpha_1}_{\beta_1}(1-)\, \bar q^{\beta_2}_{\alpha_2}(2+)
q^{\alpha_3}_{\gamma_3}(3+)\,\bar q^{\gamma_4}_{\alpha_4}(4-)
\,] &=~ 2\ g_a^2\ 
\delta^{\beta_2}_{\beta_1}\delta^{\gamma_4}_{\gamma_3}{\langle 14\rangle [23]}
\cr 
& \times {G'_{su}\over s}
\sum_{a}
\Big[(T^{a})^{\alpha_1}_{\alpha_2}
(T^{a})^{\alpha_3}_{\alpha_4}+Q^2_{A}\delta^{\alpha_1}_{\alpha_2}
\delta^{\alpha_3}_{\alpha_4}\Big]}}
\eqn\gaabc{\eqalign{\Mc[\,  q^{\alpha_1}_{\beta_1}(1-)\, \bar q^{\gamma_2}_{\alpha_2}(2-)
q^{\alpha_3}_{\gamma_3}(3+)\,\bar q^{\beta_4}_{\alpha_4}(4+)
\,] &=~2\,g_a^2\,\delta^{\beta_4}_{\beta_1}\delta^{\gamma_2}_{\gamma_3}{\langle
12\rangle [34]} \cr 
& \times {G'_{us}\over u}\sum_{a}
\Big[(T^{a})^{\alpha_1}_{\alpha_4}
(T^{a})^{\alpha_3}_{\alpha_2}+Q^2_{A}\delta^{\alpha_1}_{\alpha_4}
\delta^{\alpha_3}_{\alpha_2}\Big]~,}}
and the two amplitudes describing the helicity configurations reversed by 
$(+\leftrightarrow -)$. These can be obtained from the above by the
permutation $(1\leftrightarrow 3~,2\leftrightarrow 4)$, with the net effect of 
complex conjugation. The functions $G'_{su}=G'(s,u)$ and $G'_{us}=G'(u,s)$ are
defined in Eq. \othc. Recall that their low-energy expansions have the form 
$G(s,u)\approx G'(s,u)\approx 1+{\cal O}(M^{-2})$.

The squared moduli of the amplitudes \gabc\ and \gaabc, {\it summed\/} over initial and final gauge indices are, respectively:
\eqn\mabb{|{\cal M}( q^-_1, \bar q^+_2,q^+_3,\bar q^-_4)|^2 ~=~ g^4\,K(N_a)N_b\,{u^2\over s^2}\,G_{su}^{\prime 2}\ ,}
\eqn\mbba{|{\cal M}( q^-_1, \bar q^-_2,q^+_3,\bar q^+_4)|^2 ~=~ g^4\, K(N_a)N_b\,{s^2\over u^2}\,G_{us}^{\prime 2}\ .}
where we set $g_a=g$ and $N_c=1$. The factor $N_b=2$ takes combines the cases of same and different components (flavors) of the $SU(2)$ doublet which can be easily disentangled if flavor summation or averaging is not desirable.
We should also set
\eqn\sett{K(N_a)=N^2-1}
in order to eliminate the contributions of intermediate color singlets.

At this point, we have all ingredients at hand, ready for writing down the
squared amplitudes for quark-quark scattering and quark-antiquark annihilation, averaged over the polarizations, colors  of the incident particles, and summed over
the polarizations, colors  of the outgoing quarks and antiquarks. We will consider the cases of identical and different flavors separately. The sum over helicity configurations combines disk diagrams with various stack configurations along the boundary, with the QCD stack $a$ repeating two times and the two other being either $b$ [electroweak $SU(2)$], $c$ (right-handed $u$ quark)
or $c'$ (right-handed $d$ quark). We distinguish functions $F$ associated to disk diagram with two $b$ stacks, two $c$ stacks or two $c'$ stacks as $F^{bb}$, $F^{cc}$ and $F^{c'c'}$, respectively, see Eq. \otha. Similarly, $G$ and $G'$ need indication of two non-QCD stacks. Thus we define $G^{cc'}$, by Eq. \othb\ with $V_{acac'}$ {\it etc}.

By adding all helicity configurations contributing to unpolarized quark-quark scattering, we obtain the following squared amplitudes:
\eqn\newqq{\eqalign{|{\cal M}( qq\to qq)|^2 &= g^4\Big({N^2-1\over 4N^2}\Big){1\over t^2}\Big[\big(sF^{bb}_{tu}\big)^2  +\big(sF^{cc}_{tu}\big)^2 +\big(uG^{\prime bc}_{tu}\big)^2  +\big(uG^{\prime cb}_{tu}\big)^2     \Big] \cr & +g^4
\Big({N^2-1\over 4N^2}\Big){1\over u^2}\Big[\big(sF^{bb}_{ut}\big)^2  +\big(sF^{cc}_{ut}\big)^2 +\big(tG^{\prime bc}_{ut}\big)^2  +\big(tG^{\prime cb}_{ut}\big)^2\Big]\cr &
  -g^4\Big({N^2-1\over 2N^3}\Big) {s^2\over tu}\big(F^{bb}_{tu} F^{bb}_{ut} +F^{cc}_{tu} F^{cc}_{ut}\big)   }}
for identical flavors and
\eqn\newqqpr{|{\cal M}( qq'\to qq')|^2 ~=~ g^4
\Big({N^2-1\over 4N^2}\Big){1\over t^2}\Big[\big(sF^{bb}_{tu}\big)^2  +\big(s\,G^{cc'}_{tu}\big)^2 +\big(uG^{\prime bc}_{tu}\big)^2  +\big(uG^{\prime bc'}_{tu}\big)^2     \Big]}
for different flavors.
Similarly, for quark-antiquark annihilation:
\eqn\manfav{|{\cal M}( q\bar q\to q\bar q)|^2 ~=~ |{\cal M}( qq\to qq)|^2(s\to u,~u\to t,~t\to s)}
\eqn\mann{|{\cal M}( q\bar q'\to q\bar q')|^2 ~=~ |{\cal M}( qq'\to qq')|^2(s\leftrightarrow u)}
\eqn\manfn{|{\cal M}( q\bar q\to q'\bar q')|^2 ~=~ |{\cal M}( q q'\to q q')|^2 (s\to u,~u\to t,~t\to s) .}

\subsec{$q\bar q\to l\bar l$}

The disk amplitudes involving four D-brane stacks do not contribute to $2\to 2$
parton scattering processes, at least in the simplest realizations of intersecting
D-brane scenarios. They are relevant though to the Drell-Yan process
$q\bar q\to l\bar l$. The relevant amplitude is
\eqn\drell{\Mc[\,  q^{\alpha_1}_{\beta_1}(1-)\bar q^{\gamma_2}_{\alpha_2}(2-)\,
l^{\delta_3}_{\gamma_3}(3+)
\bar l^{\beta_4}_{\delta_4}(4+)\,]~=~ \delta^{\alpha_1}_{\alpha_2}
\delta^{\beta_4}_{\beta_1}\delta^{\gamma_2}_{\gamma_3}\delta^{\delta_3}_{\delta_4}\,
\langle 12\rangle[34]\,V_{abcd}(s,u)\ ,}
with the function $V_{abcd}(s,u)$ is defined in Eq. \tomiii.
The low-energy expansion of this amplitude is free of kinematic singularities
and begins at the order ${\cal O}(M^{-2})$.
For this process, there are also helicity amplitudes receiving
contributions from three stacks, as already discussed in the context of quark-quark
scattering:
\eqn\drella{\eqalign{\Mc[\,  q^{\alpha_1}_{\beta_1}(1-)\, \bar q^{\beta_2}_{\alpha_2}(2+)
l^{\gamma_3}_{\beta_3}(3-)&\,\bar l^{\beta_4}_{\gamma_4}(4+)
\,] ~=~ 
2\ \delta^{\alpha_1}_{\alpha_2}\delta^{\gamma_3}_{\gamma_4}\langle 13\rangle [24] \cr
& \times V_{babc}(s,u)\sum_{b}\Big[(T^{b})^{\beta_1}_{\beta_2}
(T^{b})^{\beta_3}_{\beta_4}+Q^2_{B}\delta^{\beta_1}_{\beta_2}
\delta^{\beta_3}_{\beta_4}\Big]\ .}}
It is clear from the above expressions, especially from Eq. \drella\ which
includes all gauge bosons exchanged in the $s$-channel, that the amplitudes
involving leptons are sensitive to the implementation of electro-weak symmetry
breaking mechanism in string theory. Since it is a model-dependent problem, we
stop short from computing the
production rates (averaged squared moduli) for such processes.

\vfill\break

\subsec{Tables}

In the Tables below, we collect the squared amplitudes for all parton subprocesses discussed in this Section, summed over the polarizations and colors of final particles and averaged over the polarizations and colors of incident partons. The number of colors has been set  to $N=3$. Recall that $A$ denotes the $U(1)$ gauge boson from the QCD stack, {\it i.e.} the ``quiver neighbor'' of $SU(3)$ gluons. The corresponding coupling $Q_A=1/\sqrt{6}$ is displayed explicitly. Furthermore, $B$ is a generic (massless) gauge boson from another stack, for example a $SU(2)$ boson. We assumed that $B$ couples to left-handed quarks only; the generalization to a left-right symmetric vector coupling is straightforward. We factored out the QCD coupling factor $g^4$. Thus in the amplitudes involving $B$ vector bosons, marked with $(*)$, this factor should be corrected to $g^2g_B^2$, where $g_B$ denotes the coupling of $B$ gauge group. In these amplitudes, $T^B_{q\bar q'}$ denotes the $(q\,\bar q')$
matrix element of the corresponding group generator. The $SU(3)\times SU(2)\times U(1)$ SM limit of the amplitudes, with $V=F=G=G'=1$ as $s\ll M^2$, is in agreement with Table 9.1 of \barger.
\vskip 1cm

{\centerline{\noindent{\bf Table 5: }{\it Gluon fusion processes.}}
\vbox{\ninepoint{$$
\vbox{\offinterlineskip\tabskip=0pt
\halign{\strut\vrule#
%%%%%%%%%%%%%%%%%%
&~$#$~\hfil
&\vrule$#$
&~$#$~\hfil
&\vrule$#$\cr
\noalign{\hrule}
%%%%%%%%%%%%%%%%%%
& && &\cr
&{\rm subprocess}  &&\hskip4.5cm|\Mc|^2/g^4 & \cr
& && &\cr
\noalign{\hrule}\noalign{\hrule}
%%%%%%%%%%%%%%%%%%%%%%%%%%%%%%%%%%%%%%%%%%%%%%%%
& && &\cr
& gg\to gg &&
\displaystyle\Big({1\over s^2}+{1\over t^2}+{1\over u^2}\Big)
\bigg[\,{9\over 4}\!\left(\, s^2V^2_s+ t^2V^2_t
+u^2V^2_u\,\right) - {1\over 3}\!\left(\, sV_s
+ tV_t+uV_u\,\right)^{2}\bigg]        &\cr
& && &\cr
\noalign{\hrule}
%%%%%%%%%%%%%%%%%%%%%%%%%%%%%%%%%%%%%%%%%%%%%%%%%
& && &\cr
&  gg\to gA &&\displaystyle  {5\over 6}Q_A^2\Big({1\over s^2}+{1\over t^2}+{1\over
u^2}\Big)\left(\, sV_s+ tV_t+uV_u\,\right)^{2}         &\cr
& && &\cr
\noalign{\hrule}
%%%%%%%%%%%%%%%%%%%%%%%%%%%%%%%%%%%%%%%%%%%%%%%%%
& && &\cr
&gg\to AA   &&\displaystyle
2\,Q_A^4\Big({1\over s^2}+{1\over t^2}+{1\over u^2}\Big)\left(\, sV_s
+ tV_t+uV_u\,\right)^{2}     &\cr
& && &\cr
\noalign{\hrule}
%%%%%%%%%%%%%%%%%%%%%%%%%%%%%%%%%%%%%%%%%%%%%%%%%
& && &\cr
&gg\to q\bar q&&\displaystyle
{t^2+u^2\over s^2}\bigg[\, {1\over 6}{1\over ut}(tV_t+uV_u)^2
-{3\over 8}V_tV_u\bigg]          &\cr
& && &\cr
%%%%%%%%%%%%%%%%%%%%%%%%%%%%%%%%%%%%%%%%%%%%%%
\noalign{\hrule}}}$$
\vskip-6pt
\vskip10pt}}}
\noindent
%%%%%%%%%%%%%%%%%%%%%%%%%%%%%%%%%%%%%%%%%%

\vbox{{\centerline{\noindent{\bf Table 6: }{\it Gluon-quark scattering.}}
\ninepoint{$$
\vbox{\offinterlineskip\tabskip=0pt
\halign{\strut\vrule#
%%%%%%%%%%%%%%%%%%
&~$#$~\hfil
&\vrule$#$
&~$#$~\hfil
&\vrule$#$\cr
\noalign{\hrule}
%%%%%%%%%%%%%%%%%%
& && &\cr
&{\rm subprocess}  &&\hskip1.75cm|\Mc|^2/g^4 & \cr
& && &\cr
\noalign{\hrule}\noalign{\hrule}
%%%%%%%%%%%%%%%%%%%%%%%%%%%%%%%%%%%%%%%%%%%%%%%%
& && &\cr
& gq\to gq && \displaystyle
{s^2+u^2\over t^2}\bigg[ V_sV_u-{4\over 9}{1\over su}(sV_s+uV_u)^2
\bigg]     &\cr
& && &\cr
\noalign{\hrule}
%%%%%%%%%%%%%%%%%%%%%%%%%%%%%%%%%%%%%%%%%%%%%%%%%
& && &\cr
&  \displaystyle{gq\to Aq} &&
\displaystyle{-{1\over 3}Q^2_A{s^2+u^2\over sut^2}(sV_s+uV_u)^2}   &\cr
& && &\cr
\noalign{\hrule}
%%%%%%%%%%%%%%%%%%%%%%%%%%%%%%%%%%%%%%%%%%%%%%%%%
& && &\cr
&gq\to Bq'  &&\displaystyle
-{1\over 6}|T^B_{q\bar q'}|^2{s^2+u^2\over su}V_t^2\qquad\qquad
(*)           &\cr
& && &\cr
%%%%%%%%%%%%%%%%%%%%%%%%%%%%%%%%%%%%%%%%%%%%%%
\noalign{\hrule}}}$$
\vskip-6pt
\vskip10pt}}}
\noindent
\vskip 1cm
%%%%%%%%%%%%%%%%%%%%%%%%%%%%%%%%%%%%%%%%%%
{\centerline{\noindent{\bf Table 7: }{\it Quark-quark scattering.}}
\vbox{\ninepoint{$$
\vbox{\offinterlineskip\tabskip=0pt
\halign{\strut\vrule#
%%%%%%%%%%%%%%%%%%
&~$#$~\hfil
&\vrule$#$
&~$#$~\hfil
&\vrule$#$\cr
\noalign{\hrule}
%%%%%%%%%%%%%%%%%%
& && &\cr
&{\rm subprocess}  &&\hskip5cm|\Mc|^2/g^4 & \cr
& && &\cr
\noalign{\hrule}\noalign{\hrule}
%%%%%%%%%%%%%%%%%%%%%%%%%%%%%%%%%%%%%%%%%%%%%%%%
& && &\cr
& qq\to qq&& \displaystyle
\eqalign{&{2\over 9}{1\over t^2}\Big[\big(sF^{bb}_{tu}\big)^2  +\big(sF^{cc}_{tu}\big)^2 +\big(uG^{\prime bc}_{tu}\big)^2  +\big(uG^{\prime cb}_{tu}\big)^2     \Big]  \cr +~
&{2\over 9}{1\over u^2}\Big[\big(sF^{bb}_{ut}\big)^2  +\big(sF^{cc}_{ut}\big)^2 +\big(tG^{\prime bc}_{ut}\big)^2  +\big(tG^{\prime cb}_{ut}\big)^2\Big]
  -{4\over 27} {s^2\over tu}\big(
F^{bb}_{tu} F^{bb}_{ut}+F^{cc}_{tu} F^{cc}_{ut}\big)}       &\cr
& && &\cr
\noalign{\hrule}
%%%%%%%%%%%%%%%%%%%%%%%%%%%%%%%%%%%%%%%%%%%%%%%%%
& && &\cr
&  \displaystyle{qq'\to qq'} && \displaystyle
{2\over 9}{1\over t^2}\Big[\big(sF^{bb}_{tu}\big)^2  +\big(s\,G^{cc'}_{tu}\big)^2 +\big(uG^{\prime bc}_{tu}\big)^2  +\big(uG^{\prime bc'}_{tu}\big)^2     \Big]&\cr
& && &\cr
\noalign{\hrule}}}$$
\vskip-6pt
\vskip10pt}}}
\noindent\vfill\break
{\centerline{\noindent{\bf Table 8: }{\it Quark-antiquark annihilation.}}\br
\vbox{\ninepoint{$$
\vbox{\offinterlineskip\tabskip=0pt
\halign{\strut\vrule#
%%%%%%%%%%%%%%%%%%
&~$#$~\hfil
&\vrule$#$
&~$#$~\hfil
&\vrule$#$\cr
\noalign{\hrule}
%%%%%%%%%%%%%%%%%%
& && &\cr
&{\rm subprocess}  &&\hskip3.5cm|\Mc|^2/g^4 & \cr
& && &\cr
\noalign{\hrule}\noalign{\hrule}
%%%%%%%%%%%%%%%%%%%%%%%%%%%%%%%%%%%%%%%%%%%%%%%%
& && &\cr
& q\bar  q\to g g &&
\displaystyle
{8\over 3}{t^2+u^2\over s^2}\Big[ {4\over 9}{1\over ut}
(tV_t+uV_u)^2-V_tV_u\Big]       &\cr
& && &\cr
\noalign{\hrule}
%%%%%%%%%%%%%%%%%%%%%%%%%%%%%%%%%%%%%%%%%%%%%%%%%
& && &\cr
&  q\bar q\to gA &&\displaystyle {8\over 9}Q^2_A{t^2+u^2\over tus^2}
(tV_t+uV_u)^2            &\cr
& && &\cr
\noalign{\hrule}
%%%%%%%%%%%%%%%%%%%%%%%%%%%%%%%%%%%%%%%%%%%%%%%%%
& && &\cr
&  q\bar q\to AA &&\displaystyle
{2\over 3}Q^4_A{t^2+u^2\over tus^2}(tV_t+uV_u)^2            &\cr
& && &\cr
\noalign{\hrule}
%%%%%%%%%%%%%%%%%%%%%%%%%%%%%%%%%%%%%%%%%%%%%%%%%
& && &\cr
&q\bar q'\to gB&&\displaystyle {4\over 9}|T^B_{q\bar q'}|^2{t^2+u^2\over tu}V_s^2\qquad\qquad\quad~ (*)         &\cr
& && &\cr
%%%%%%%%%%%%%%%%%%%%%%%%%%%%%%%%%%%%%%%%%%%%%%
\noalign{\hrule}
%%%%%%%%%%%%%%%%%%%%%%%%%%%%%%%%%%%%%%%%%%%%%%%%%
& && &\cr
&  q\bar q'\to BA&&\displaystyle
{1\over 3}|T^B_{q\bar q'}|^2 Q^2_A{t^2+u^2\over tu}V_s^2 \qquad\qquad (*)          &\cr
& && &\cr
\noalign{\hrule}
%%%%%%%%%%%%%%%%%%%%%%%%%%%%%%%%%%%%%%%%%%%%%%%%%
& && &\cr
&  q\bar q\to q\bar q &&\displaystyle
|{\cal M}( q\bar q\to q\bar q)|^2 ~=~ |{\cal M}( qq\to qq)|^2(s\to u,~u\to
t,~t\to s)          &\cr
& && &\cr
\noalign{\hrule}
%%%%%%%%%%%%%%%%%%%%%%%%%%%%%%%%%%%%%%%%%%%%%%%%%
& && &\cr
&  q\bar q'\to q\bar q'&&\displaystyle
|{\cal M}( q\bar q'\to q\bar q')|^2 ~=~ |{\cal M}( qq'\to
qq')|^2(s\leftrightarrow u) &\cr
& &&  &\cr
\noalign{\hrule}
%%%%%%%%%%%%%%%%%%%%%%%%%%%%%%%%%%%%%%%%%%%%%%%%%
& &&  &\cr
&  q\bar q\to q'\bar q' &&\displaystyle
|{\cal M}( q\bar q\to q'\bar q')|^2 ~=~ |{\cal M}( q q'\to q q')|^2
(s\to u,~u\to t,~t\to s)            &\cr
& && &\cr
\noalign{\hrule}}}$$
\vskip-6pt
\vskip10pt}}}
\noindent
\vfill\break

\newsec{Summary}

This article is intended to provide some information useful in the upcoming searches
for the signals of string physics at the LHC, assuming that the fundamental scale
determining the masses of string excitations is of order few TeVs. While on the
theoretical side,  low mass scenarios face many challenges, it is an experimental
question whether string theory describes the physics beyond the SM.
Needless to say, since low mass strings require the existence of large extra
dimensions, the discovery of fundamental strings at the LHC would revolutionize our
understanding of space and time.

The search for string signals should focus on Regge excitations \ie the resonances
created by vibrating strings. The main message of this work is that string theory
provides very clear, model-independent, universal predictions not only for the masses
and spins of these particles\foot{The decay widths of lowest Regge recurrences have
been recently computed in Ref. \AnchordoquiHI.}, but also for their couplings to
gluons and quarks.
These predictions do not depend on details of compactification, D--brane configurations
and hold even if supersymmetry is broken in four dimensions.
The reason why certain amplitudes are universal, independent of the spectrum of
Kaluza--Klein excitations is very simple. At the disk level, the gluon scattering
amplitudes involve only one stack of D-branes, thus the momentum components  along
the compactified D-brane directions are conserved and as a consequence, Kaluza--Klein 
states carrying such momenta cannot appear as intermediate states. From all $2\to 2$
parton scattering amplitudes, only four-fermion processes are model-dependent, but
these are suppressed by group-theoretical factors and usually occur at luminosities
lower than gluon collisions. The model-dependence of these amplitudes,
and also the necessity to avoid FCNC's or proton decay
via four-fermion amplitudes, could be useful for some ``precision tests''
that would distinguish between various compactification scenarios.

The resonant character of parton cross sections should not be difficult to observe.
In Refs. \AnchordoquiDA\ and \AnchordoquiAC, the process $gg\to g\gamma$, which is
absent in the SM at the tree-level, but
appears in string theory as a QCD process involving strongly interacting resonances,
has been examined to that effect. It turns out that a string mass as high as 3 TeV is
observable in this process. More recently \aglnst, we examined the dijet invariant
mass spectrum which is sensitive to even higher mass scales.
The resonant behavior of stringy cross section at the
parton center of mass
energies equal to the masses of Regge states is a signal that cannot be missed at the LHC,
unless the string scale is too high or the theory does not describe  the
physics beyond the SM correctly.

We have computed the full--fledged string four-particle scattering amplitudes for
the SM fields, as they occur at the (leading) disk level in a large
class of orientifold compactifications
on an internal manifold with large volume and low string scale. The SM
fields arise in these models as open strings ending on a set
of intersecting D-branes.
Here are the basic characteristics of these amplitudes:

\vskip0.3cm\noindent
{\sl (i) Two gluon/two SM gauge boson processes:}
\vskip0.2cm

\noindent These amplitudes are given in terms of one kinematic function
$V(s,t,u)$, given in \vexpdef,
and can be computed in a completely model independent way. The poles
of $V(s,t,u)$ are due to the exchange of massless SM gauge bosons and heavy string Regge
excitations. In some particular processes, like $gg\to gY$ or
$gg\to YY$ ($Y=\gamma,Z_0$),  the poles due to
massless gauge bosons are absent, and the leading contribution originates from
heavy string states.

\vskip0.3cm\noindent
{\sl (ii) Two SM gauge boson/two SM fermion processes:}
\vskip0.2cm

\noindent
As before these processes can be computed in a completely model independent
way and are given in terms of the same  function $V(s,t,u)$. Hence they receive contributions
from the exchange of SM gauge bosons and heavy string excitations.
We find that low scale string theory at the LHC leads to model
independent string contributions to
processes such as $q\bar q\rightarrow gW^\pm$ or $q\bar q\rightarrow gZ$, which should be
a clear signal for new physics.
Likewise, exchanges of Regge excitations contribute to
processes like $gq\rightarrow qW^\pm$ and $gq\rightarrow qZ$
in a model independent way.

\vskip0.3cm\noindent
{\sl (iii) Four SM fermion processes:}
\vskip0.2cm

\noindent
The four quark or two quark/two lepton amplitudes like
the Drell--Yan process $q\bar q\rightarrow l\bar l$ are  model-dependent and can be
expressed in terms of three functions, $V_{abab}(s,u),\ V_{abac}(s,u)$ and
$V_{abcd}(s,u)$, given in \tomi,\ \WITH\ and \tomiii, respectively.
In general, the latter depend on the string scale and on the parameters describing the
internal manifold
and the cycles around which D-branes are wrapped.
Here one finds poles not only due to exchanges of
SM gauge bosons and Regge excitations thereof, but also poles
due to internal Kaluza--Klein and winding modes, and open string states with masses
depending on the intersection angles.

\noindent
All parton subprocesses receive string contributions which should
be separable from the SM background if the string scale is
not too high.

The squared amplitudes, summed over the polarizations and colors of final
particles and averaged over the polarizations and colors of incident partons,
are collected in Tables 5--8.
They are presented in a form suitable for the computations of the respective
cross sections and are ready to be implemented in the LHC data analysis.

\vskip1cm
\goodbreak
\centerline{\noindent{\bf Acknowledgments} }\vskip 2mm

We wish to thank Michael Peskin and Andreas Ringwald
for inspiring discussions. We are grateful to Luis Anchordoqui, Haim Goldberg and
Satoshi Nawata for their collaboration on Ref. \aglnst.
We also thank Ignatios Antoniadis for his permission to use \anton.
Most of other diagrams have been created by the program JaxoDraw \Jaxo.
This work  is supported in part by the European Commission under
Project MRTN-CT-2004-005104.
St.St. thanks the Theory Division of CERN for
hospitality and financial support during preparation of this work.
The research of T.R.T.\ is supported in part by the U.S.
National Science Foundation Grants
PHY-0600304 and PHY-0757959, and by the Cluster of Excellence ``Origin and Structure of
the Universe'' in M\"unchen, Germany.  He is grateful
to Arnold Sommerfeld Center for Theoretical Physics at
Ludwig--Maximilians--Universit\"at, and to Max--Planck--Institut f\"ur Physik
in M\"unchen, for their kind hospitality. Any opinions, findings, and
conclusions or recommendations expressed in this material are those of
the authors and do not necessarily reflect the views of the National
Science Foundation.

\listrefs
\end